\documentclass[letterpaper,12pt]{article}
\pdfoutput=1

\usepackage{jheppub}
\usepackage{amsmath}
\usepackage{graphicx}
\usepackage{subfig}
\usepackage{cancel}
\usepackage[dvipsnames]{xcolor}
\usepackage{color}
\usepackage{mathrsfs}
\usepackage{enumitem}
\usepackage{amssymb}
\usepackage{epstopdf}
\def\({\left(}
\def\){\right)}
\def\[{\left[}
\def\]{\right]}

\newpage

\title{Bag-of-gold spacetimes, Euclidean wormholes, and inflation from domain walls in AdS/CFT}

\author{Zicao Fu}
\author{and Donald Marolf}

\affiliation{Department of Physics, University of California, Santa Barbara, CA 93106, USA}

\emailAdd{zicaofu@physics.ucsb.edu}
\emailAdd{marolf@physics.ucsb.edu}

\abstract{We use Euclidean path integrals to explore the set of bulk asymptotically AdS spacetimes with good CFT duals. We consider simple bottom-up models of bulk physics defined by Einstein-Hilbert gravity coupled to thin domain walls and restrict to solutions with spherical symmetry. The cosmological constant is allowed to change across the domain wall, modeling more complicated Einstein-scalar systems where the scalar potential has multiple minima. In particular, the cosmological constant can become positive in the interior. However, in the above context, we show that inflating bubbles are never produced by smooth Euclidean saddles to asymptotically AdS path integrals. The obstacle is a direct parallel to the well-known obstruction to creating inflating universes by tunneling from flat space. In contrast, we do find good saddles that create so-called ``bag-of-gold'' geometries which, in addition to their single asymptotic region, also have an additional large semi-classical region located behind both past and future event horizons. Furthermore, without fine-tuning model parameters, using multiple domain walls we find Euclidean geometries that create arbitrarily large bags-of-gold inside a black hole of fixed horizon size, and thus at fixed Bekenstein-Hawking entropy. Indeed, with our symmetries and in our class of models, such solutions provide the unique semi-classical saddle for appropriately designed (microcanonical) path integrals. This strengthens a classic tension between such spacetimes and the CFT density of states, similar to that in the black hole information problem.}

\begin{document}
\maketitle

\section{Introduction}
\label{Introduction}

Recent years have seen great progress in understanding the anti-de Sitter/conformal field theory (AdS/CFT) dictionary, especially in the limit of small perturbations about a given classical background \cite{Jafferis:2015del,Dong:2016eik,Faulkner:2017vdd}.  However, it remains to fully understand which set of classical backgrounds are in fact allowed by the AdS/CFT correspondence.  Said simply, does every Lorentz signature solution of the low energy bulk theory correspond to a well-defined state in the dual CFT?

\begin{figure}[t]
\centerline{
	\includegraphics[width=0.5\textwidth]{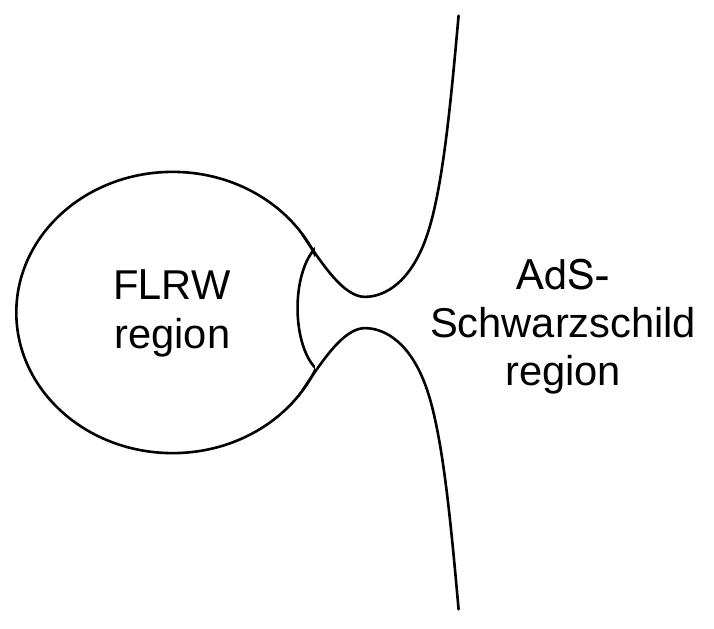}
	}
\caption{A moment of time symmetry in a bag-of-gold spacetime.  The region outside the minimal surface is precisely AdS-Schwarzschild, as is a small part of the interior.  The rest of the interior is a (say, radiation dominated) FLRW universe.}
\label{fig:bog}
\end{figure}

At least at first glance, the answer would appear to be negative.  Indeed, particular suspicion falls on so-called bag-of-gold spacetimes which consist of an eternal black hole exterior attached to an interior that is essentially an large Friedman-Lema\^itre-Robertson-Walker (FLRW) cosmology, which we take here to be filled with whatever bulk radiation is natural in the theory\footnote{Such spacetimes were introduced by Wheeler in \cite{Wheeler}.  That work also introduced term ``bag
of gold,'' but for a somewhat different purpose.  Over the years, the term has come to mean a spacetime of the form shown in figure \ref{fig:bog}.}; see figure \ref{fig:bog}.  Such solutions exist in essentially any theory of gravity, and the interior FLRW region can be arbitrarily large.  In particular, one can take the interior large enough that the entropy in its radiation exceeds the Bekenstein-Hawking entropy $S_{\rm BH}$ of the black hole.  As a result, the entropy also exceeds the density of states in the dual CFT.  This prohibits each microstate of the bulk ensemble from mapping to a linearly independent CFT state.

This issue has been discussed several times in the literature; see e.g. \cite{Marolf:2008tx,Hsu:2009kv,Almheiri:2018xdw} as well as the related discussion in \cite{Freivogel:2005qh} of large cosmological horizons inside bags-of-gold.  And it has long been of interest as a close analogue of the black hole information problem (see e.g. \cite{Harlow:2014yka,Marolf:2017jkr} for recent reviews).  But the full resolution of the problem remains to be understood.  In addition, the connection was recently reemphasized in \cite{Penington:2019npb,Almheiri:2019psf} where it was noted that both are associated with novel phase transitions of quantum extremal surfaces.

In addition to their overly large entropy, another reason to suspect that solutions with large bags-of-gold may lack CFT duals is that they cannot be constructed from the AdS vacuum by applying sources at the boundary and evolving forward in (Lorentz signature) time.  Indeed, at least with spherical symmetry the focusing theorem (see e.g. \cite{Wald:1984}) of general relativity prohibits the causal past of the interior from including any piece of the AdS boundary when the matter fields in the spacetime satisfy the null energy condition.  In contrast, bags-of-gold {\it can} be constructed from the AdS vacuum using time folded sources (this is identical to the construction of long wormholes in section 2.3 of \cite{Shenker:2013yza}) that add shock waves.  But each shock wave adds positive energy, and this energy is less than the temperature $T$ of the black hole one must worry that grey body factors will prohibit its full absorption.   So it is not clear that this can yield large bags-of-gold at fixed black hole mass, which is what we need to create tension with $S_{\rm BH}$.

It is thus natural to explore the construction of bags-of-gold via Euclidean path integrals.  In parallel with \cite{Maldacena:2001kr}, if we can find a Euclidean CFT path integral for which the dominant Euclidean saddle of the bulk dual Wick rotates to a large bag-of-gold, then slicing open the CFT path integral will yield a CFT state dual to this bag-of-gold.  Indeed, this basic argument has been used to construct black hole interiors in \cite{Krasnov:2000zq,Maldacena:2001kr,Krasnov:2003ye,Balasubramanian:2014hda,Maxfield:2014kra,Marolf:2015vma,Maxfield:2016mwh,Marolf:2017shp,Kourkoulou:2017zaj,Marolf:2017vsk}. However, the interior geometries in \cite{Kourkoulou:2017zaj} again have sizes bounded by a (in this case logarithmic) function of their total mass.  And while the other references above consider 2+1 dimensional solutions determined by a Riemann surface of genus $g$ and find interior geometries with size proportional to $g$, past works \cite{Maxfield:2016mwh,Marolf:2017shp,Marolf:2017vsk} have found that there are always lower action Euclidean saddles that lead to solutions with small genus and thus small interiors.  While apparently not discussed in the prior literature, at least in the case where the matter in the bag-of-gold is pressureless dust it is similarly straightforward to Wick rotate AdS versions of the original aribtrarily large FLRW bags-of-gold \cite{Wheeler} of figure \ref{fig:bog} to Euclidean signature.  But the dust case turns out to be somewhat degenerate, having a continuum of distinct bulk solutions (not all having bags-of-gold) with a given density of dust particles exiting through each point of a fixed Euclidean boundary.  While worth investigating further in the future, determining whether bags-of-gold dominate is thus non-trivial.

Below, we return to the basic issue of finding large Euclidean bags-of-gold that dominate path integrals by studying gravity coupled to thin domain walls.  For simplicity, we assume spherical symmetry throughout and also forbid domain wall intersections. This system models a more complete Einstein-scalar theory where the scalar potential has multiple minima in the limit where the domain walls become thin.  Since it is thus natural to allow the cosmological constant to change across the wall, this framework also provides an opportunity to study the possibility of inflating bubbles in AdS/CFT (see \cite{Alberghi:1999kd,Freivogel:2005qh} for studies of Lorentzian such solutions) and perhaps thus to better understand answers to fundamental questions \cite{Banks:2000fe,Bousso:2000nf,Banks:2001yp,Witten:2001kn,Fischler:2001yj,Hellerman:2001yi} concerning holography and inflation; see e.g. \cite{Strominger:2001pn,Strominger:2001gp,Alishahiha:2004md,Dong:2010pm,Freivogel:2006xu} for other approaches to this issue.  Some of our results overlap with those reported in \cite{Cooper:2018cmb,Antonini:2019qkt,deAlwis:2019dkc}, though the bags-of-gold discussed in these references are not large enough to create tensions with the Bekenstein-Hawking entropy.

In addition, as we will see, finding Euclidean solutions that create bags-of-gold is closely related to (but subtly different than) constructing Euclidean solutions with multiple disconnected boundaries known as Euclidean wormhole solutions.  In the context of Jackiw-Teitelboim gravity \cite{Jackiw:1984je,Teitelboim:1983ux} such solutions have recently played a key role in understanding features of quantum chaos and black holes \cite{Saad:2018bqo,Saad:2019lba}, though in general such solutions are associated with their own array of conceptual issues for AdS/CFT; see e.g. \cite{Maldacena:2004rf,Betzios:2019rds} as well as those that follow from \cite{Coleman:1988cy,Giddings:1988cx,Giddings:1988wv}.  Both features make them interesting objects of study in their own right.

The plan of this paper is as follows.  We begin in section \ref{TwoPiecesSpacetimes} with a brief review of the thin wall formalism, specializing to the case of spherically symmetric solutions with asymptotically locally AdS (AlAdS) boundary conditions and taking the opportunity to fix notation for later use.  We then investigate potentially inflating settings in section \ref{subsec:dS-SAdS} and show that such cases do not admit Euclidean solutions with AlAdS boundaries that are smooth apart from the domain walls\footnote{This part of the paper overlaps with the essentially simultaneous work \cite{deAlwis:2019dkc}.  However, that work takes a very different view of the singular/degenerate constructions of \cite{Farhi:1989yr,Fischler:1989se,Fischler:1990pk}.}.  As a result, they do not define bulk saddles for CFT path integrals.  We also address suggestions \cite{Farhi:1989yr,Fischler:1989se,Fischler:1990pk}  (see also \cite{Bachlechner:2016mtp,deAlwis:2019dkc}) that in similar contexts one should include certain singular saddles, provide counterarguments, and finally argue that allowing domain wall intersections would not change the conclusion.

The second part of this work then turns to the construction of bags-of-gold.  This begins with a general analysis in section \ref{subsec:AdS-SAdS} of when such solutions can arise.  We then study bags-of-gold in more detail in section \ref{sec:search}.   After first considering settings with a single domain wall, we progress to showing that Euclidean solutions with large numbers of domain walls can create correspondingly large bags-of-gold inside a black hole of fixed horizon size.  In particular, one can create bags-of-gold large inside a black hole with Bekenstein-Hawking entropy $S_{\rm BH}$ where the bag is large enough to hold entropy much greater than $S_{\rm BH}$.  Here we find it useful to add magnetic charge to the solutions, though the domain walls remain uncharged.  We also argue in our models that these are the only spherically symmetric bulk saddles for properly chosen ``microcanonical'' Euclidean path integrals similar to those used in \cite{Marolf:2018ldl}.  Since symmetry breaking is typically suppressed, we thus expect these saddles to dominate even if less symmetric saddles also exist.  Finally, we close with a discussion of open issues and future directions.

\section{Thin wall spacetimes in AdS}
\label{TwoPiecesSpacetimes}

Our spacetimes are spherically symmetric and vacuum except for a thin relativistic domain wall of tension $\frac{D-2}{8\pi G_D} \kappa \ge 0$ that separates two distinct vacua.  Here $G_D$ is the bulk Newton constant,  we have required positive tension in order to enforce the null energy condition at the wall, and $D \ge 3$ is the bulk spacetime dimension.
After reviewing the relevant formalism, we investigate general features Euclidean solutions in sections \ref{subsec:dS-SAdS} and \ref{subsec:AdS-SAdS} below.  Detailed investigation of bag-of-gold solutions will be deferred to section \ref{sec:search}. In this section, we also confine ourselves to cases with a single (connected) domain wall in the Euclidean section.  Additional walls can often be added but, at least in stable theories, are associated with additional sources of domain walls at the Euclidean boundary.  We will return to this issue in section \ref{sec:search} as well.

When two walls collide, interactions between the walls become important and the simple thin wall approximation tends to fail.  We thus restrict attention to non-intersecting domain walls in the bulk of this paper, though we comment briefly on interactions in section \ref{subsec:selfint}.  We also require our solutions to be smooth (up to the presence of our thin domain walls).  In contrast, in related settings, references \cite{Fischler:1989se,Fischler:1990pk} used the Hamiltonian formalism to advocate the use of certain a priori singular spacetimes.  Arguments against the use of these configurations will be presented in section \ref{subsec:degenerate}.

Our domain walls will separate two vacua which we call interior and exterior.  The analysis is much like that in the classic work \cite{Coleman:1980aw}, though we are explicitly interested in cases where the domain walls reach the Euclidean boundary.  Since our spacetimes have spherical symmetry, Birkhoff's theorem with cosmological constant implies our metric to take the form
\begin{equation}
\label{eq:metric}
ds^2_{i,e}=-f_{i,e}(r)dt^2_{i,e}+\frac{dr^2}{f_{i,e}(r)}+r^2d\Omega^{2}_{D-2},
\end{equation}
where $i,e$ refer to the interior/exterior vacua and $f_{i,e}$ define either a Schwarzchild de Sitter (SdS) solution or a Schwarzschild anti-de Sitter (SAdS) solution.  Following reference \cite{Freivogel:2005qh}, we write
\begin{equation}
\label{eq:setting}
\begin{aligned}
f_i(r)&=1-\lambda r^2-\frac{\mu_i}{r^{D-3}},\\
f_e(r)&=1+r^2-\frac{\mu_e}{r^{D-3}},
\end{aligned}
\end{equation}
taking the exterior to have an AdS cosmological constant with unit AdS length scale.  The interior cosmological constant is parametrized by $\lambda$ and allowed to take either sign.  Since we are interested in Lorentz signature spacetimes having only one boundary and a regular origin, we will shortly set $\mu_i =0$, though we will return to the case $\mu_i >0$ in section \ref{sec:search}. The theory is fixed by a choosing $\lambda, \kappa$, while the allowed solutions within the theory are specified by the scales $\mu_{e}, \mu_i$.  Note that $\mu_e, \mu_i$ have dimensions of $(\text{length})^{D-3}$ and so should be thought of as setting the horizon sizes of SAdS black holes rather than their mass. Note also that when the external vacuum is stable we can find non-trivial solutions in the $(e)$ vacuum near an asymptotically locally anti-de Sitter (AlAdS) boundary only for  $\mu_e>0$.

We will see below that these data uniquely specify the trajectory of the domain wall.  Our focus will be on cases in which the exterior solution extends to an asymptotically AdS boundary, and in particular where it does so on the moment of time symmetry $(t=0)$.  Indeed, for cases with a single asymptotic region we take this to be the defining property of exterior versus interior.

The domain wall equation of motion follows from the Israel junction conditions \cite{Israel:1966rt}.  The analysis is standard, and can be read off from, e.g., reference \cite{Blau:1986cw}.  The Lorentz signature result takes the form
\begin{equation}
\label{eq:junction}
\alpha_i-\alpha_e=\kappa r,
\end{equation}
where $\alpha_{i,e}=  rK_{\theta\theta}$ with $K_{ab}$ the extrinsic curvature of the wall computed in the appropriate (interior/exterior) region using a normal pointing from the interior region toward the exterior.  Our $\alpha_{i,e}$ are traditionally called $\beta_{i,e}$, but we use $\alpha_{i,e}$ to avoid confusion with the period $\beta$ of Euclidean time.  Following \cite{Blau:1986cw} one finds
\begin{equation}
\label{eq:beta}
\alpha_{i,e} = \pm \sqrt{\dot r^2+f_{i,e}(r)},
\end{equation} where $\dot r\equiv \frac{dr}{d\tau }$ denotes the derivative with respect to proper time $\tau$ along the wall.  The sign of $\alpha_{i,e}$ is determined by whether the size $r$ of the spheres increases or decreases as one approaches the wall from the interior side ($i$) or as one moves away from the wall in the exterior ($e$).  In particular, $r$ is monotonic near the wall when $\alpha_{i,e}$ have identical signs but is locally extremized at the wall when their signs differ.

For any signs in \eqref{eq:beta}, squaring \eqref{eq:junction} twice yields
\begin{equation}
\label{eq:EOM}
\dot r^2+V_\text{eff}(r)=0
\end{equation}
in terms of the effective potential
\begin{equation}
\label{eq:potential}
V_\text{eff}(r)=
f_e(r)- \frac{\left(f_i(r)-f_e(r)-\kappa ^2r^2\right)^2}{4\kappa ^2r^2} \\
= A r^2 + 1 + \frac{B}{r^{D-3}} - \frac{C}{r^{2D-4}}
\end{equation}
with
\begin{equation}
\label{eq:ABC}
A = 1-\frac{\left(1+\kappa ^2+\lambda \right)^2}{4\kappa ^2},\text{ }
B = \frac{1}{2\kappa^2} \left[\left(1+\lambda -\kappa ^2\right)\mu _e-\left(1+\lambda +\kappa ^2\right)\mu _i \right],\text{ }
C =\frac{\left(\mu _e-\mu _i\right)^2}{4\kappa ^2}.
\end{equation}
Other components of the Israel junction conditions on the sphere are related to \eqref{eq:junction} by spherical symmetry, and as usual the $\tau\tau$ component is proportional to $d/d\tau$ of \eqref{eq:EOM}.  Thus \eqref{eq:junction} fully specifies the dynamics of the wall.

Euclidean solutions are obtained by substituting $\tau_E = i\tau$ into \eqref{eq:EOM}, or equivalently by changing the sign of $V_\text{eff}$.  In particular, Euclidean solutions can exist only for choices of parameters that allow $V_\text{eff}$ to take non-negative values for some $r$.

Since our solutions will be constructed by cutting and pasting pieces of the exterior and interior metrics specified by \eqref{eq:setting}, it will be useful to directly find the curves defined by the domain wall in both the $(r,t_e)$ and $(r_,t_i)$ planes.  Since we will focus on Euclidean solutions below, we now introduce the Euclidean time coordinates $t_{Ee} = i t_e$, $t_{Ei} = i t_i$ and note that \eqref{eq:metric} and \eqref{eq:EOM} can be combined to yield
\begin{equation}
\label{eq:tEEOM}
\frac{dt_{Ei}}{dr} = \pm \left(\frac {\alpha_i}{f_i\sqrt{V_\text{eff}}} \right),\text{ }
\frac{dt_{Ee}}{dr} = \pm \left(\frac{\alpha_e}{f_e\sqrt{V_\text{eff}}} \right).
\end{equation}
Here we make explicit that the interior and exterior generically define two different time coordinates $t_{Ei}, t_{Ee}$ along the wall as the coordinates of \eqref{eq:metric} are not guaranteed to be continuous at the domain wall. A careful check of the signs shows that the sign $\pm$ in \eqref{eq:tEEOM} changes only on the surface $t_{Ei}= t_{Ee} =0$ of time reflection symmetry and, in particular, not at other possible zeros of $\alpha_e$.

For later use, we also note that combining \eqref{eq:beta}, \eqref{eq:EOM}, and \eqref{eq:potential} yields
\begin{equation}
\label{eq:betas}
\begin{aligned}
\alpha_i(r)= \frac{f_i(r)-f_e(r)+\kappa ^2r^2}{2\kappa r}
=  -\left(\frac{1+\lambda -\kappa ^2}{2\kappa }\right)r+\left(\frac{\mu _e-\mu _i}{2\kappa }\right)\frac{1}{r^{D-2}},\\
\alpha_e(r) = \frac{f_i(r)-f_e(r)-\kappa ^2r^2}{2\kappa r}=  -\left(\frac{1+\lambda +\kappa ^2}{2\kappa }\right)r+\left(\frac{\mu _e-\mu _i}{2\kappa }\right)\frac{1}{r^{D-2}},
\end{aligned}
\end{equation}
and
\begin{equation}
\label{eq:V2}
V_\text{eff} = f_e -\alpha_e^2 = f_i - \alpha_i^2.
\end{equation}
Since $\kappa \ge 0$, one also finds $\alpha _i \ge \alpha_e$ from either \eqref{eq:EOM} or \eqref{eq:betas}.

In sections \ref{subsec:dS-SAdS} and \ref{subsec:AdS-SAdS}, as well as the first part of section \ref{sec:search}, we will set $\mu_i=0$. Before proceeding, it is worth noting that even in this case, by tuning $\kappa $, $\lambda \in {\mathbb R}$ and $\mu_e >0$ one can realize all values $A$, $B \in {\mathbb R}$ and $C>0$ of the coefficients in $V_{\text{eff}}$ (see equation \eqref{eq:ABC}).  This can be argued by writing $A$ in terms of $B$, $C$, $\kappa $ and noting that for any fixed $B$, $C$ with $C>0$ we have $A \rightarrow +\infty$ as $\kappa \rightarrow 0$ and that $A\rightarrow -\infty$ as $\kappa \rightarrow +\infty$.  Thus for $\mu_i=0$ dS or AdS interiors, $C>0$ is the only constraint on $A$, $B$, $C$.

\section{No AlAdS spacetimes with inflating interiors}
\label{subsec:dS-SAdS}

We will now show that for de Sitter interior regions ($\lambda >0, \mu_i=0$)  there can be no smooth Euclidean solutions with asymptotically AdS boundaries.  Indeed, the identical argument will also apply to cases with $0 > \lambda > - (\kappa-1)^2$, so we assume only $\lambda > - (\kappa-1)^2$ in this section, and we refer to this as case $(I)$ in the rest of this work.    All parameter choices in case $(I)$ give $A<0$, and since $V_{\text{eff}} \approx A r^2$ at large $r$ the Lorentz signature solutions expand exponentially with respect to proper time.  For $0 > \lambda > - (\kappa-1)^2$, such solutions are examples of domain wall inflation rather than inflation driven by a cosmological constant.

The case $(I)$ Euclidean solutions without self-intersections that are smooth up to the presence of out thin walls are constructed in section \ref{subsec:nobndy} and shown to have no AlAdS boundaries.  A brief argument in section \ref{subsec:unstable} then shows that all other cases $(\lambda < - (\kappa+1)^2)$ where the Lorentz signature wall can grow to infinite size in fact lead to instabilities of the vacuum.
Section \ref{subsec:degenerate} argues against using analogues of the degenerate solutions of references \cite{Farhi:1989yr,Fischler:1989se,Fischler:1990pk}, and section \ref{subsec:selfint} argues that allowing self-intersections alone cannot produce case $(I)$ Euclidean solutions with AlAdS boundaries.  So at least subject to our symmetries, within our class of models there are no stable theories that allow AlAdS path integrals with good Euclidean saddles that create bubbles of inflating spacetime.

\subsection{Euclidean solutions for case $(I)$}
\label{subsec:nobndy}

The detailed formulas given in section \ref{TwoPiecesSpacetimes} are not needed for our main argument.  Instead, we require only the following three properties of $V_\text{eff}$, $\alpha_e$:
\begin{enumerate}[label=\textnormal{(\roman*)}]
\item \label{p1} $V_\text{eff} \rightarrow -\infty$ as $r \rightarrow 0$ and as $r \rightarrow \infty$,
\item \label{concavedown} $V_\text{eff}'' < 0$ at points with $V_{\text{eff}}'=0$,
\item \label{p3} $\alpha_e$ decreases monotonically from $+\infty$ to $-\infty$.
\end{enumerate}
Properties \ref{p1} and \ref{p3} follow immediately from inspection of \eqref{eq:potential} and \eqref{eq:betas}, while property \ref{concavedown} is derived in appendix \ref{app:Vcd}. We will also need the observation that any allowed solution defines a (not self-intersecting) domain wall $r(s), t_E(s)$ (say, parametrized by the proper distance $s$) in the relevant (locally) dS, AdS, or SAdS Euclidean geometries. Here an important point is that the equations of motion require the exterior to be {\it locally} equivalent to \eqref{eq:metric}.  But there is no harm in introducing conical singularities in regions of the SAdS spacetime that will be excised.  As a result, the period of the exterior Euclidean Killing time $t_{Ee}$ need not always agree with that of the standard Euclidean SAdS black hole.

We will also need that, at each $r$, whether $r$ decreases or increases as one moves away from the wall toward the interior/exterior is determined by the signs of $\alpha_{i,e}(r)$.
We are then to cut the exterior (locally Euclidean SAdS black hole) geometry along the above curve, discard any pieces that do not satisfy the sign constraint, and glue the remaining piece to a corresponding piece of the interior (locally Euclidean dS or AdS) geometry so that the full metric is continuous.

\begin{figure}[t]
\centerline{
	\includegraphics[width=0.45\textwidth]{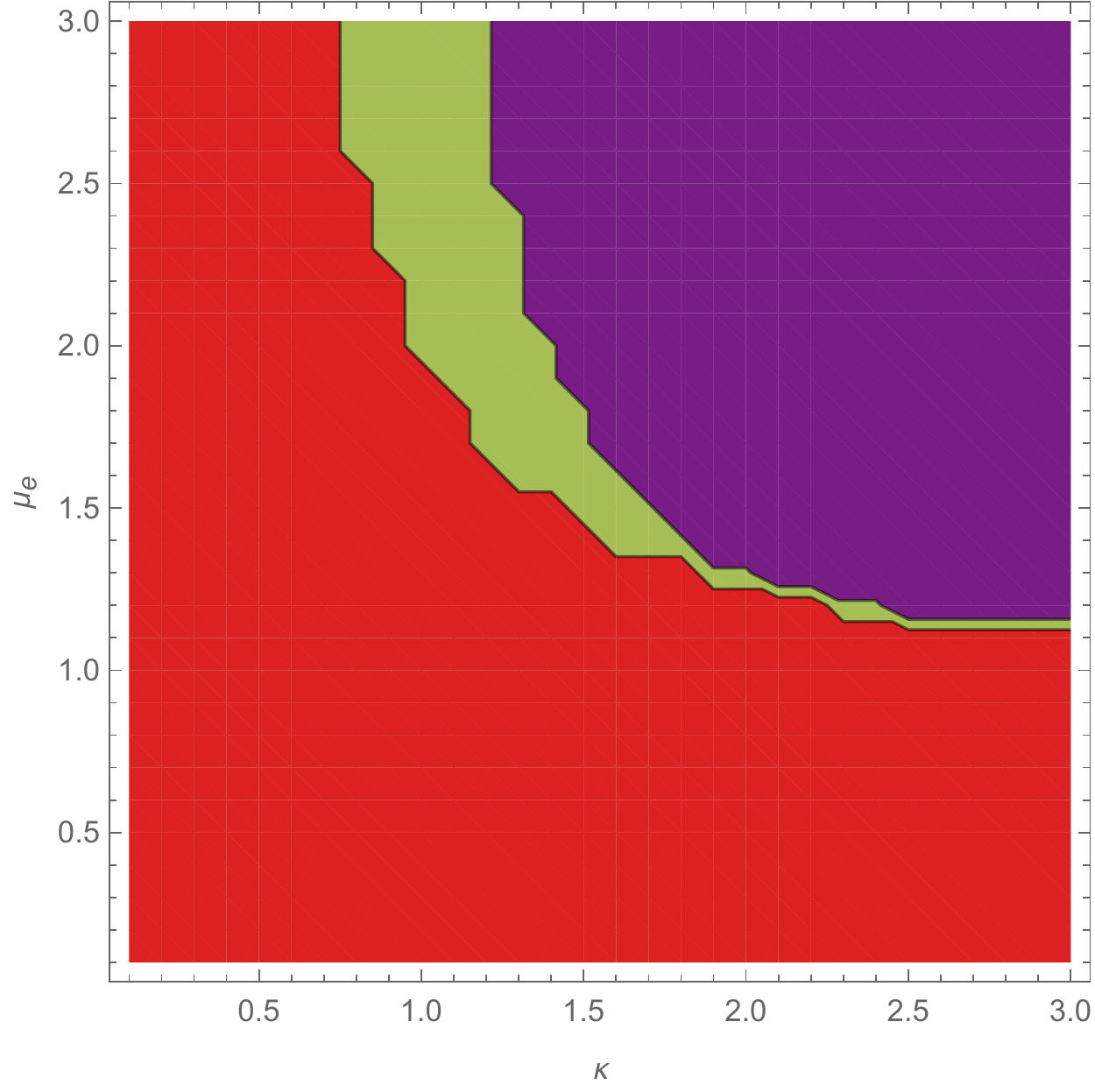}
	\hspace{.5cm}
	\includegraphics[width=0.45\textwidth]{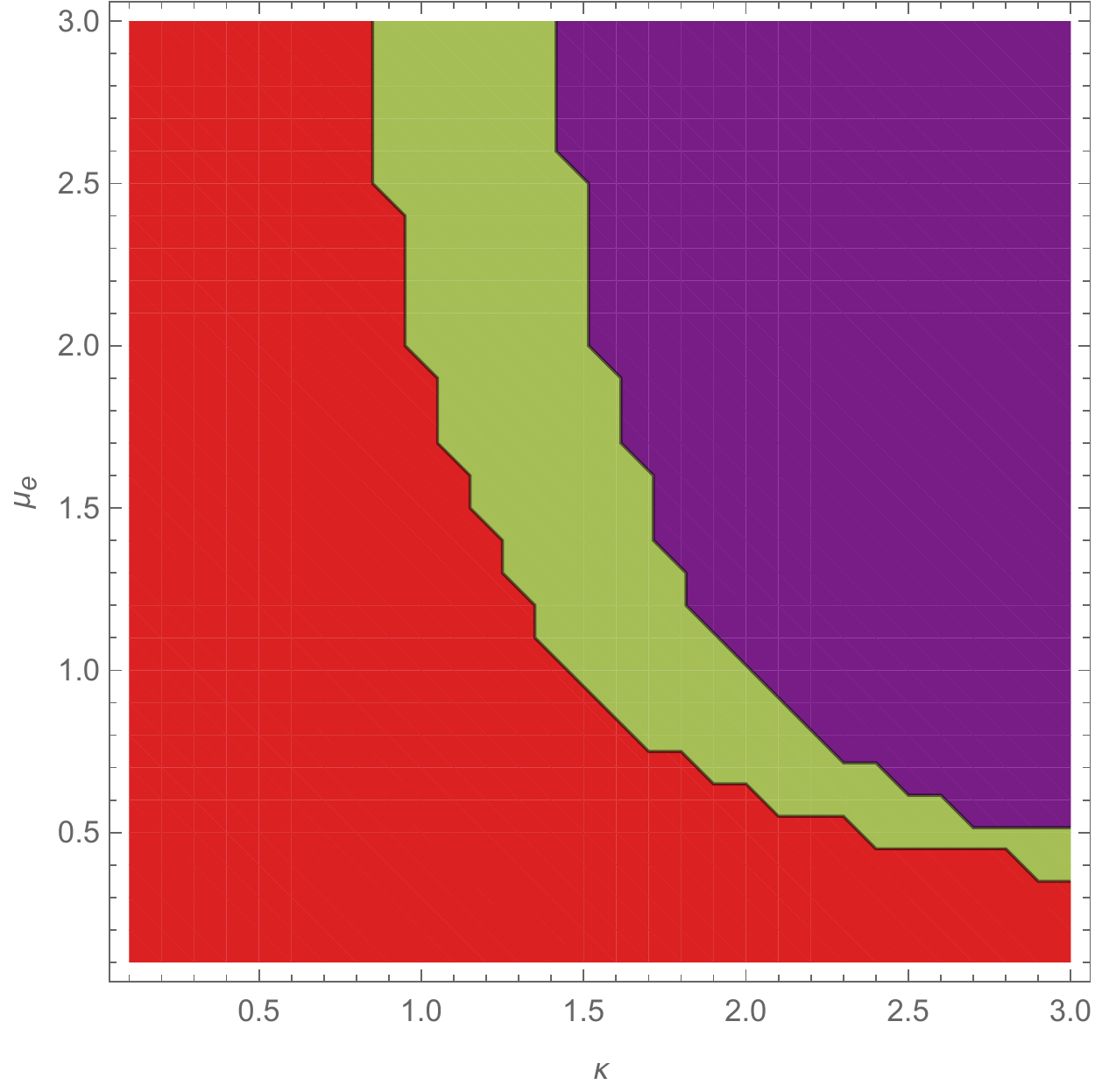}
}
\caption{Both signs of $\alpha_e(r_{\min{}})$ can be found for $\lambda>0$ and for $\lambda <0$. The cases $D=3$ (left) and $D=4$ (right) are shown.   The figure displays results at $\lambda=0$, but $\alpha_e(r_{\min{}})$ is continuous so the results for small positive or negative $\lambda$ are essentially identical. Regions with $\alpha_e(r_{\min{}}) >0$ are shaded red while those with $\alpha_e(r_{\min{}}) <0$ are shaded green. In the purple regions $V<0$ at all $r$ and Euclidean solutions do not exist. }
\label{fig:signbermindS}
\end{figure}

Property \ref{concavedown} implies $V_\text{eff}$ to have no local minima and no zeros of order higher than 2.  Indeed, even a second order zero must be a local maximum, which must in fact be a global maxium since there are no local minima to separate multiple local maxima.

We may thus divide the parameter space into three regimes:  If $V_\text{eff}$ is negative everywhere, there can be no Euclidean solutions.  If the maximum of $V_\text{eff}$ vanishes, the only Euclidean solution is $r=\text{constant}$ at this zero (call its location $r_{\min{}} = r_{\max{}}$).  If the maximum is positive, then $V_\text{eff}$ has two simple zeros that occur at precisely two values $r_{\min{}}, r_{\max{}}$ between which the solution oscillates periodically.\footnote{That all 3 cases occur can be seen from \eqref{eq:potential} and \eqref{eq:ABC}.  Since $B$ and $C$ vanish for $\mu_e=0$, the maximum of $V_\text{eff}$ approaches $1$ at small $r$ as $\mu_e \rightarrow 0$ at any fixed $\lambda $, $ \kappa$.  In such cases the maximum of $V_\text{eff}$ is clearly positive.  But fixing $\lambda $, $C$ and taking $\kappa $ large yields $A$, $B \rightarrow -\infty$ so that $V_\text{eff}$ becomes very negative everywhere.  In this case  the maximum is clearly negative.  Since the maximum is a continuous function of $A$, $B$, $C$, it must vanish in between.}

Furthermore, at the maximum of $V_\text{eff}$ we must have $0 = V'_\text{eff}=f_e' -2\alpha_e\alpha_e'$ so that $\alpha_e = f_e'/\left(2\alpha_e'\right)$.  But clearly $f_e'>0$ and $\alpha_e'<0$, so $\alpha_e$ is already negative at this maximum.  Monotonicity of $\alpha_e$ then requires $\alpha_e(r_{\max{}}) < 0$ as well.  In contrast, $\alpha_e$ can take either sign at $r_{\min{}}$; see figure \ref{fig:signbermindS}.

Let us suppose that the sign of $\alpha_e$ tells us to use a region ${\cal R}_e$ of the exterior SAdS geometry containing an asymptotic boundary. Since the wall satisfies $r \le r_{\max }$, curves of constant $r$ near the boundary will remain entirely within ${\cal R}_e$.  Let us deform such curves inward (preserving the property that each curve has constant $r$) until some such curve first contacts the wall.  Such a contact can only occur at a local maximum of $r$, but all local maxima on the wall occur at $r=r_{\max{}}$ where $\alpha_e(r_{\max{}}) <0$.

This contradiction thus requires us to instead discard the exterior SAdS solution on the $r>r_{\max{}}$ side of our wall.  The physical origin of this requirement may be explained using the fact that $\dot{r}=0$ at $r_{\max{}}$, which allows us to use this configuration as initial data for a Lorentz signature solution.  If the solution is SAdS in the neighboring region with $r>r_{\max }$, both the negative de Sitter pressure in the interior and the positive tension of the black hole pull the wall inward, so the Lorentzian solution must have $\ddot{r} <0$ at this point.  But then $\frac{d^2r}{d\tau_E^2} <0$ in Euclidean signature (with $\tau_E$ Euclidean proper time, also known as Euclidean proper distance), so such configurations cannot occur at $r_{\max{}}$.

We conclude that all (connected) exterior solutions with $\lambda > - (\kappa-1)^2$ are instead of the form at left shown in figure \ref{fig:dS-SAdS} below.  The details of the interior solutions depend on the sign of $\lambda$ and other features.  For the case $\lambda >0$ (shown at center in figure \ref{fig:dS-SAdS}), they follow from the fact that $\alpha_i(r_{\min{}}) >0$ in all cases.  This may be argued by noting that $\alpha_i$ monotonically decreases from $+\infty$ for $1 + \lambda - \kappa^2 >0$ and is positive definite for $1 + \lambda - \kappa^2 \le 0$.  Since $\alpha_i(r_{\min }) >0$ is then clear in the second case, we may concentrate on showing it in the first.  There $\alpha_i^2$ must decrease monotonically from $+\infty$ to zero before monotonically increasing back to $+\infty$.    Now note that $f_i$ is positive everywhere for $\lambda \le 0$, and is positive at small $r$ for $\lambda >0$.  And while for $\lambda >0$ the function $f_i$ eventually becomes negative, it cannot do so until after $r_{\min{}}$ as it decreases monotonically and $V_\text{eff}(r_{\min{}}) = 0$ requires $f_i = \alpha_i^2 \ge 0$. Since $V_\text{eff} = f_i - \alpha_i^2$, for $r \in [0,r_{\min{}}]$ we must have $\alpha_i^2 \ge f_i > 0$.  It follows that the first zero of $V_\text{eff}$ must occur before $\alpha^2_i$ reaches zero and thus that $\alpha_i(r_{\min{}}) >0$ as claimed.

Sewing any allowed exterior to any allowed exterior yields a solution with no AlAdS boundary.
Note that in solutions where $\dot{r}$ does not vanish identically, the curves tracing $(r, t_E)$ need not necessarily close in either the AdS interior or the SAdS exterior.    But smooth solutions without self-intersections exist only when they do.  Imposing this condition in the exterior imposes a relation between $\lambda $, $\kappa $, $\mu _e$, but in the interior (center panels in figure \ref{fig:dS-SAdS}) the domain wall curve can be made to close by inserting an appropriate conical singularity in the non-physical (shaded) region.    Smooth solutions thus typically exist on a co-dimension surface in parameter space, or for discrete values of $\mu_e$ within a given theory (fixed $\lambda $, $\kappa $).

\begin{figure}[t]
\centerline{\includegraphics[width=0.3\textwidth]{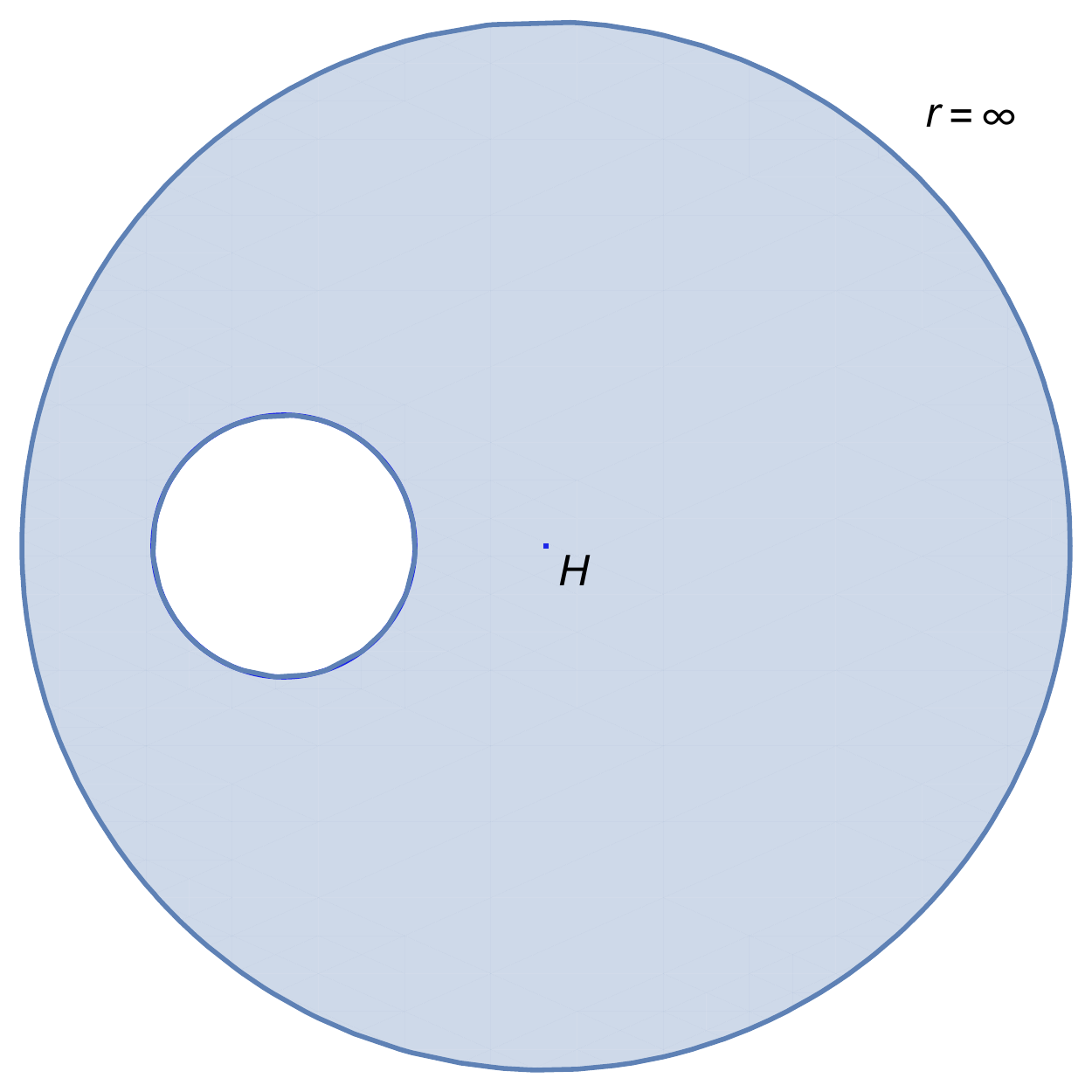} \hspace{.1cm}
\includegraphics[width=0.3\textwidth]{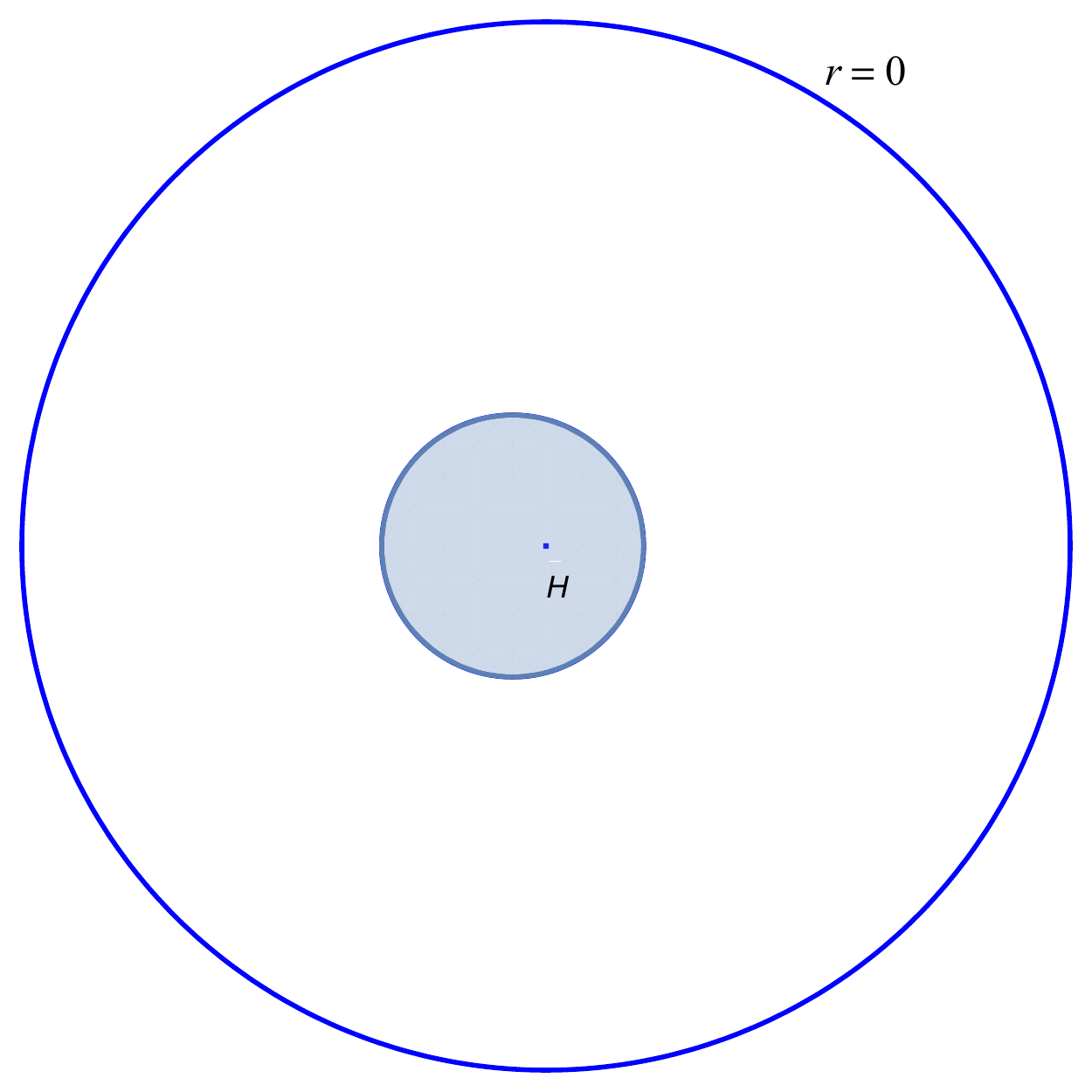}\hspace{.1cm}
	\includegraphics[width=0.3\textwidth]{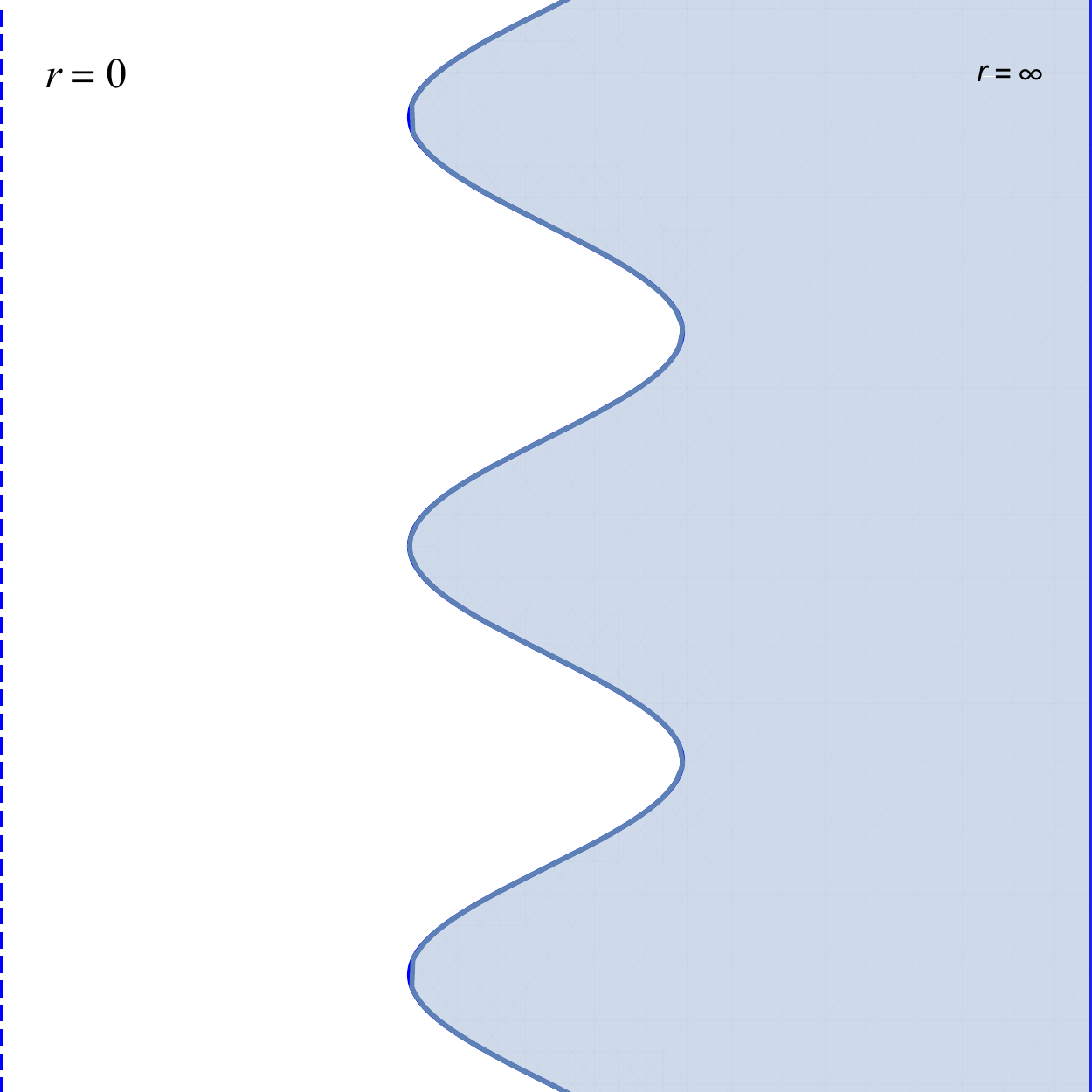}}
\centerline{\includegraphics[width=0.3\textwidth]{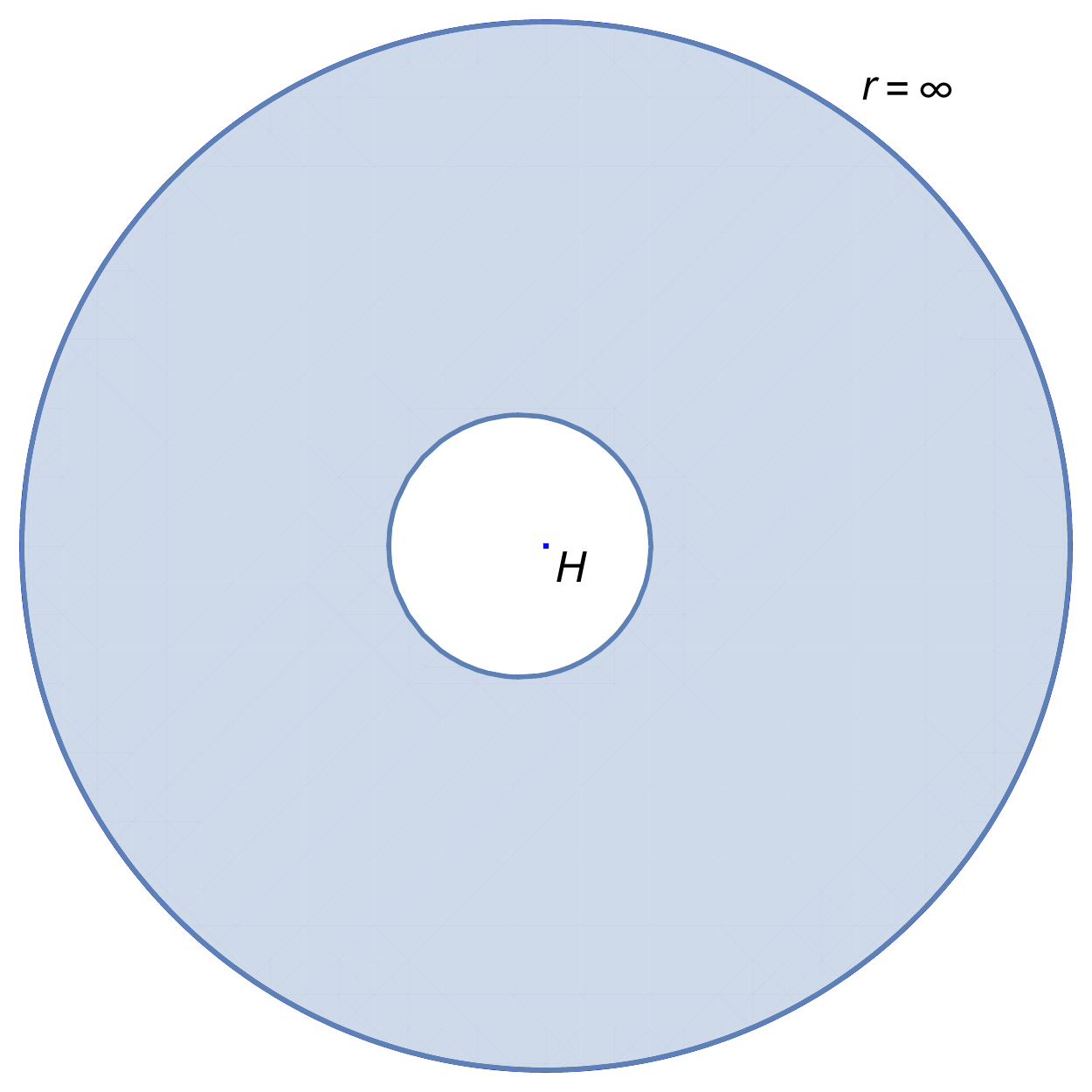} \hspace{.1cm}
\includegraphics[width=0.3\textwidth]{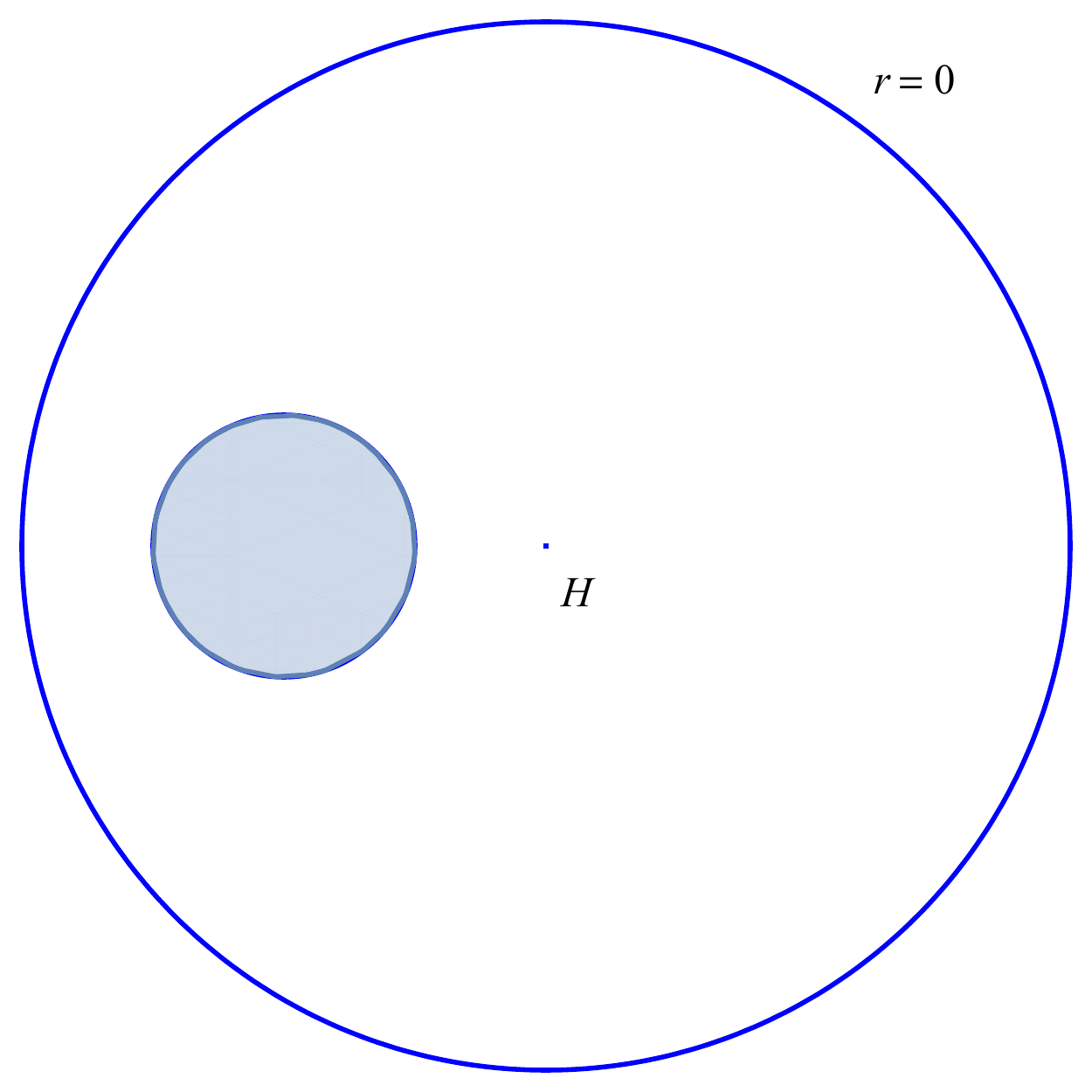} \hspace{.1cm}
	\includegraphics[width=0.3\textwidth]{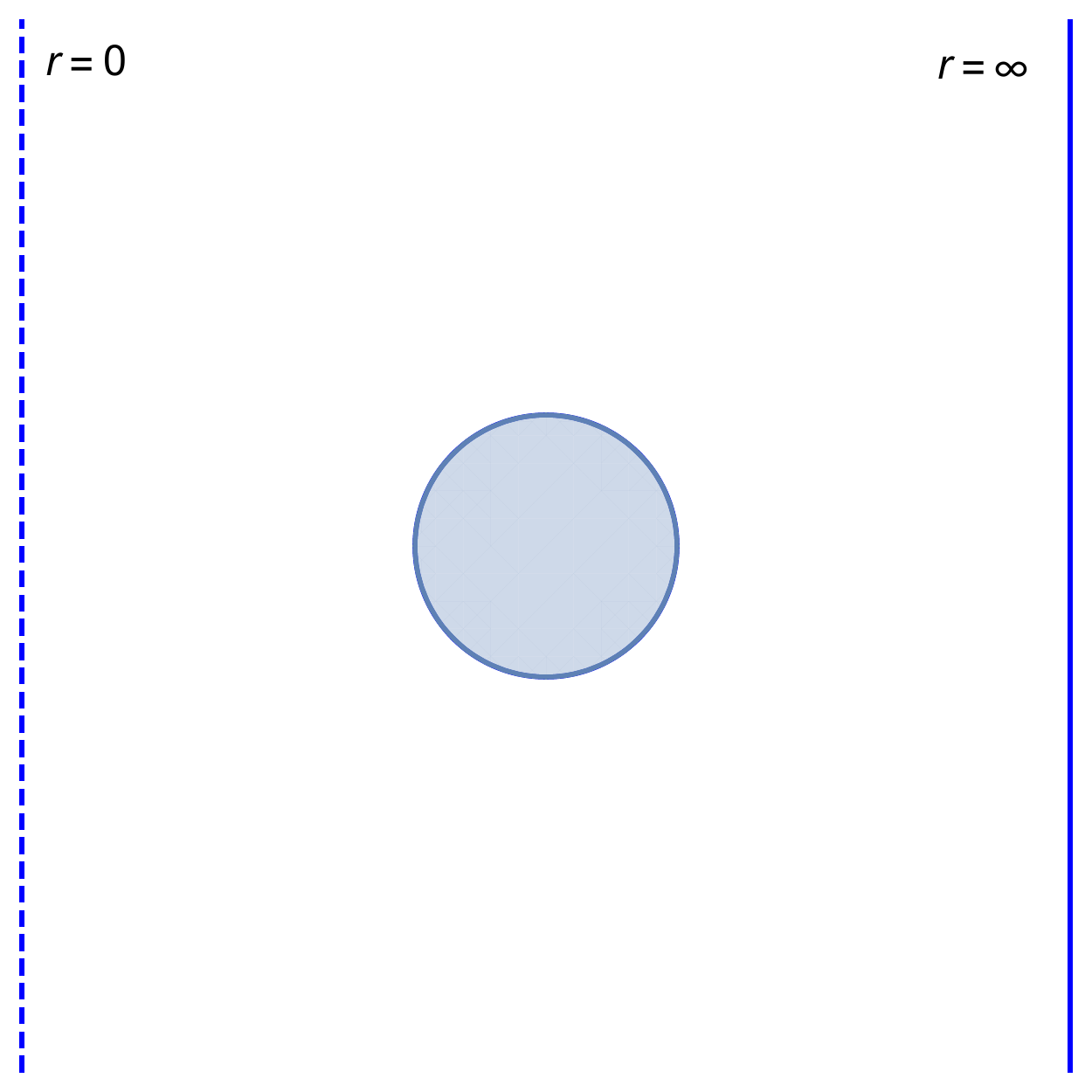}}
\caption{{\bf Left:} Unshaded regions provide allowed Euclidean SAdS exteriors with $r$ increasing outward for $\lambda > - (\kappa-1)^2$ (case ($I$)) with $\mu_i=0$. The angular direction is Euclidean time. Since $\alpha_e(r_{\max{}}) <0$, one must excise the (shaded) region $r> r_{\max }$. The sign of $\alpha_e(r_{\min{}})$ is positive in the top panel and negative in the bottom panel. {\bf Center:} Allowed interiors for $\lambda >0$ shown as (unshaded) regions of Euclidean dS with $r$ increasing inward. Since $\alpha_i(r_{\min{}}) >0$, one must excise the (shaded) region $r> r_{\min }$. The sign of $\alpha_i(r_{\max{}})$ is also positive at top but is negative at bottom.
{\bf Right:} Allowed interiors for $0 > \lambda > - (\kappa - 1)^2$ shown as (unshaded) regions of Euclidean AdS, with $r$ increasing to the right and Euclidean time running vertically. Here  $\alpha_i(r_{\min{}}) >0$ requires one to excise the (shaded) region $r> r_{\min }$.  The sign of $\alpha_i(r_{\max{}})$ is also positive at top but is negative at bottom.
Sewing any allowed interior to any allowed exterior shows the full solution to have no asymptotic region.
}
\label{fig:dS-SAdS}
\end{figure}

As noted above, for $0 > \lambda > - (\kappa-1)^2$ the exterior solutions are again of the form shown at left in figure \ref{fig:dS-SAdS}.  And again we have $\alpha_i(r_{\min }) >0$, though the form of the interior solution depends on the sign of  $\alpha_i(r_{\max })$ as shown at right.  As for $\lambda >0$, the interior curves can be made to close without tuning parameters. Here this is due to the arbitrary period of Euclidean time at top right, and the ability to insert conical singularities in the non-physical (shaded) region at bottom right.
Again, there is no AlAdS boundary in any solution.

\subsection{No other inflating cases in stable theories}
\label{subsec:unstable}

The case $\lambda < - (\kappa+1)^2$ is similar to case $(I)$ in that $A< 0$ so that the Euclidean domain wall is again confined to some finite range $r$, so it again defines a curve in SAdS that does not reach any boundary.  But one may now check that $\alpha_i > \alpha_e >0$, so Euclidean solutions with the desired asymptotics can always be found, and the exterior curve can be made to close without self-intersections by inserting a conical singularity into the excluded non-physical region of SAdS inside the curve.  But the fact that such solutions have no boundary source suggests that the theory is unstable.  Indeed, setting $\mu_e=\mu_i =0$ yields $V_\text{eff} = -Ar^2 +1$ with $A<0$, so  Lorenzian zero energy domain walls expand to $r=\infty$ in finite time.  Wick rotating to Euclidean signature gives a smooth instanton describing the nucleation of the corresponding bubbles and thus the expected vacuum instability. We will refer to $\lambda < - (\kappa+1)^2$ as case $(IV)$ below.

\subsection{Good saddles are non-degenerate}
\label{subsec:degenerate}

Although the context is slightly different, the mechanism that excludes AlAdS saddles for $\lambda > - (\kappa-1)^2$ is directly analogous to the classic obstruction \cite{Farhi:1989yr} to finding smooth instantons mediating nucleation of false vacuum bubbles by quantum tunneling from flat space.  In that context, it has been suggested \cite{Farhi:1989yr,Fischler:1989se,Fischler:1990pk} that the process may nevertheless take place and that it is instead mediated by certain non-smooth saddles (see also \cite{Bachlechner:2016mtp,deAlwis:2019dkc}).  As noted in reference \cite{Fischler:1990pk}, this view can be motivated by considering formulations that extend Einstein-Hilbert gravity to include solutions where the metric can become degenerate.  Reference \cite{Fischler:1990pk} in particular emphasized the Hamiltonian framework, but it is perhaps simpler to consider covariant first order tetrad formulations in which all dynamical fields are differential forms.

For example, in four dimensions one may use the covariant Palatini action
\begin{equation}
\label{eq:1storder}
S = \frac{1}{2} \int \left( e^a \wedge e^b \wedge R^{cd} + \frac{\Lambda}{6} e^a  \wedge e^b \wedge e^c \wedge e^d \right) \epsilon_{abcd},
\end{equation}
where $R =  d\omega + \omega \wedge \omega$, $a,b,c,d$ are internal SO(3,1) indices, the fundamental variables with independent variations are the one-forms $e^a$ and $\omega^{ab}$, and $\epsilon_{abcd}$ is the constant antisymmeric tensor in the internal space but is a spacetime scalar.  As reviewed in \cite{Horowitz:1991fr} for $\Lambda=0$, for non-degenerate tetrads the dynamics is that of Einstein-Hilbert gravity.  But given {\it any} smooth map $f$ from any manifold $\tilde M$ to a given spacetime $M$, one may pullback any solution $e,\omega$  through $f$ to define a new solution $\tilde e, \tilde \omega$ on $\tilde M$.  Indeed, when $f$ has degree one this pullback leaves the action invariant and it may be argued \cite{Horowitz:1991fr} (see also \cite{Marolf:1993ij}) that the two solutions are gauge equivalent and should be physically identified.

\begin{figure}[t]
\centerline{\includegraphics[width=0.3\textwidth]{Fig2L2} \hspace{.5cm}
\includegraphics[width=0.3\textwidth]{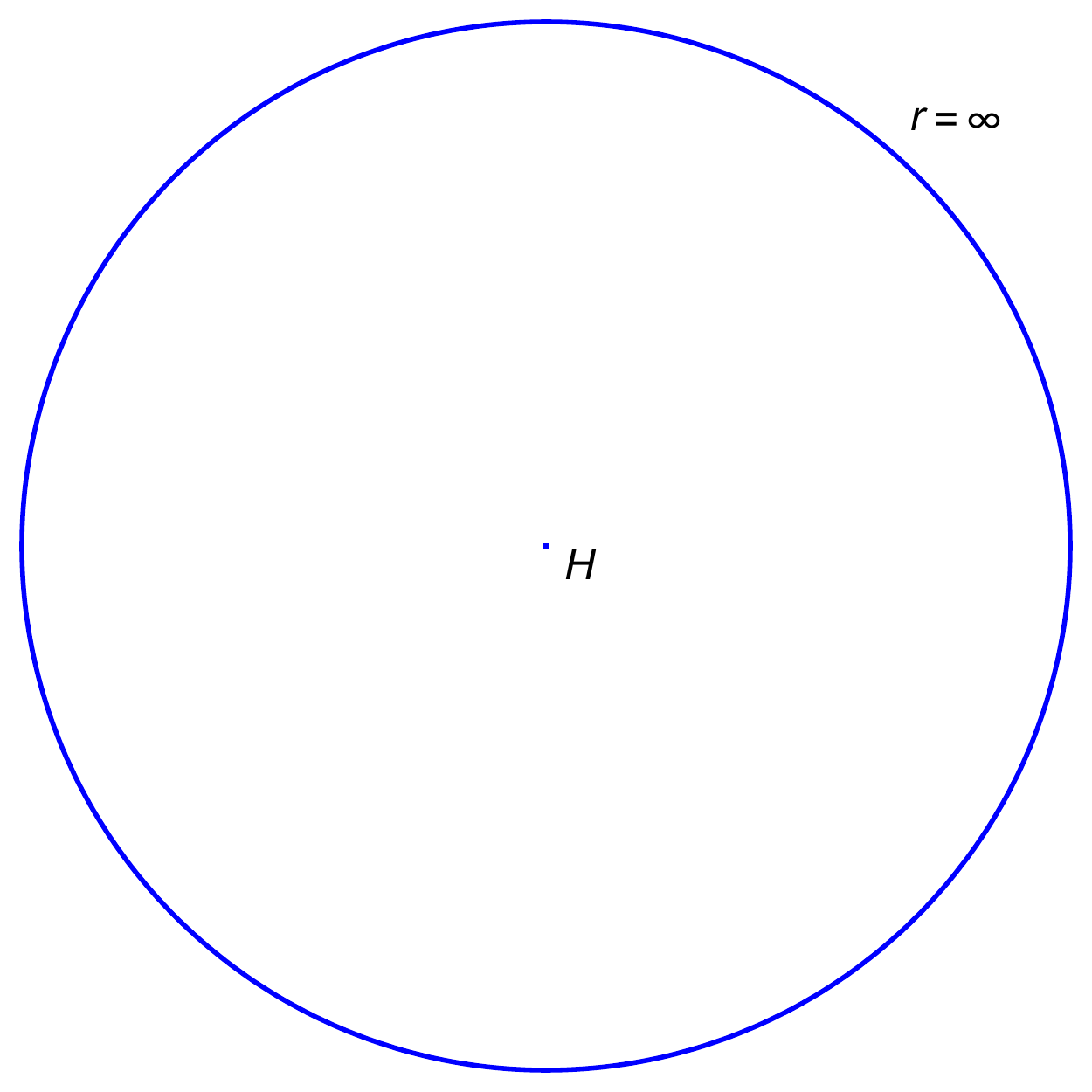}}
\caption{A degenerate case $(I)$ exterior (unshaded region) with AlAdS boundary.  Two copies of Euclidean SdS are shown, with an allowed case $(I)$ domain wall trajectory shown inside the left copy. The sign of $\alpha_e$ requires us to discard the shaded region.  Gluing the remaining unshaded regions together along the co-dimension 2 horizons ($H$) gives a singular (degenerate) solution analogous to those advocated in \cite{Farhi:1989yr,Fischler:1989se,Fischler:1990pk}.
}
\label{fig:degencut}
\end{figure}

We can now use this pullback construction to explain the analogue in our context of the degenerate solutions advocated in references \cite{Farhi:1989yr,Fischler:1989se,Fischler:1990pk}.  As in section \ref{subsec:dS-SAdS}, the solutions are to the constructed by a ``cut and paste'' procedure.  But now, instead of taking the exterior solution to be cut to be the standard form of Euclidean SAdS on which $r \ge r_h$ for horizon radius $r_h$, we instead pull back some $e,\omega$ associated with this solution through the map $r - r_h = \tilde r^2$ to yield a solutions with two copies of Euclidean SAdS, one with $\tilde r >0$ and one with $\tilde r <0$, joined by a degenerate metric at $\tilde r=0$.  We may then cut the solution along a domain wall in the region $\tilde r <0$ while preserving the AlAdS boundary at $\tilde r >0$; see figure \ref{fig:degencut}.   The relevant piece can then be sewn onto an AdS interior as in section \ref{subsec:dS-SAdS}.  In particular, by placing any resulting conical singularity in the region where the metric degenerates, we can avoid any need to tune the parameters $(\lambda, \kappa, \mu_e)$ described in the construction of smooth solutions above.  In summary, in this way we can glue onto the original Euclidean SAdS solution any compact Euclidean spacetime which satisfies the equations of motion up to having a codimension-2 conical singularity\footnote{And by allowing multiple regions where the metric degenerates we can allow further conical singularities as well.}.

However, we see two reasons why such constructions should be excluded. One is simply that, so long as the solution preserves time reflection symmetry, the $t=0$ surface of time reflection symmetry where one would glue our Euclidean solution to its Lorentzian analogue is also degenerate.  So the instanton should be viewed as creating a Lorentzian solution where the metric again degenerates, and one would need to understand the physics of degenerate metrics in Lorentz signature as well.

Perhaps more critically one may also note that when two regions $R_1$, $R_2$ are separated by a region of degenerate metrics, there is in general no correlation at all between the metrics in $R_1$ and $R_2$.  Indeed, consider a map $f$ from $\tilde M$ to $M$ that maps the boundaries of both $R_1$ and $R_2$ to a common point $x_0 \in M$ and which also maps all of $\tilde M$ outside $R_1 \cup R_2$ to the same point $x_0$.  Then the pullback through $f$ of some non-degenerate solution on $M$ is identically zero on $\tilde M\setminus (R_1 \cup R_2)$.  As a result, given any two non-degenerate solutions $(e_1,\omega_1)$, $(e_2,\omega_2)$ on $M$, we may cut each solution anywhere in the region between $R_1$ and $R_2$ and glue the pullback $(\tilde e_1, \tilde \omega_1)$ of the first solution in $R_1$ to the pullback $(\tilde e_2, \tilde \omega_2)$ of the second  solution in $R_2$.  The result clearly satisfies the equation of motion in both $R_1$ and $R_2$, and also in $\tilde M\setminus (R_1 \cup R_2)$ where it continues to vanish identically.

In this way, given an arbitrary compact Euclidean solution (perhaps with conical singularities where the equations of motion would naively fail),  we can use degenerate metrics to glue the compact solution to any other solution using a degenerate metric at a single point $x_0$.  Perhaps even worse, since the sign of \eqref{eq:1storder} depends on a choice of orientation, one is free to choose the sign of the action in such bubbles at will.  Adding many bubbles of the proper sign then shows the Euclidean action of this class of saddles to be unbounded below
\footnote{If one nevertheless chooses to sum over them in the path integral one finds that they add a contribution to the action at each point $x_0$ that is independent of $x_0$.  While it is unclear if this contribution is well-defined, after regularization it should serve only to renormalize the various local couplings in the original action.  One could thus consistently neglect such bubbles in a low energy effective field theory treatment of the path integral in which one simply uses the renormalized values of the couplings.  While more complicated bubbles might induce nonlocal couplings, they again renormalize any nonlocal couplings that might already be present, in which case the resulting renormalized couplings are experimentally constrained to be small. We may thus simply consider the situation where the renormalized effective theory is Einstein-Hilbert and then ignore any possible further contributions of such degenerate solutions.  This may provide a  consistent interpretation of such path integrals.  Another consistent interpretation is to insist that any single partition function be associated with a smooth saddle, and to interpret degenerate geometries defined by sewing together such saddles as computing a product of such partition functions $Z_i$ for each smooth component (or perhaps inverse partition functions $\frac{1}{Z_j}$ for saddles taken to contribute to the action with the opposite of the usual sign).}.

Now, the Hamiltonian framework as used in \cite{Fischler:1989se,Fischler:1990pk} is a priori more restrictive than the fully first-order framework of \eqref{eq:1storder}.  However, as already discussed in section VIII of \cite{Fischler:1990pk}, it nevertheless allows a similar gluing when the compact Euclidean solution defining the bubble coincides on a codimension-2 surface with the solution to which it will be glued.  The particular example discussed in \cite{Fischler:1990pk} involved bubbles of empty Euclidean de Sitter space, the effects of which were described as providing ``a possibly fatal divergence'' for the Hartle-Hawking wavefunction of quantum cosmology. But we see that the compact Euclidean domain-wall solutions (with or without conical singularity) give equally problematic bubbles which can also contribute to the original tunneling amplitudes of \cite{Farhi:1989yr,Fischler:1989se,Fischler:1990pk}.  Indeed, the point here is that while the WKB tunneling amplitudes are invariant under continuous deformations of the path connecting the initial and final points (so long as such deformations avoid turning points where the WKB approximation breaks down), they can change significantly under discontinuous changes associated with changes of topology.  In the Euclidean action formalism, this is the statement that a given process can receive contributions from multiple saddles with different Euclidean actions.  Here the addition of each bubble defines a topologically distinct path through configuration space connecting the initial and final tunneling configurations, which can thus alter the predictions of \cite{Farhi:1989yr,Fischler:1989se,Fischler:1990pk} and which again lead to difficulties.

In summary, we see many issues with following \cite{Farhi:1989yr,Fischler:1989se,Fischler:1990pk} in the use of degenerate saddles to compute tunneling amplitudes.  Lacking a satisfactory resolution of such issues, we tentatively conclude that only smooth saddles should be used in such computations.  However, it would be useful to explore this issue further in the future.  While there may be much room for subtlety, at a concrete level it would be very interesting to understand in detail the effect of moving beyond the thin-wall approximation and considering such issues in a context where the domain wall arises from smooth scalar fields.

\subsection{Comments on self-intersecting walls}
\label{subsec:selfint}

We have limited ourselves to not self-intersecting solutions to the thin wall equations of motion.  But some of our results readily generalize to at least some contexts with intersecting thin walls.

Consider for example the argument that there are no Euclidean dS-SAdS and consider a set of walls where the wall locations can be approximated by some (perhaps self-intersecting) solution to the thin wall equation of motion \eqref{eq:EOM}.  Since equation \eqref{eq:EOM} determines $\dot{R}$ from $R$ only up to an overall sign, two-wall intersections where $\dot{R}_1 = - \dot{R}_2$ are common in such solutions at points where $\dot{R}\neq 0$. But intersections cannot occur at $r_{\max{}}$ or $r_{\min{}}$, as two solutions to \eqref{eq:EOM} that coincide at such points must agree everywhere; two such walls that meet at $r_{\max{}}$ or $r_{\min{}}$ will never separate.

\begin{figure}[t]
\centerline{\includegraphics[width=0.3\textwidth]{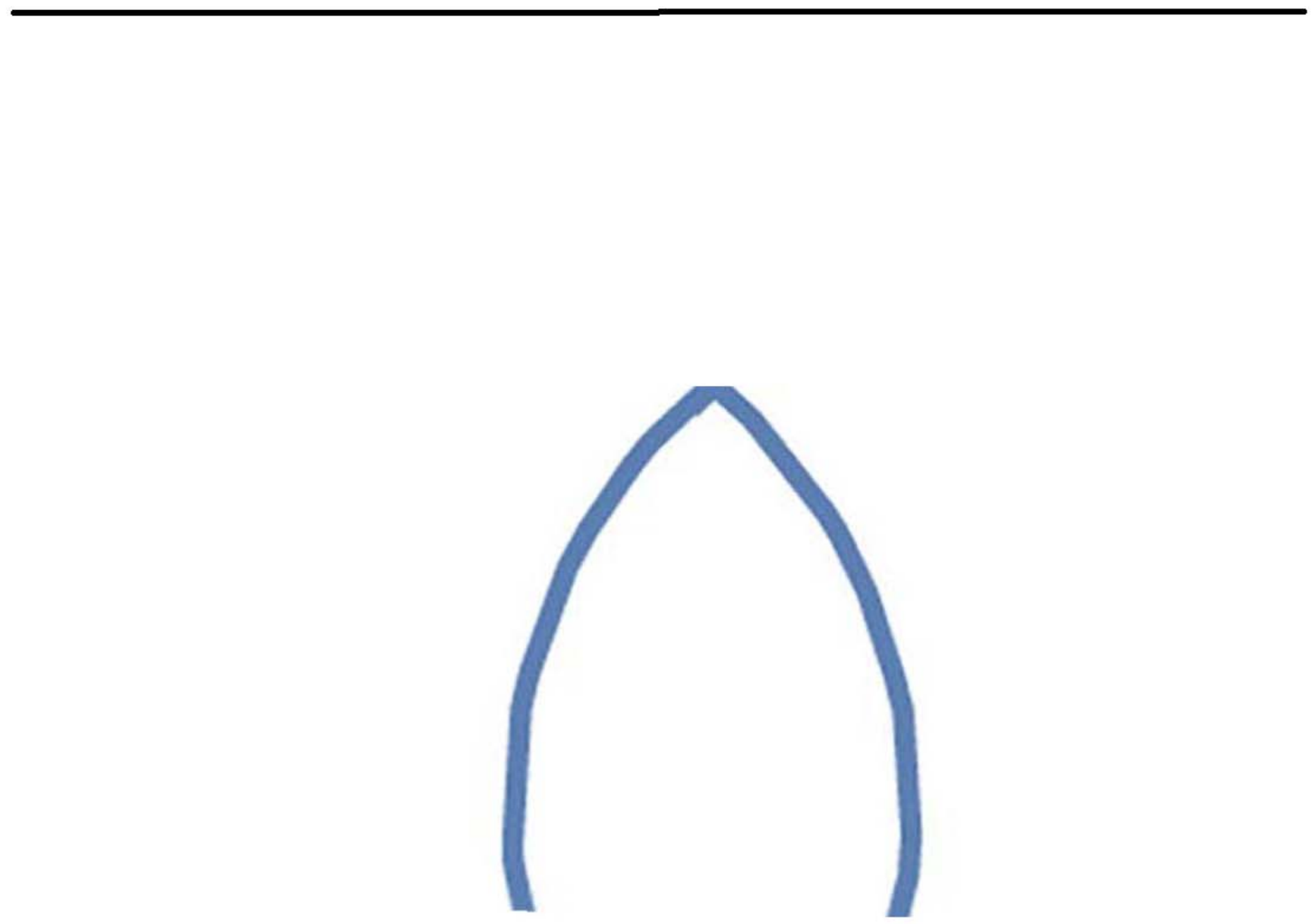} \hspace{.5cm}
\includegraphics[width=0.3\textwidth]{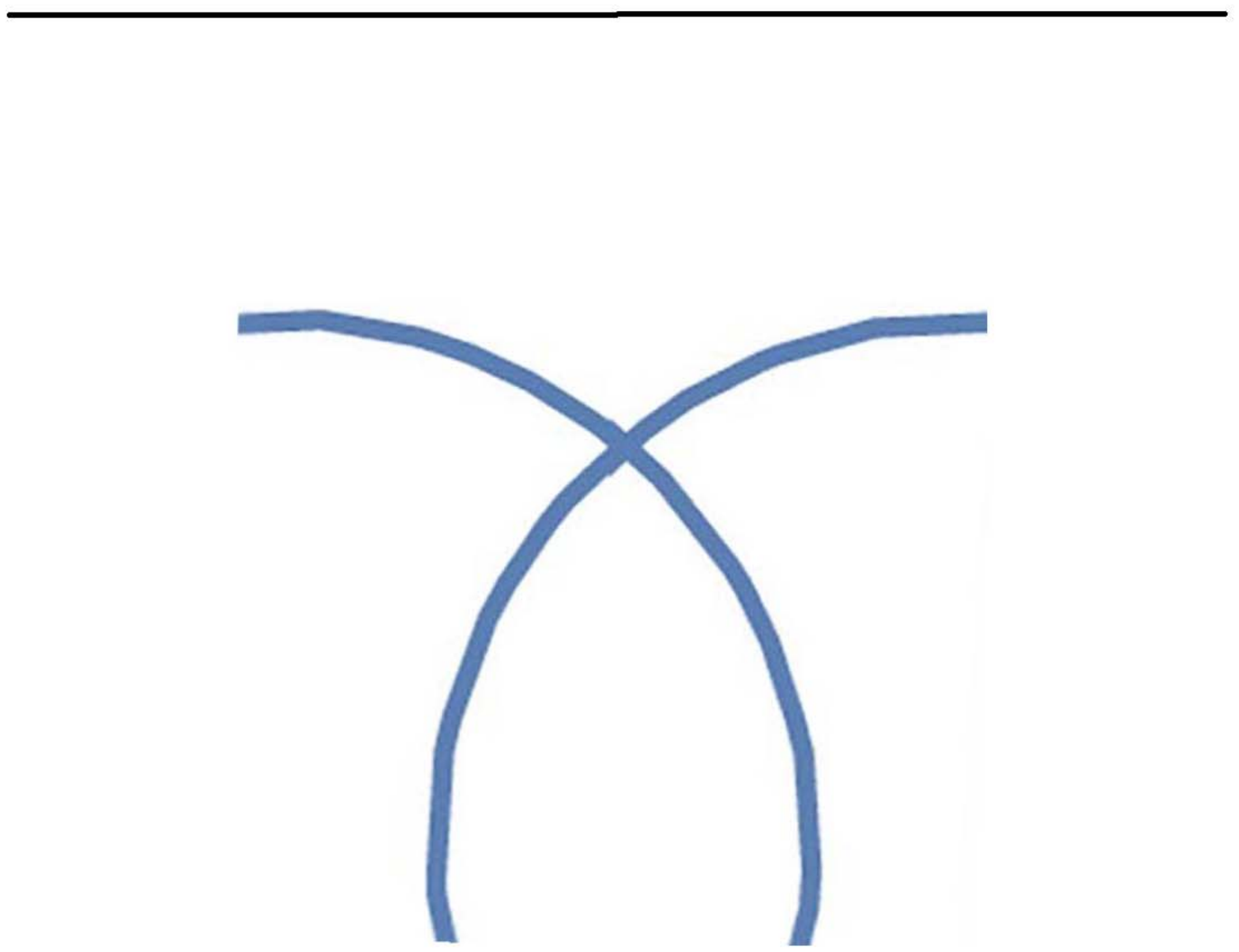}}
\caption{{\bf Left:} A domain wall intersection that would allow both an $r_{\max}$ and an AlAdS boundary (straight line at top of figure), but which does not satisfy conservation of the stress-energy tensor.  {\bf Right:} A domain wall intersection that conserves the stress-energy tensor.
}
\label{fig:intB}
\end{figure}

It is therefore natural to take a Euclidean solution with interacting thin walls to have a locally SAdS exterior region  ${\cal R}_e$ such that whose boundary $r(s), t_E(s)$ satisfies the Euclidean version of \eqref{eq:EOM} (perhaps with discontinuities in $\dot{R}$), and in particular with $r \le r_{\max{}}$. Such spacetimes now exist, though the points closest to the asymptotic boundary must have $\alpha >0$ and so cannot have $r=r_{\max{}}$.  Indeed, the exterior region must look something like that shown above in figure \ref{fig:intB} (left), where the points with locally maximal $r$ have discontinuities in $\dot{r}$.   But conservation of stress energy will require solutions in which two thin walls approach each other and interact on some sphere $p$ should take the very different form near $p$ shown in figure \ref{fig:intB} (right).  In other words, the domain wall world line cannot execute a sharp turn like that shown in figure \ref{fig:intB} (left) without some injection of Euclidean momentum from another sources.

We conclude that adding intersections to our thin wall model will not by itself allow Euclidean path integrals to create spacetimes with inflation.  However, it remains to investigate other generalizations involving walls with internal dynamical degrees of freedom, multiple types of walls,  or more general Einstein-scalar systems using either analytic or numerical techniques.   The former may be of particular interest as an internal degree of freedom might allow models with configurations that inflate to also have configurations where $V_\text{eff} >0$ as $r\rightarrow \infty$ so that the wall can reach the Euclidean boundary.  If the two above configurations can be continuously connected in a Euclidean solution, one might imagine that inflating bubbles could in fact be created by sourcing the correct wall configurations at the Euclidean boundary.  We leave such studies for future work.

\section{Non-inflating interiors}
\label{subsec:AdS-SAdS}

While strictly de Sitter interiors yield no Euclidean solutions with asymptotically AdS boundaries, AdS interiors (negative $\lambda$) can be more interesting.  We will divide the space of models into four regimes in accordance with the behaviors of $\alpha_e, V_\text{eff}$.  Since $\mu_i=0$, for $\lambda \ge-(1+\kappa^2)$ we find $\alpha_e$ to be positive everywhere while for $\lambda > -(1+\kappa^2)$ the derivative $\alpha_e' $ is everywhere negative.  Furthermore, $V_\text{eff} \rightarrow + \infty$ at large $r$ for $\lambda > -(\kappa -1)^2$ or $\lambda < - (\kappa +1)^2$, while $V_\text{eff} \rightarrow - \infty$ in between these regimes.  The relevant cases are thus
\begin{align}
  (I) & \ \hphantom{ -(\kappa -1)^2 \ge} \ \  \lambda > -(\kappa -1)^2 & \alpha_e' <0, V_\text{eff} \rightarrow -\infty \ {\rm at \ large} \ r \\
  (II) &  \ -(\kappa -1)^2 \ge \lambda > - (\kappa^2 +1) &  \alpha_e' <0, V_\text{eff} >0  \ {\rm at \ large} \ r \\
  (III) & \ - (\kappa^2 +1) \ge \lambda \ge -(\kappa +1)^2  &  \alpha_e > 0, V_\text{eff} >0  \ {\rm at \ large} \ r \\
  (IV)& \ -(\kappa +1)^2 > \lambda   &  \alpha_e > 0, V_\text{eff} \rightarrow -\infty \ {\rm at \ large} \ r.
\end{align}
In all cases  $V_\text{eff} \rightarrow -\infty$ as $r\rightarrow 0$ and $\alpha_i(r_{\min{}}) >0$, with $r_{\min{}}$ the smallest zero of $V_\text{eff}$.   The latter result may be argued much as at the end of section \ref{subsec:dS-SAdS} but now using the fact that $f_i$ is manifestly positive for all $r$.  The statement  $\alpha_i(r_{\min{}}) >0$ is vacuously true in cases where $V_\text{eff}$ is always negative and $r_{\min{}}$ fails to exist.

Having with cases $(I)$ and $(IV)$ in section \ref{subsec:dS-SAdS}, we now focus on cases $(II)$ and $(III)$.  These latter ranges of parameters allow our domain walls can reach the Euclidean boundary.  In addition, some algebra (see appendix \ref{app:uniquezero}) shows that for $\mu_e >0$ we have $V'_{\text{eff}} > 0$ at any zero of $V_{\text{eff}}$, so there is a unique zero $r_{\min{}}$.  The Euclidean domain wall thus defines a curve along which $r$ decreases from $r=\infty$ to $r_{\min{}}$  and then returns to $r=\infty$ in a manner consistent with time reflection symmetry.

\begin{figure}[t]
\centerline{\includegraphics[width=0.3\textwidth]{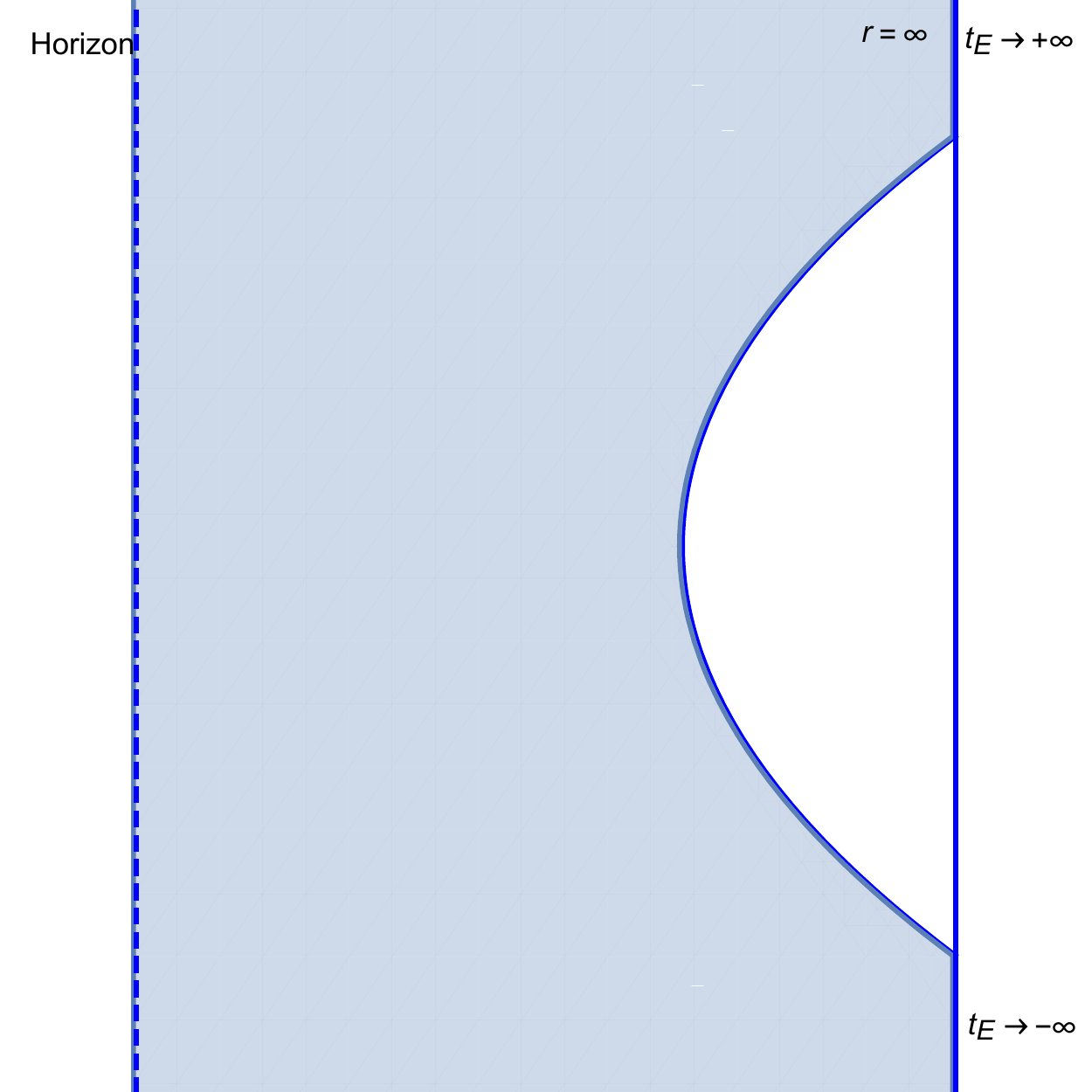}
	\hspace{.5cm}
	\includegraphics[width=0.3\textwidth]{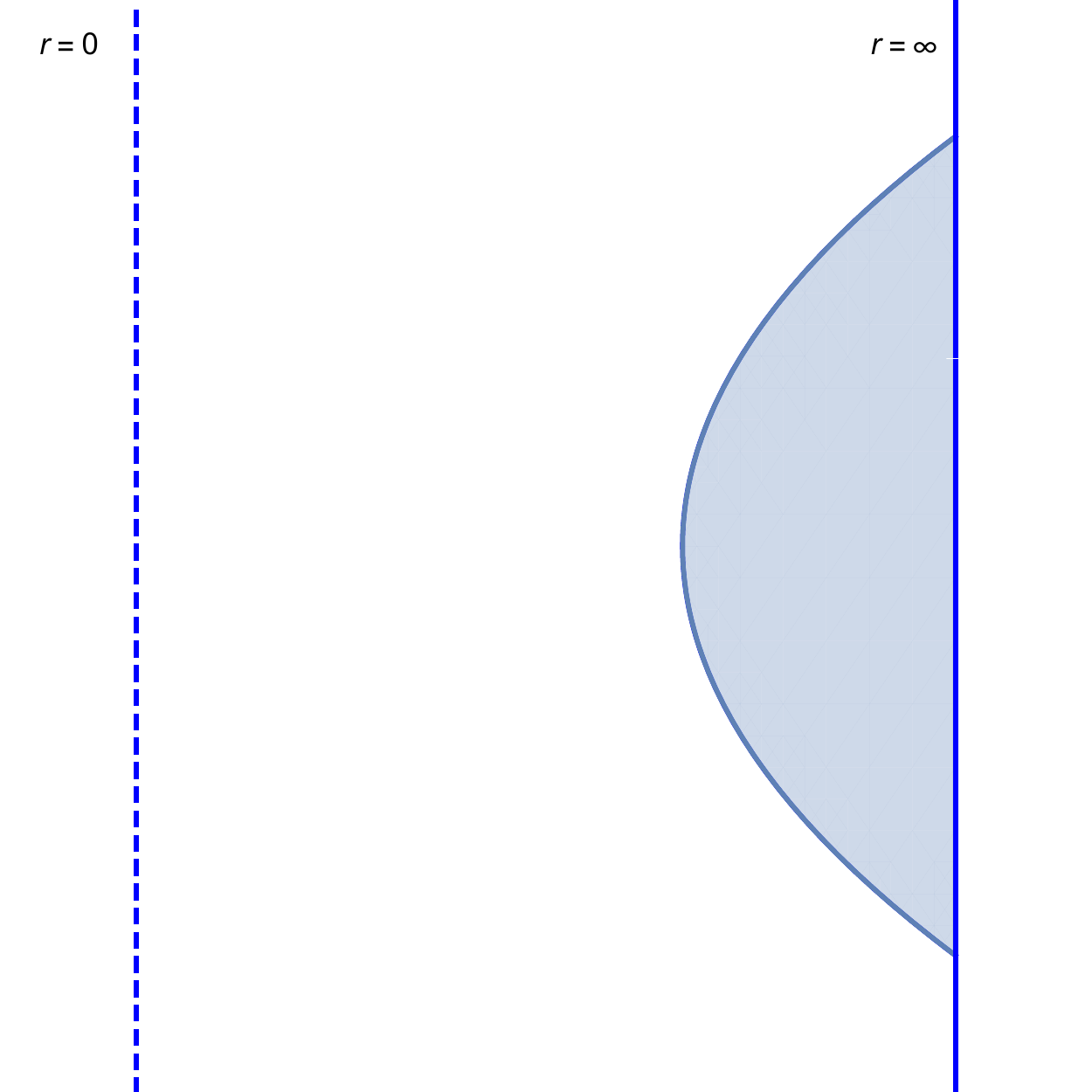}}
\caption{
 Unshaded regions provide allowed exteriors (left) and interiors (right) for  case (III), where $\alpha_i > \alpha_e >0$ at all $r$.  Since the Euclidean horizon is always excised, we take Euclidean time to run vertically over $(-\infty, +\infty)$. Wick rotating to Lorentz signature yields  time symmetric collapsing-shell solutions, where a domain wall outside its horizon at $t=0$ collapses to form a black hole at late times.}
\label{fig:caseIII}
\end{figure}

We can now address case $(III)$, where $\alpha_i > \alpha_e > 0$ at all $r$.  From \eqref{eq:tEEOM}, we see that the exterior solution moves outward from $r=r_{\min{}}$. As shown in figure \ref{fig:caseIII}, self-intersections of the wall in the exterior SAdS region may avoided by taking $t_{Ee}$ to be defined on $(-\infty, +\infty)$ (without identifications).  Such solutions are free of conical singularities since the condition $\alpha_e >0$ forces us to discard the piece of SAdS containing the Euclidean horizon.

Although there is no minimal surface at $t=0$, after Wick rotation to Lorentz signature the walls will accelerate inward from their initial location on the surface of time symmetry and form black holes at large $|t|$.  Such Lorentzian solutions thus represent black holes that form from time symmetric collapse.  We use the term collapsing-shell solutions to refer to them below.

It remains only to address case $(II)$, which turns out to have
various subcases associated with the signs of $\alpha_{i,e}$ at $r_{\min}{}$ and at $r = \infty$.  Since case $(II)$ has $\alpha_e$ negative at large $r$, and since the beginning of section \ref{subsec:AdS-SAdS} showed $\alpha_i(r_{\min{}})$ to be positive in all cases, four subcases remain:

\begin{align}
    (IIA+)\  & \kappa^2 -1 \ge \lambda , \alpha_e(r_{\min{}}) >0  & \alpha_i >0 \ {\rm at \ all} \ r, \alpha_e \ {\rm changes \ sign}, \\
    (IIA-)\ &  \kappa^2 -1 \ge \lambda, \alpha_e(r_{\min{}}) \le 0  & \alpha_i >0 \ {\rm at \ all} \ r, \ {\rm no} \ \alpha_e \ {\rm sign \ change}, \\
      (IIB+)\  & \kappa^2 -1 < \lambda, \alpha_e(r_{\min{}}) >0  & \alpha_i \rightarrow -\infty \ {\rm at \ large} \ r, \alpha_e \ {\rm changes \ sign}, \\
    (IIB-)\  & \kappa^2 -1 < \lambda, \alpha_e(r_{\min{}}) \le 0  & \alpha_i \rightarrow -\infty \ {\rm at \ large} \ r, \ {\rm no} \ \alpha_e \ {\rm sign \ change}.
\end{align}
Here we have emphasized that \eqref{eq:betas} clearly shows the sign of $\alpha_i$ at large $r$ to agree with the sign of $-(1+\lambda-\kappa^2)$.   The sign of $\alpha_e(r_{\min{}})$ is more complicated to determine and will be studied numerically below.

None of these subcases allow obstructions from the interior.  In the $(IIA\pm)$ cases,  the interior solution behaves just as in case $(III)$ and thus remains smooth for all parameters in this regime.  The situation is more interesting for cases $(IIB\pm)$ where $\alpha_i$ changes sign at its unique zero.  Via \eqref{eq:tEEOM}, this sign change entails a maximum value of $t_{Ei}$ and the interior takes one of the forms shown in figure \ref{fig:caseIIB} below.  As described in the figure caption, any apparent self-intersections can then be removed by inserting a conical singularity in the unphysical excised region of the interior solution, so the interior is smooth for all parameters.

\begin{figure}[t]
\centerline{\includegraphics[width=0.3\textwidth]{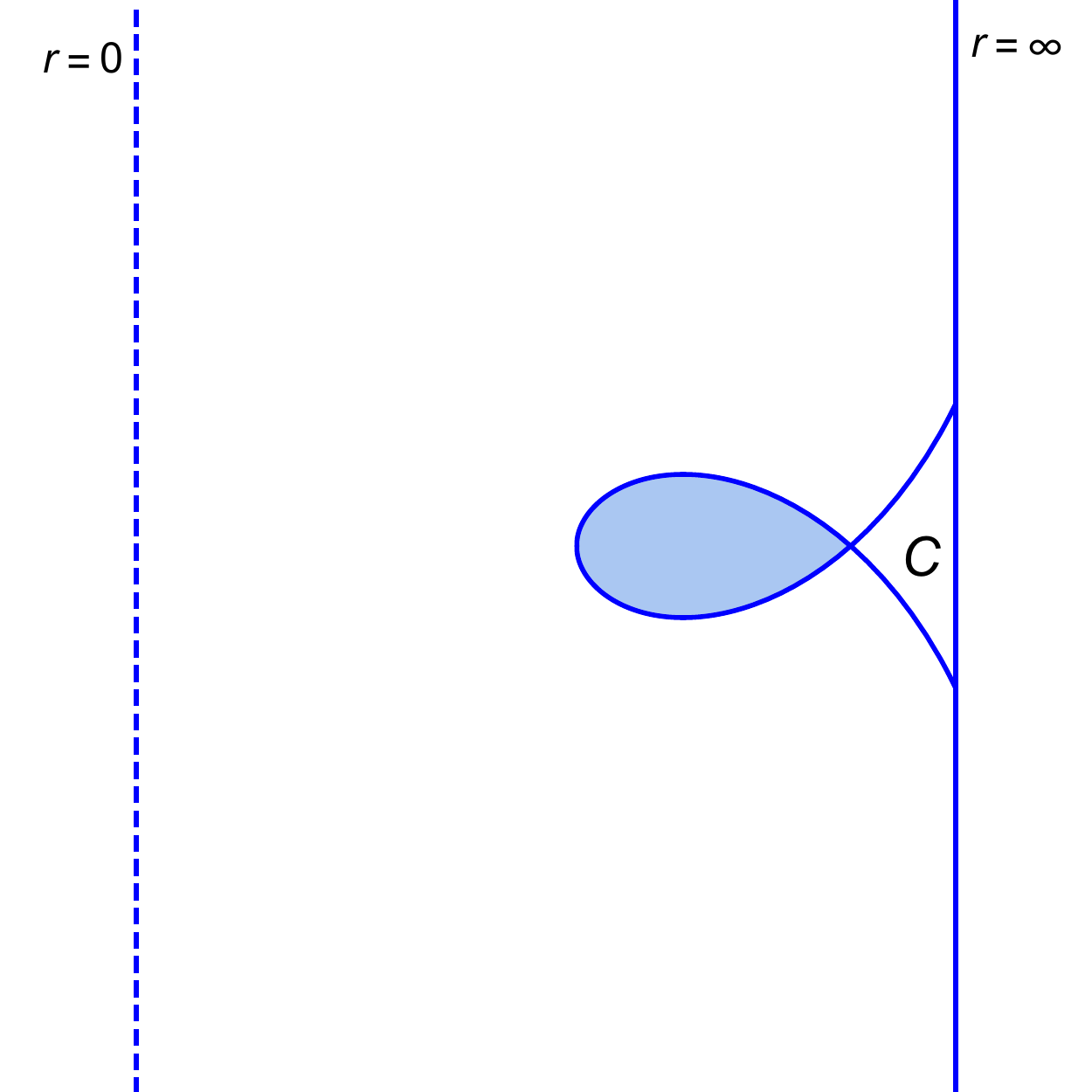}
	\hspace{.5cm}
	\includegraphics[width=0.3\textwidth]{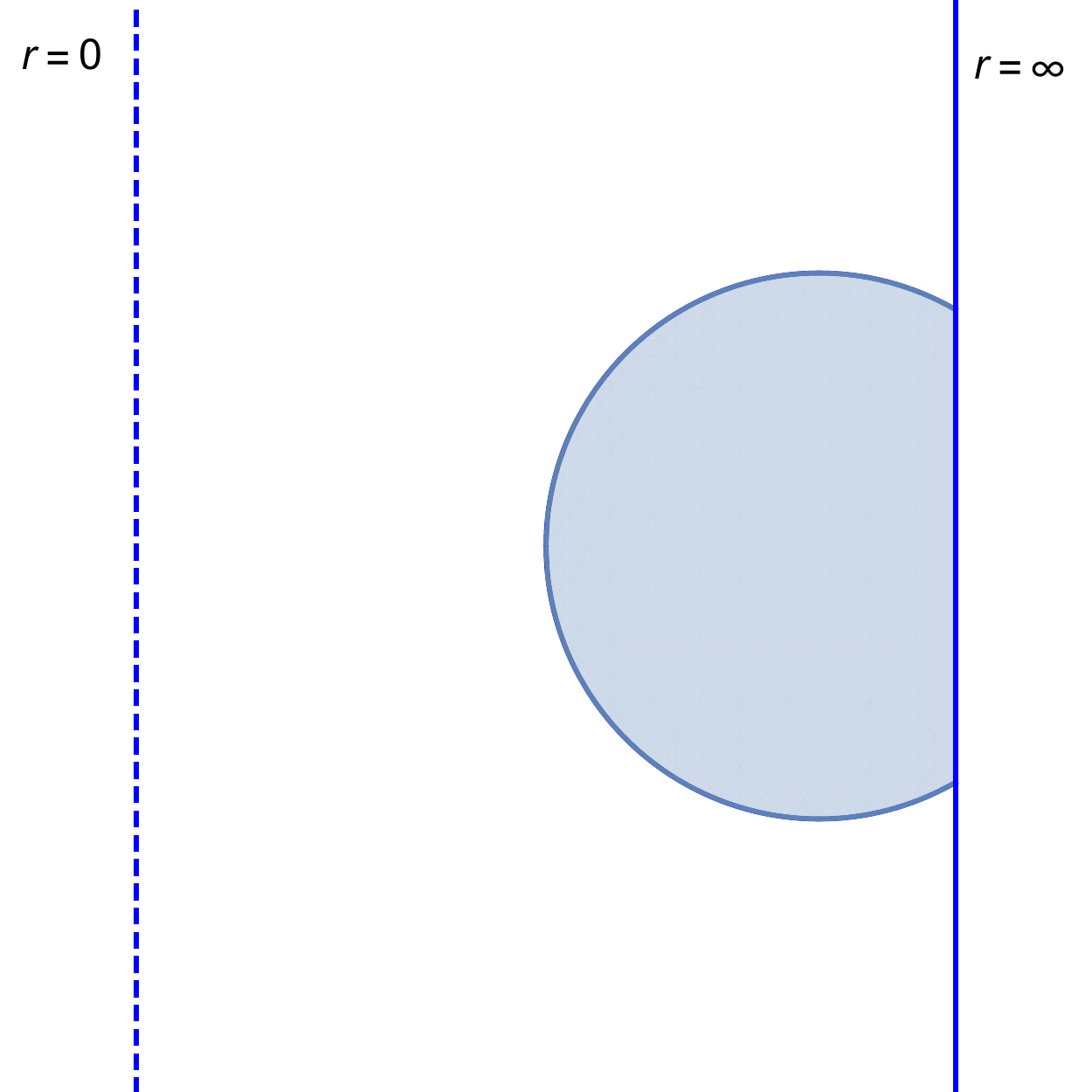}}
\caption{Interior solutions for cases $(IIB+), (IIB-)$.  For these cases, $\alpha_i$ is negative at $r_{\min{}}$ but positive at large $r$.  The change of sign means that $t_{Ei}$ has a maximum so that the interior solution takes one of the forms shown.  Since $\alpha_i (r_{\min }) > 0$, in each case we must keep a piece of the solution containing $r=0$. When there are no self-intersections (right) this presents no problems.  And since $\alpha_i$ can vanish at only one value of $r > r_{\min }$, self-intersections are always of the form shown at left.  Such self-intersections can be removed by inserting an appropriate conical singularity into the shaded (unphysical) region.  In terms of the physical (unshaded) region, the effect is to include two copies of the region marked $C$ near $t_E =0$ and $r=\infty$.  The first copy $C_U$ of $C$ is attached smoothly to the rest of the interior solution along the {\it upper} left boundary $C$, while the lower left boundary of $C_U$ is a domain wall junction with the exterior.  In contrast, the second copy $C_L$ of $C$ is attached smoothly to the rest of the interior solution along the {\it lower} left boundary $C$, while the upper left boundary of $C_U$ is a domain wall junction with the exterior.}
\label{fig:caseIIB}
\end{figure}

\begin{figure}[t]
\centerline{\includegraphics[width=0.3\textwidth]{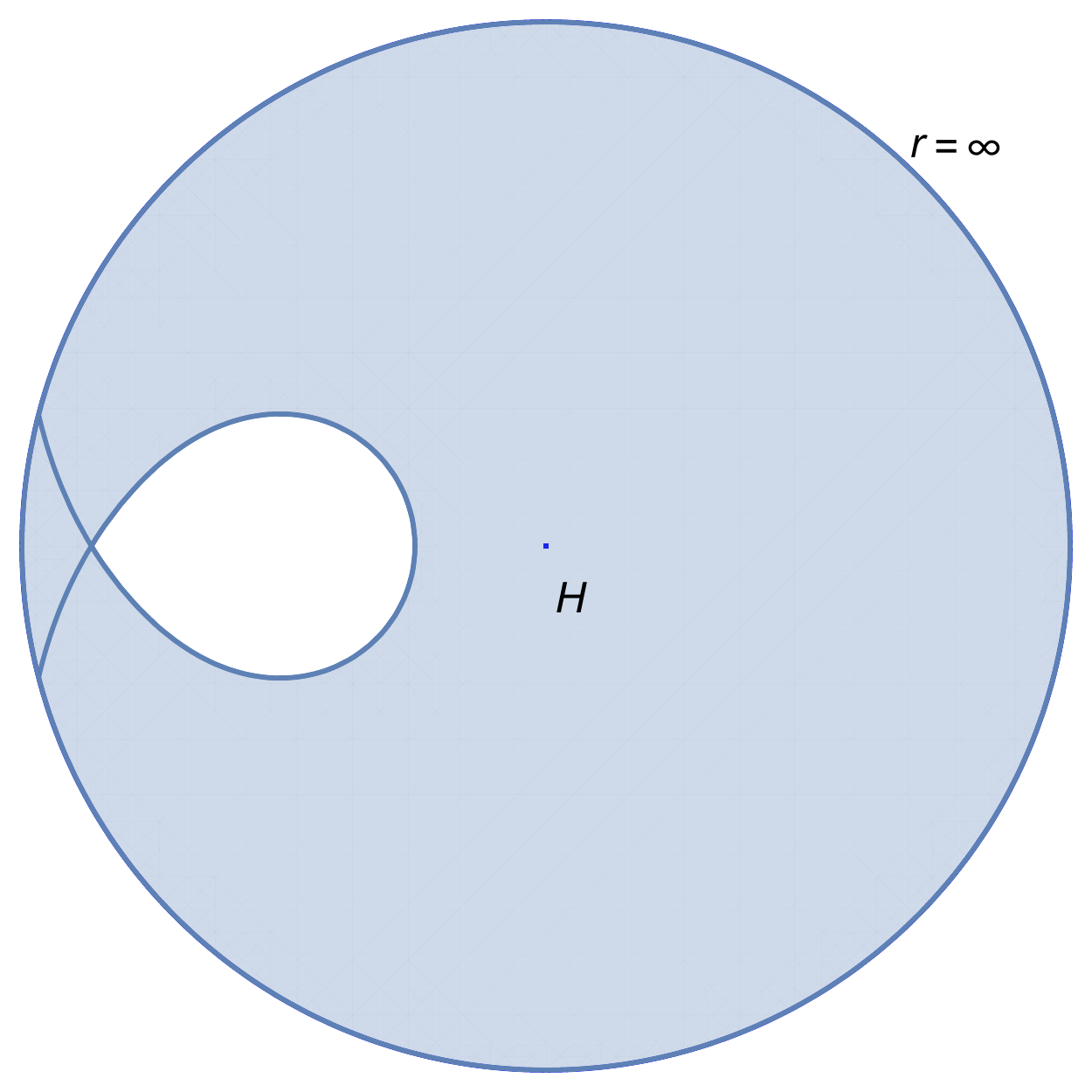}
	\hspace{.5cm}
	\includegraphics[width=0.3\textwidth]{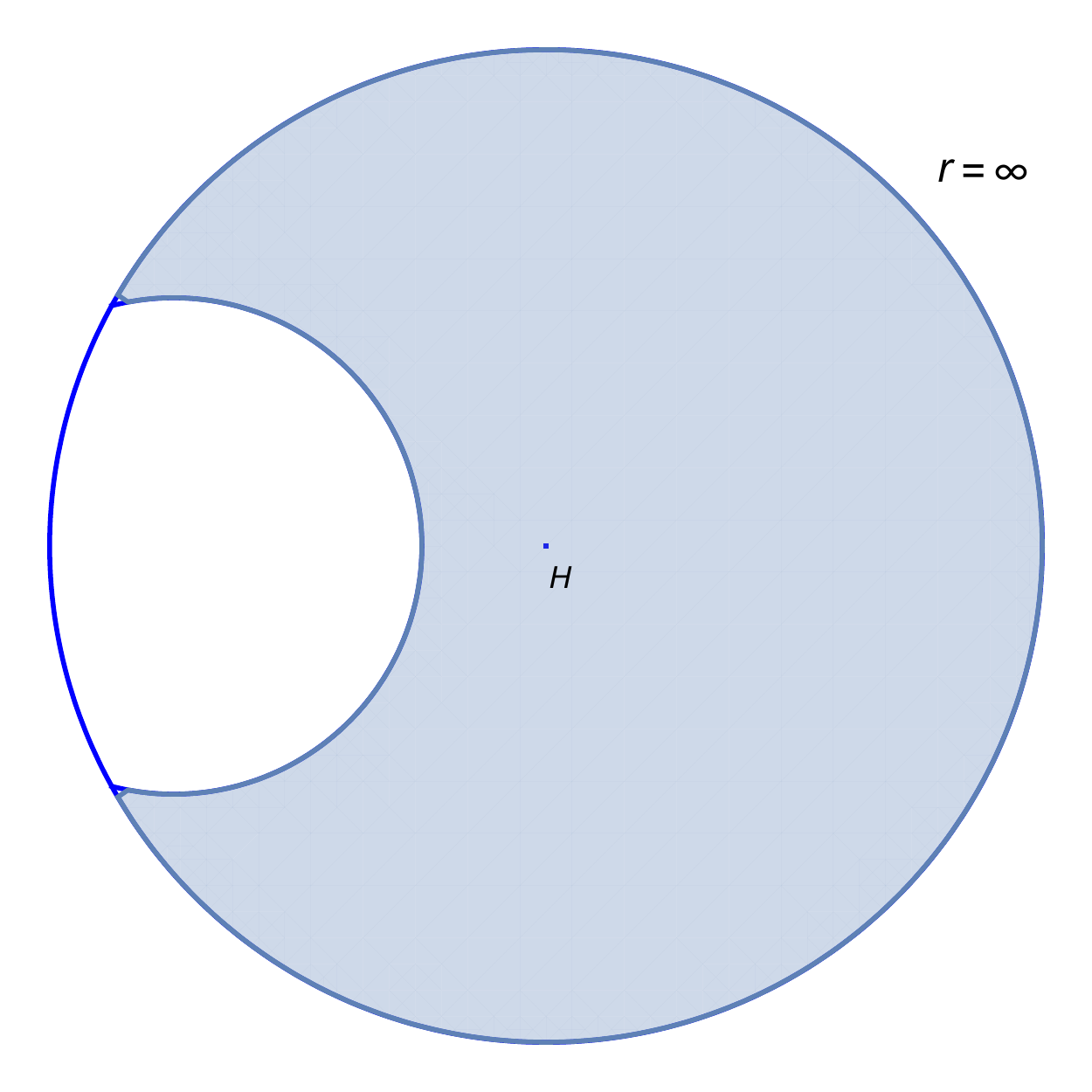}}
\caption{Cases $(IIA+), (IIB+)$ have $\alpha_e$ is positive at $r_{\min{}}$ but negative at large $r$.  The change of sign means that $t_{Ee}$ has a maximum, and the solution must take one of the forms shown.  Since $\alpha_e(r_{\min{}}) >0$ requires us to keep the (unshaded) piece of the solution containing the larger $r$ side of the wall at $r_{\min}$, any point with $t_{Ee}=0$ and $r\neq r_{\min{}}$ would yield a self-intersection (left).  Avoiding such self-intersections (right) thus requires $t_{Ee} > 0$ for $r > r_{\min{}}$ along one entire branch of \eqref{eq:tEEOM} with $t_{Ee} <0$ on the other.}
\label{fig:caseII+}
\end{figure}

The $(-)$ cases have $\alpha_e$ negative at all $r$, and in particular at $r_{\min }$.  Such situations hold the potential to create bags of gold.  We will show that this potential is realized by studying these settings in detail in section \ref{sec:search} below.

The remaining cases $(IIA+)$ and $(IIB+)$ may now be handled quickly.  There the function $\alpha_e$ changes sign, and since $\alpha_e$ is non-negative at $r_{\min{}}$, we now keep the external SAdS piece in which $r$ approaches  $r_{\min{}}$ from the outside.  These solutions thus create collapsing shells much as in case $(III)$.  As shown in figure \ref{fig:caseII+}, the condition to avoid self-intersections is now that $t_{Ee}$ is positive for $r > r_{\min{}}$ along one entire branch of \eqref{eq:tEEOM} with $t_{Ee} <0$ on the other.

\section{Creating bags of gold}
\label{sec:search}

\begin{figure}[t]
\centerline{\includegraphics[width=0.3\textwidth]{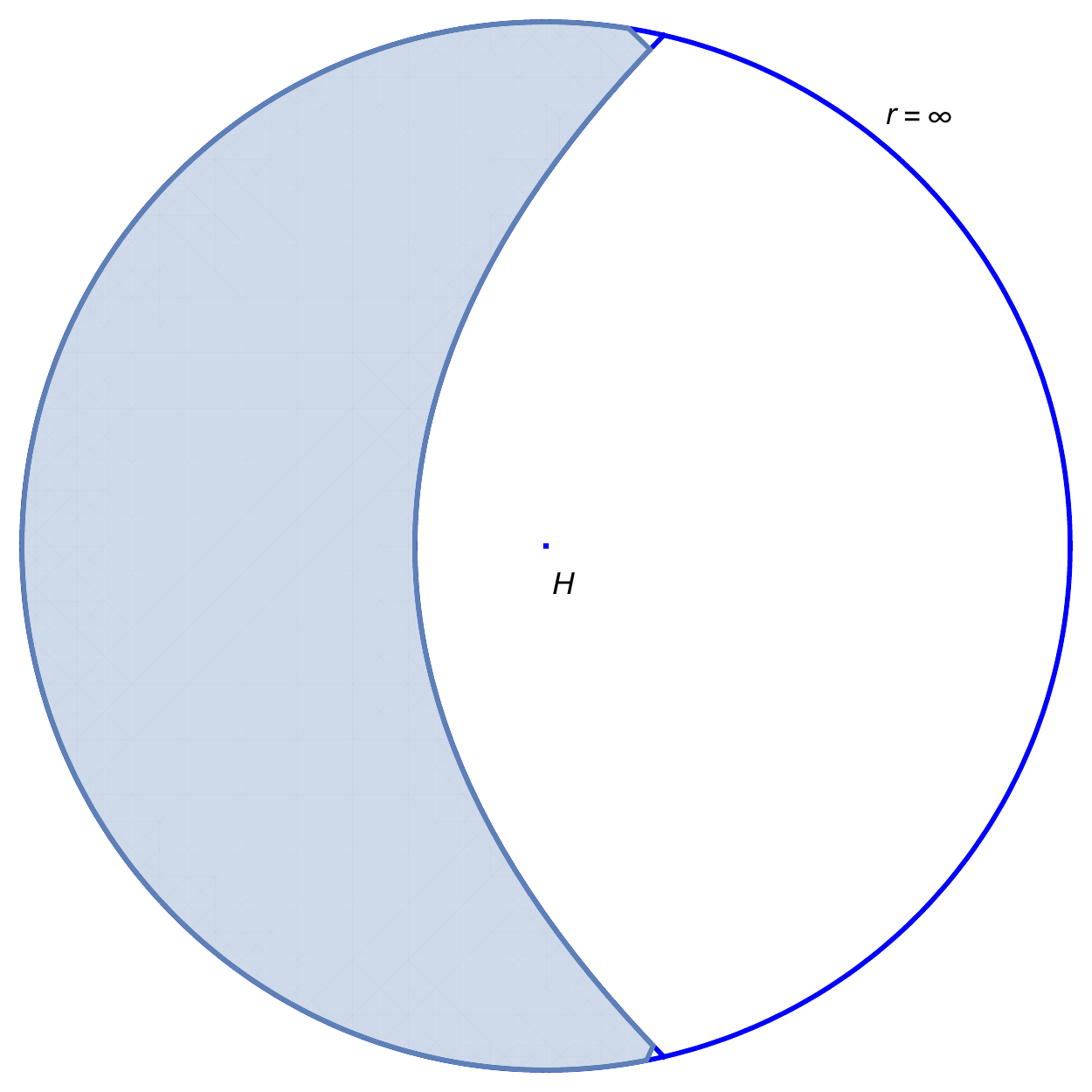}
	\hspace{.5cm}
	\includegraphics[width=0.3\textwidth]{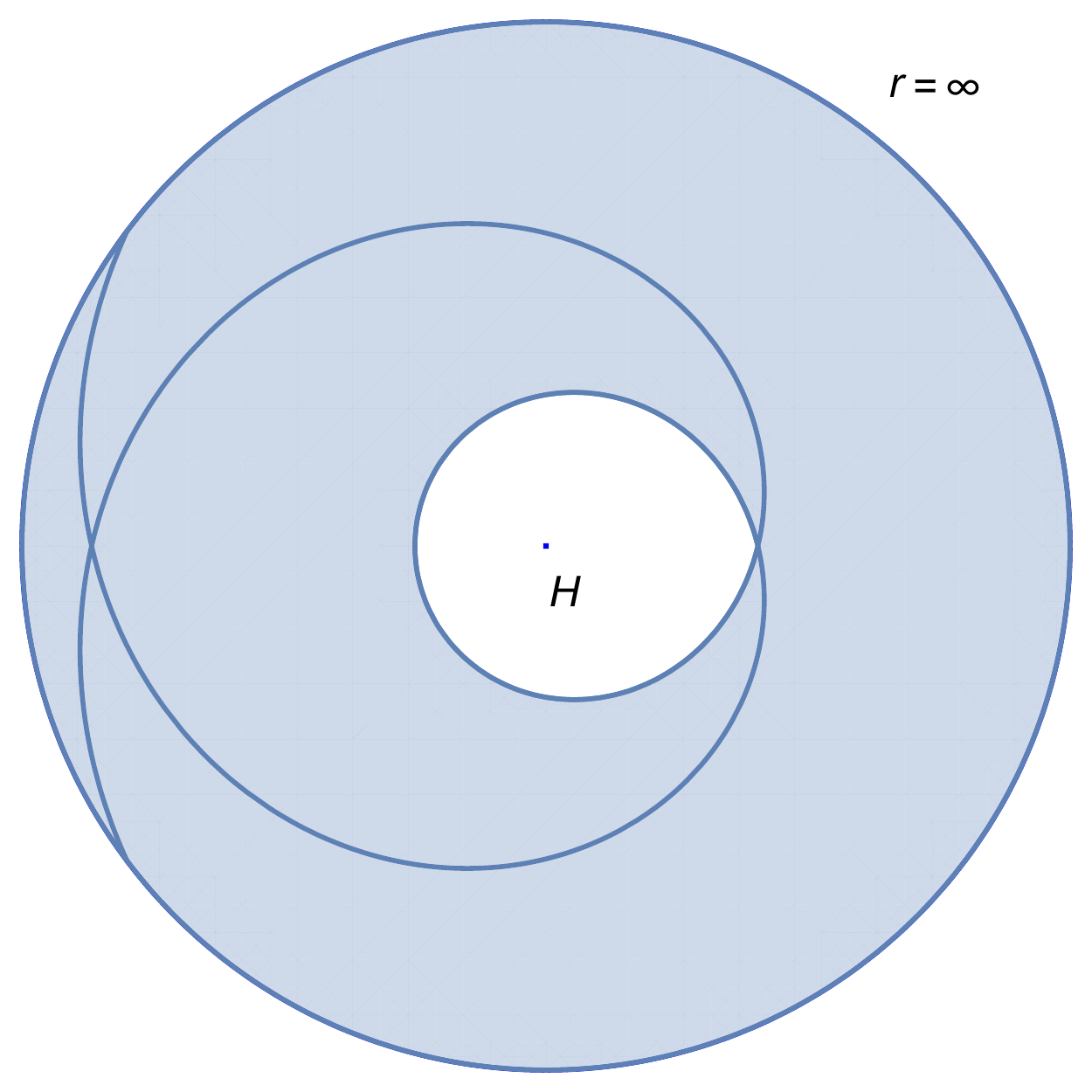}}
\caption{Cases $(IIA-)$ and $(IIB-)$ have $\alpha_e$ negative at all $r$ so that exterior solutions tend to spiral outward from $r_{\min{}}$.  {\bf Left:} In some cases the wall reaches the Euclidean AdS boundary without self-intersections.  We may then keep the unshaded region and potentially build a smooth solution, depending on the interior.  Recall that the interior imposes no restrictions in case $(IIA-)$, where interior solutions are as in case $(III)$, shown at right in figure \ref{fig:caseIII}. {\bf Right:} In other cases, one or more self-intersections occur before the wall reaches the Euclidean AdS boundary and smooth solutions do not exist.}
\label{fig:caseII-}
\end{figure}

We now carefully examine cases $(IIA-)$ and $(IIB-)$ to show that we can create bags of gold. In either such case we have $\alpha_e$ negative at all $r$, so from \eqref{eq:tEEOM} the sign of $dt_{Ee}/dr$ changes only at $r_{\min{}}$. This means that the exterior solutions spiral outward from $r_{\min{}}$ and take one of the forms shown in figure \ref{fig:caseII-}. Any solutions without self-intersections will describe bags of gold\footnote{If it is free of self-intersections, the special case where $\alpha_e$ vanishes at $r_{\min{}}$ describes a degenerate bag-of-gold of zero size where the domain wall is located at its Schwarzschild radius on the surface of time symmetry ($t=0$).}.  

As can be seen from figure \ref{fig:caseII-}, such solutions occur precisely when the world line of the wall wraps less than once around the origin.  This is the condition that the range  $\Delta t_{Ee}$ of $t_{Ee}$ over the domain wall world line be less than the natural Euclidean period of the external SAdS solution.  One can explore this condition numerically in detail, but we show in section \ref{subsec:nose} that there is at least a regime with large $\kappa, \mu_e$  where the condition is satisfied.  At the level of our bottom-up analysis, this will then establish that at least some bags of gold (i.e., those in that regime) have CFT duals.

Now, as shown in appendix \ref{subsec:bogssmall}, for $D >3$ the bags of gold that can be created from Euclidean path integrals using a single domain wall have their size bounded by a power of $\mu_e$ and so produce no immediate tension with the dual CFTs Bekenstein-Hawking density of states.  For $D=3$ one can in fact create arbitrarily large bags of gold at fixed $\mu_e$, but only by tuning the parameter $A$ to be small and taking $\kappa > 4/3$ (see again appendix \ref{subsec:bogssmall}).  As a result, the bag of gold is subject to an additional IR cutoff associated with the finite value of the internal ($i$) cosmological length scale $\ell_i = \ell_e /\sqrt{-\lambda} \approx
 \ell_e/(\kappa-1) < 3 \ell_e$ which again limits the entropy such bags may contain.

However, analyzing cases with multiple domain walls in section \ref{subsec:bogslarge} will show that arbitrarily large bags of gold can be created at fixed $\mu$, and that no fine-tuning of model parameters is required. Furthermore, we argue in section \ref{subsec:dominate} that in our models these are the only spherically-symmetric bulk saddles for appropriately chosen path integrals.  We thus expect them to dominate, so that such bag-of-gold geometries do indeed have good CFT duals.

\subsection{External solutions without self-intersections}
\label{subsec:nose}

In particular, we can identify a regime free of self-intersections by studying the limit $\mu_e \rightarrow +\infty$.  Numerical results show that such solutions can also exist at moderate-to-small $\mu_e$, but at large $\mu_e$ the treatment simplifies and more can be said analytically. As usual, this simplification is associated with the fact that large SAdS black holes can be approximated as planar, so that the term of $1$ can be dropped from both $f$ and $V_{\rm eff}$.  Since this large $\mu_e$ limit will also play an important role in later sections, we take the opportunity to develop it carefully here, and in particular to do so in a way that will also allow the case $\mu_i >0$ to be considered later.

To be specific, we consider any large $\mu_e$ limit in case $(II)$ in which $B <0$, the quantity $\gamma:=-\mu_e A/B$ is bounded away from one, the positive quantities $A$ and $B^2/4C$ are all bounded away from zero, and in which
\begin{equation}
\label{eq:alphalim}
\frac{\kappa \sqrt{C}}{\mu_e (1+
\lambda + \kappa^2)}  = O(\mu_e^{-\eta_1}), \  \ \ {\rm and} \ \ \
\frac{AC}{B^2}  = O(\mu_e^{-\eta_2})
\end{equation}
for some $\eta_1, \eta_2 >0$.

For example, with $\mu_i=0$ one may take $\mu_e$ large holding fixed (finite) values of
\begin{equation}
\label{eq:samplelim}
\nu := \frac{\mu_e}{\kappa} \text{, }\rho:=  \frac{1 + \lambda + \kappa^2}{\kappa},
\end{equation}
so that (even at finite $\mu_e$) we have
\begin{equation}
\label{eq:ABCrn}
A = 1 - \frac{\rho^2}{4} \text{, }B = - \mu_e + \frac{\nu \rho }{2} \text{, }C = \frac{\nu^2}{4}.
\end{equation}
For $2 > \rho > 0$ the limit lies in case $(II)$ as desired.
Furthermore, $B \rightarrow - \infty$ at large $\mu_e$ while $A, C$ are held fixed.  This satisfies all of the conditions above, and in particular \eqref{eq:alphalim} yields
\begin{equation}
\label{eq:alphalimworks}
\frac{\kappa \sqrt{C}}{\mu_e (1+
\lambda + \kappa^2)}  = \frac{\nu }{2\rho \mu _e} = O(\mu_e^{-1}),\text{ }\frac{AC}{B^2} =\frac{{{\nu ^2}\left( {4 - {\rho ^2}} \right)}}{{4{{\left( {2{\mu _e} - \nu \rho } \right)}^2}}}= O(\mu_e^{-2}).
\end{equation}
Other limits satisfying the above conditions will also be of interest in later sections.

The first step in showing the above limits to be free of self-intersections is to estimate $r_{\min }$.  At large $\mu_e$ the horizon radius $r_h$ at which $f=0$ is $r_h = \mu_e^{-\frac{1}{D-1}} (1 + O(1/\mu_e))$. So since $V_{\rm eff} = f_e-\alpha_e^2$ from \eqref{eq:V2} implies $V_{\rm eff}(r_h) <0$, we must have $r_{\min } > r_h = \mu_e^{-\frac{1}{D-1}} (1 + O(1/\mu_e))$ so that $r_{\min }$ must become large as well.  It is then useful to rewrite \eqref{eq:potential} in the form
\begin{equation}
V_{\rm eff} = r^2\left[A + \frac{B^2}{4C}  + \frac{1}{r^2} - C\left(r^{-(D-1)} - \frac{B}{2C} \right)^2 \right].
\end{equation}
Using $r_{\min }^{-(D-1)} > 0$ one then finds
\begin{align}
\label{eq:sqrt}
r_{\min }^{-(D-1)} & = \frac{B}{2C} + \sqrt{\frac{B^2}{4C^2} + \frac{A}{C} + \frac{1}{Cr^2_{\min }}}.
\end{align}
This quantity must approach zero since $r_{\min }$ becomes large.  But $\frac{B^2}{4C^2}$ is bounded below, so the two terms must nearly cancel.  This requires $\frac{A}{C} + \frac{1}{Cr^2_{\min }} \ll
\frac{B^2}{4C^2}$. Furthermore, $A \gg 1/r_{\min }^2$.  Thus we find
\begin{equation}
\label{eq:sqrtapproximation}
r_{\min }^{-(D-1)}  = -\frac{A}{B}(1 + O(\mu_e^{-\frac{1}{D-1}})) = \frac{\gamma}{\mu_e}(1 + O(\mu_e^{-\frac{1}{D-1}})).
\end{equation}
Using the condition that $\gamma $ is bounded away from one, we thus find
\begin{equation}
\label{eq:rhrmin}
r_{\min } - r_h = O(r_{\min }).
\end{equation}
For future use, we also note that a short computation from \eqref{eq:ABC} yields
\begin{equation}
\label{eq:ratios}
\frac{B}{2C} = \frac{1+\lambda-\kappa^2}{\mu_e},\text{ } \frac{B^2}{4C^2} + \frac{A}{C} = \frac{-4\lambda \kappa^2}{\mu_e^2}.
\end{equation}

Understanding $t_{Ee}(r)$ requires controlling the three ingredients $V_{\rm eff}, \alpha_e, f_e$ in \eqref{eq:tEEOM}.  The last two are straightforward, as for $r > r_{\min } \sim \mu^{\frac{1}{D-1}}$ our \eqref{eq:rhrmin} yields
\begin{equation}
\label{eq:fest}
f_e = r^2 + 1 - \frac{\mu_e}{r^{D-3}} = r^2 \left(1 - \frac{\mu_e}{r^{D-1}}\right)\left(1 + O(r_{\min }^{-2}) \right)
\end{equation}
and \eqref{eq:alphalim} gives
\begin{equation}
\label{eq:alphaest}
\alpha_e = - \frac{r \rho}{2} \left(1 + O(\mu_e^{-\eta}) \right),
\end{equation}
where we have used the definition of $\rho$ from \eqref{eq:samplelim} whether or not $\rho$ is held constant in our limit and $\eta$ is the smallest of $\left\{\eta_1, \frac{\eta_2}{2}, \frac{2}{D-1}\right\}$. In particular, we see that bag-of-gold condition $\alpha_e(r_{\min })<0$ holds as a consequence of our assumptions.

To control $V_{\rm eff}$, it is useful to build on \eqref{eq:sqrt} by defining
\begin{align}
r_{-}^{-(D-1)} & := -\frac{B}{2C} + \sqrt{\frac{B^2}{4C^2} + \frac{A}{C} + \frac{1}{Cr^2_{\min }}},
\end{align}
and also
\begin{equation}
\tilde V_{\rm eff} (r) := A r^2 \left(1 - \frac{r_{\min }^{D-1}}{r^{D-1}} \right)\left(1 + \frac{r_-^{D-1}}{r^{D-1}} \right) = \frac{A r_{\min }^2}{A r_{\min }^2+1}\left(V_{\rm eff}(r)-1+\frac{r^2}{r_{\min }^2}\right).
\end{equation}
Note that in case $(II)$ the functions $\tilde V_{\rm eff}, V_{\rm eff}$ are both positive for $r > r_{\min }$ and negative for $0< r< r_{\min }$.  Since we found above that $A \gg r^{-2}_{\min }$, the above equation implies
\begin{equation}
\label{eq:Vders}
\tilde V_{\rm eff}' = V_{\rm eff}'\left(1 + O\left(\frac{1}{r^{2}_{\min }}\right)\right) .
\end{equation}
Integrating \eqref{eq:Vders} from the common zero at $r_{\min}$ then yields
\begin{equation}
\label{eq:2Vs}
V_{\rm eff} = \tilde V_{\rm eff} \left(1 - O\left(\frac{1}{r^{2}_{\min}}\right)\right).
\end{equation}
On the other hand,
\eqref{eq:alphalim} implies $\frac{r_-^{D-1}}{r_{\min }^{D-1}} = O(\mu_e^{-\eta_2/2})$ so for $r > r_{\min }$ we have
\begin{equation}
\label{eq:tVs}
\tilde V_{\rm eff} (r) = A r^2 \left(1 - \frac{r_{\min }^{D-1}}{r^{D-1}} \right)\left(1 + O(\mu_e^{-\eta_2/2}) \right).
\end{equation}

Introducing $\tilde r = r/r_{\min }$,  we may now combine \eqref{eq:tEEOM}, \eqref{eq:sqrtapproximation}, \eqref{eq:fest}, \eqref{eq:alphaest}, \eqref{eq:2Vs}, and \eqref{eq:tVs} to find
\begin{equation}
\begin{aligned}
\label{eq:tEelargemu}
\frac{dt_{Ee}}{d\tilde r} &= -r_{\min} \frac{\alpha_e}{f_e\sqrt{V_{\text{eff}}}} \\  &= \frac{\rho}{ 2\tilde r^2 r_{\min} \sqrt{ A}}
 \left(1 - \frac{\gamma}{\tilde r^{D-1}} \right)^{-1}
  \left(1-\frac{1} {\tilde r^{D-1}} \right)^{-\frac{1}{2}} \left( 1 + O(\mu_e^{-\eta}) \right)
\end{aligned}
\end{equation}
for all $r > r_{\min }$, where $\eta$ is the smallest of $1/(D-1), \eta_1, \eta_2/2$.   Note that the physically interesting parameter is the ratio of
the range of $t_{Ee}$ to the Euclidean period $\beta = \frac{4\pi}{(D-1)\mu_e^{1/{(D-1)}}}(1 + O(\mu_e^{-1/(D-1)})$, and that the factor of $\mu_e^{\frac{1}{D-1}}$ in $r_{\min }$ in the denominator on the right-hand-side of \eqref{eq:tEelargemu} cancels in this ratio. For simplicity, we set $t_{Ee}=0$ at $r_{\min }$, so that the desired ratio is $\frac{2t_{Ee}(r=\infty)}{\beta}$.

The factor of $\tilde r^{-2}$ on the right-hand side means that $t_{Ee}(r)$ is finite as $r \rightarrow \infty$. Indeed, if one drops the $O(\mu_e^{-\frac{1}{D-1}})$ corrections the integral over the full curve can be done explicitly for $D=3,4,5$.  The case $D=3$ is simplest, as then
\begin{align}
\label{eq:3Dlimt}
\frac{t_{Ee}}{\beta} &=  \int _{1}^{\tilde r} d\tilde r \frac{\sqrt{\gamma} \rho \tilde r}{ 4 \pi  \sqrt{A}(\tilde r^2 -\gamma)\sqrt{ \tilde r^{2} -1 }}
\left( 1 + O(\mu_e^{-\eta})\right) \\
&
=\frac{\rho}{4\pi \sqrt{1-\gamma }}  \sqrt{\frac{\gamma}{A}} \arctan \left( \sqrt{\frac{\tilde r^2 -1}{1-\gamma}}\right) \left( 1 + O(\mu_e^{-\eta})\right).
\end{align}
Thus we find
\begin{equation}
\label{eq:3DDt}
\frac{2t_{Ee} (r = \infty)}{\beta} =   \frac{\rho}{ 4  \sqrt{1-\gamma}} \sqrt{\frac{\gamma}{A}}
\left( 1 + O(\mu_e^{-\eta})\right).
\end{equation}
In particular, for $\mu_i=0$ tracing through the various definitions gives
\begin{equation}
\label{eq:3DDtnomuint}
\frac{2t_{Ee} (r = \infty)}{\beta} =   \frac{1}{ 2\sqrt{1 - \frac{2}{1 + \lambda + \kappa^2}}}
\left( 1 + O(\mu_e^{-\eta})\right).
\end{equation}
So as long as $1 + \lambda + \kappa^2 > 8/3$ we find $\frac{2t_{Ee} (r = \infty)}{\beta} < 1$ and there are no intersections at large $\mu$.  In particular, this holds when the parameters in \eqref{eq:ABCrn} are held fixed at $\mu_e\rightarrow \infty$.

\begin{figure}[t]
\centerline{\includegraphics[width=0.5\textwidth]{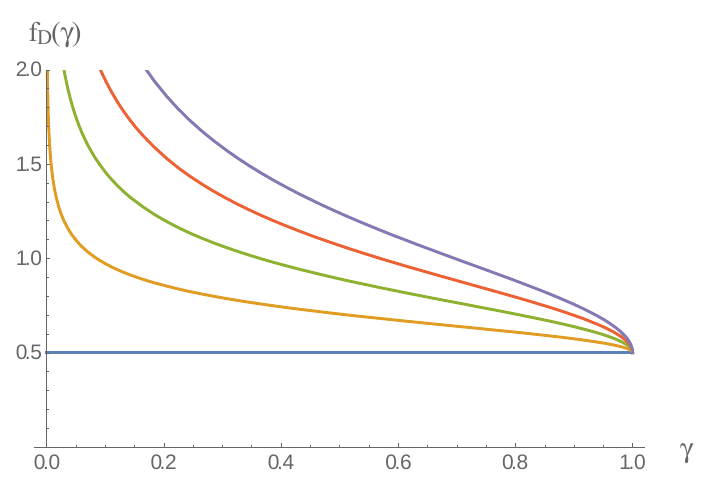}}
\caption{The functions $f_D(\gamma)$ from \eqref{eq:fDdef} for $D=3,4,5,6,7$ on the interval $\gamma\in (0,1)$.  The constant function $f_3= \frac{1}{2}$ (lowest curve) is included for reference and, moving upward in the figure, successive curves have increasing values of $D$.}
\label{fig:fD}
\end{figure}
The results for $D\ge 4$ are more complicated, but it is useful to write them in the form
\begin{equation}
\label{eq:Dg4Dt}
\frac{2t_{Ee} (r = \infty)}{\beta} =  \frac{\rho}{2\sqrt{1-\gamma}} \sqrt{\frac{\gamma}{A}} f_D(\gamma)
\left( 1 + O(\mu_e^{-\eta})\right).
\end{equation}
in terms of the functions
\begin{equation}
\begin{aligned}
\label{eq:fDdef}
f_D(\gamma) & := (D-1)\frac{\gamma^{\frac{1}{D-1}}\sqrt{1-\gamma} }{2 \pi \sqrt{\gamma}}  \int _1^\infty \frac{d\tilde r}{\tilde r^2} \left(1 - \frac{\gamma}{\tilde r^{D-1}} \right)^{-1} \left(1 - \frac{1}{\tilde r^{D-1}} \right)^{-\frac{1}{2}} \\
& = \frac{\Gamma\left(\frac{1}{D-1} \right)}{2 \sqrt{\pi} \gamma^{\frac{D-3}{2(D-1)}} \Gamma\left( \frac{D+1}{2(D-1)}\right)}{}_2 F_1\left(-\frac{D-3}{2(D-1)},
     \frac{1}{2}, \frac{D+1}{2(D-1)}, D-1\right).
\end{aligned}
\end{equation}
In \eqref{eq:fDdef},  the function ${}_2F_1$ is the standard hypergeometric function and the final expression is a conjecture that we have checked using Mathematica for all integer $D$ in the range $3 \le D \le 50$.  The function $f_D(\gamma)$ is naturally defined for $\gamma \in (0,1]$.  There the choice $D=3$ yields $f_3(\gamma) =\frac{1}{2}$ in agreement with our results above.  In contrast, for $D>3$ the functions $f_D$ monotonically decrease from positive infinity at $\gamma=0$ to $\frac{1}{2}$ at $\gamma =1$.  Interestingly, however, the divergence as $\gamma\rightarrow 0$ is fairly slow if $D$ is not too large; see figure \ref{fig:fD}.  Indeed we find the expansion
\begin{equation}
\label{eq:fDexp}
f_D(\gamma) =  \frac{\Gamma\left(\frac{1}{D-1}\right)}{2 \sqrt{\pi} \Gamma\left(\frac{D+1}{2(D-1)}\right) } \frac{\gamma^{\frac{1}{D-1}}}{\sqrt{\gamma}}  (1 + O(\gamma)).
\end{equation}

In particular, for $\mu_i =0$ the prefactor in \eqref{eq:Dg4Dt} is
\begin{equation}
 \frac{\rho}{2\sqrt{1-\gamma}} \sqrt{\frac{\gamma}{A}} = \frac{1}{\sqrt{1-\frac{2\rho}{\kappa}}} =  \frac{1}{ \sqrt{1 - \frac{2}{1 + \lambda + \kappa^2}}},
\end{equation}
and $\gamma = (1 - \rho^2/4)/(1 - \rho/2\kappa)$, so
$\frac{2t_{Ee} (r = \infty)}{\beta} \rightarrow \frac{1}{2}$ as $\rho \rightarrow 0$ as long as $\rho/\kappa$ also vanishes in that limit.  Since this ratio is less than 1, we find not self-intersecting solutions that create bags of gold in all bulk spacetime dimensions $D$.

\subsection{Euclidean wormholes and large bags of gold from multiple domain walls}
\label{subsec:bogslarge}

As stated in the introduction, it is of interest to understand whether the bags of gold we create can become large inside a black hole of fixed surface area (here, fixed $\mu_e$).   But for solutions with a single domain wall of the sort we have studied thus far, the size of a bag-of-gold is largely dictated by $r_{\min{}}$, which from \eqref{eq:metric} is in fact the {\it maximum} radius of any $S^{D-2}$ of spherical symmetry inside the black hole.  And as noted above, at large $\mu_e$ the Euclidean solutions tends to have $r_{\min } \propto \mu_e^{\frac{1}{D-1}}$.  This makes it difficult to find Euclidean solutions that create bags of gold that become parametrically large at fixed $\mu_e$.  This is especially so if, in order to avoid models that one might consider less likely to have CFT duals, we wish to exclude models in which $\kappa $, $\lambda $ are fine-tuned to some order in $\ell_p/\ell$.   Of course, Lorentz-signature solutions in which the bag-of-gold becomes large are easy to construct by taking the interior to inflate, but such models lie in case $(I)$ so as shown in section \ref{subsec:dS-SAdS} they cannot be created by smooth saddle-points of AlAdS path integrals.

Some specific bounds on $r_{\min }$ are established in appendix \ref{subsec:bogssmall}.  In short, for $D>3$ we find $r_{\min }$ to be uniformly bounded by a power of $\mu_e$. For $D=3$ one can in fact create arbitrarily large bags of gold at fixed $\mu_e$, but only by tuning the parameter $A$ to be small and taking $\kappa > 4/3$ (see again appendix \ref{subsec:bogssmall}).  As a result, the bag of gold is subject to an additional IR cutoff associated with the finite value of the internal ($i$) cosmological length scale $\ell_i = \ell_e /\sqrt{-\lambda} \approx \ell_e/(\kappa-1) < 3 \ell_e$ which again limits the entropy such bags may contain.

However, rather than provide an exhaustive search for highly-entropic single-domain-wall bags of gold, we instead turn to creating bags of gold with multiple concentric domain walls, nested one inside the other.  In this context, by slightly deforming the models discussed thus far we will construct saddles describing {\it arbitrarily} large   bags of gold with fixed surface area for the black hole horizon.  Essentially the same construction will also lead directly to Euclidean wormholes -- defined here as Euclidean saddles for which the AlAdS boundary consists of two or more smooth compact connected components.

We will maintain spherical symmetry as well as time-reflection symmetry.  It is clear that many different models can be studied.  It will be most useful to consider a large number of domain walls, but we wish to avoid possible complications associated with models having large numbers of vacuua.  We thus suppose that there are only two vacuua which alternate between successive domain walls all having the same tension $\kappa$. We may continue to call the vacuua $(e)$ and $(i)$, though at every other wall the $(e)$ vacuum will lie on the inside of the wall and the $(i)$ vacuum will lie on the outside.

\begin{figure}[t]
\centerline{\includegraphics[width=0.3\textwidth]{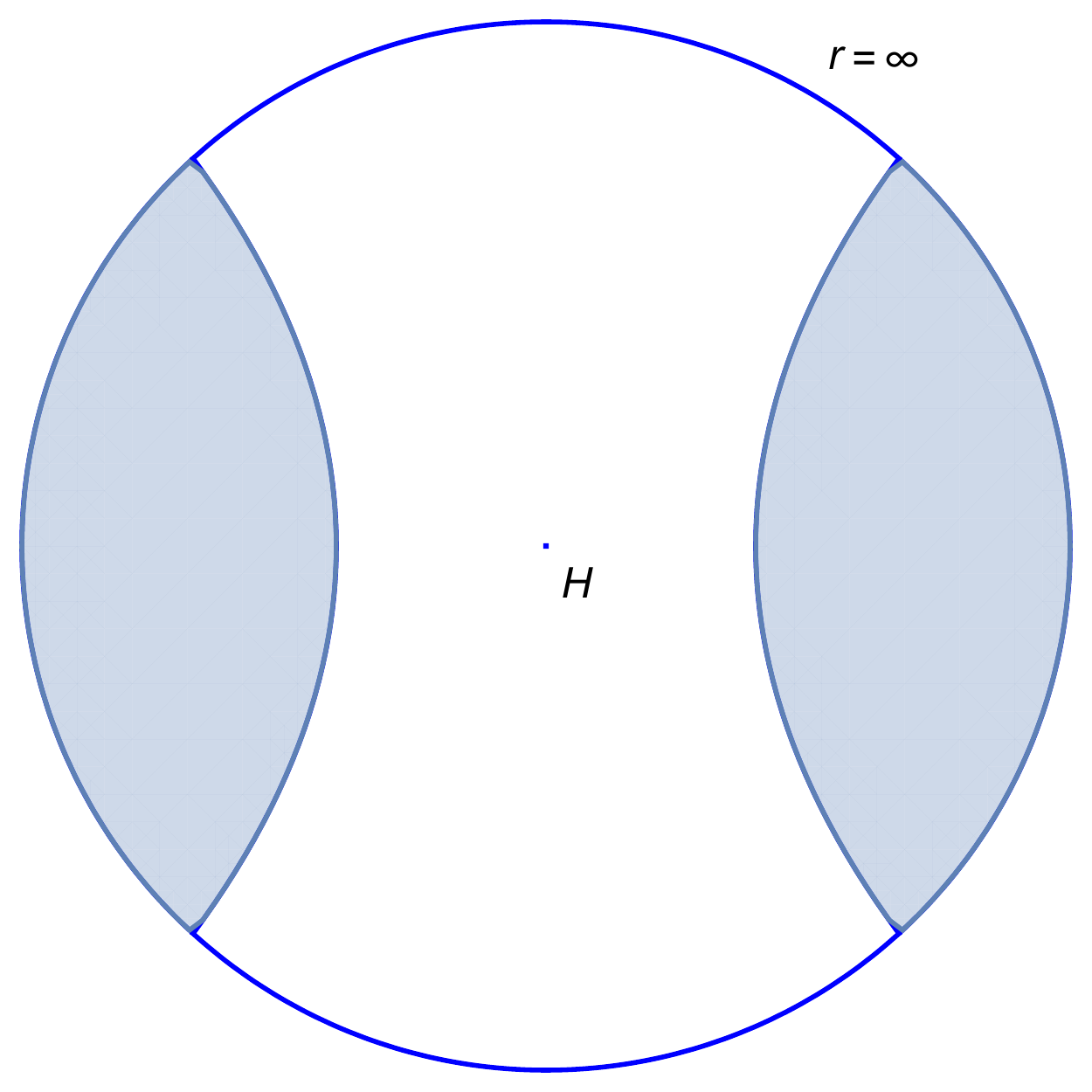}
	\hspace{.5cm}
	\includegraphics[width=0.3\textwidth]{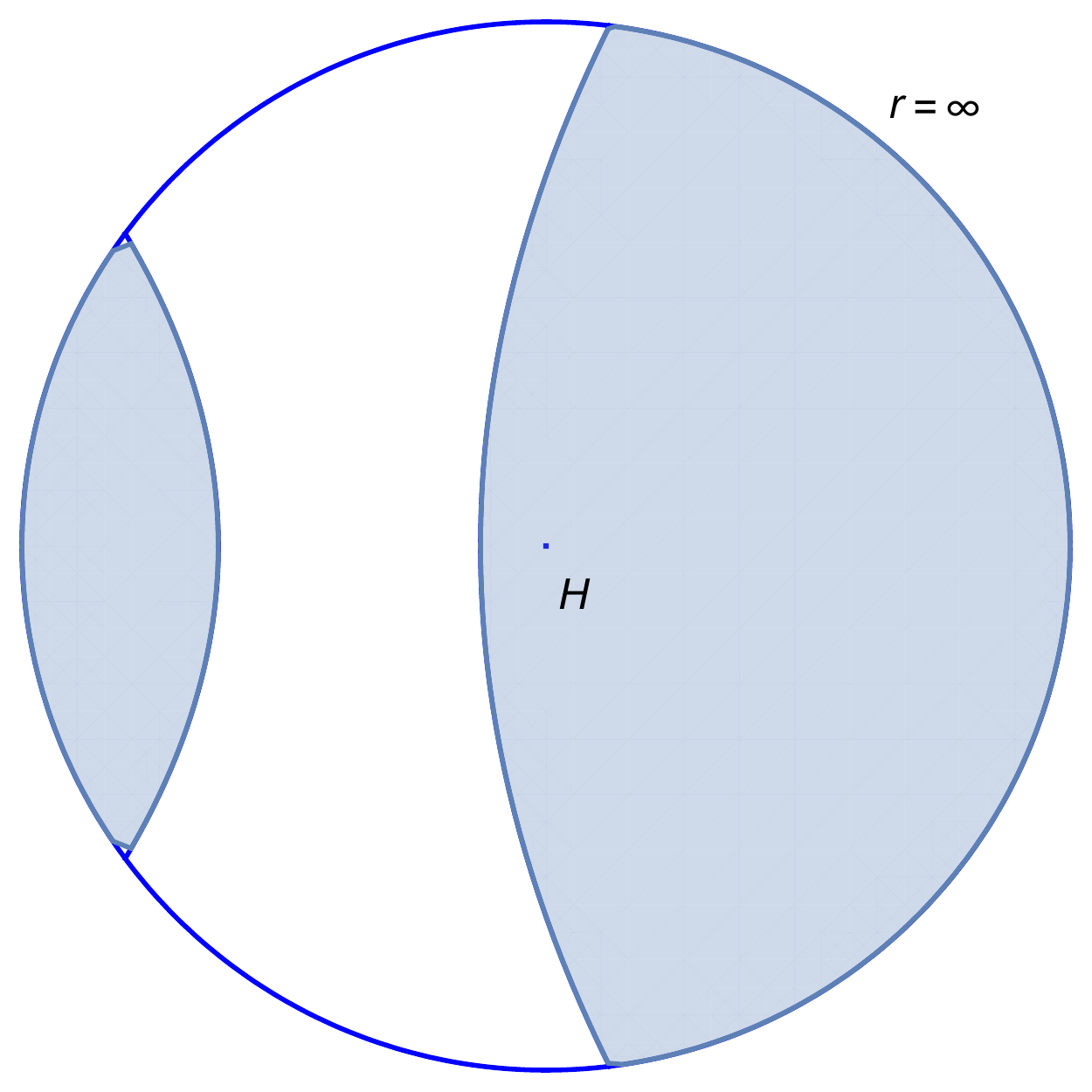}}
\caption{We consider the case where the (unshaded) SAdS region between two consecutive domain walls contains an SAdS horizon (left).  The alternative case where it does not is shown at right. In particular we will focus spacetimes that have a ${\mathbb Z}_2$ reflection symmetry exchanging the two domain walls as shown in the left panel. }
\label{fig:twochoices}
\end{figure}

The region between two successive domain walls will again be a piece of Euclidean SAdS.  In principle, we may consider cases where the two walls bounding a given SAdS piece lie on opposite sides of the Euclidean horizon or where they lie on the same; see figure \ref{fig:twochoices}.  We shall explore only the first case (left in the figure), as this clearly makes the given SAdS region larger than when the two walls lie on the same side of the horizon.  In addition, even at large $r_{\min }$ this provides a low redshift regime in which it might be possible to hold large entropy at small cost in energy.

To organize the discussion, let us note that if we can choose the mass parameters $\mu$ for each SAdS region to be identical, then each such region would be exactly the same.  We could then construct an arbitrarily large bag-of-gold by chaining together many copies of the same fundamental unit consisting of an SAdS region like that shown at fight in figure \ref{fig:twochoices}, bounded on each side by identical domain walls, with each wall described in the formalism of section \ref{TwoPiecesSpacetimes} by setting $\mu_i = \mu_e = \mu$.  And a periodic such chain would yield a Euclidean spacetime with two disconnected boundaries (i.e., it would yield a Euclidean wormhole).  We thus focus on this important special case\footnote{Setting $\lambda =-1$ and including only a single such domain wall behind the horizon gives a spacetime that is precisely the ${\mathbb Z}_2$ cover of the end-of-the-world-brane spacetimes of \cite{Cooper:2018cmb}.}.

Setting $\mu_i = \mu_e = \mu$ simplifies the analysis in several ways.  First, note from \eqref{eq:betas} and \eqref{eq:samplelim} that it yields
\begin{equation}
\alpha_e(r)= -\frac{\rho }{2}r,\text{ }\alpha_i(r) = \left(-\frac{\rho }{2}+\kappa \right)r.
\end{equation}
Recalling the sign conventions of section \ref{TwoPiecesSpacetimes}, the condition to keep the pieces of both $(e)$ and $(i)$ Euclidean SAdS solutions containing the respective horizon (so that both regions can take the form of the left panel in figure \ref{fig:twochoices}) is $2\kappa > \rho >0$. Case $(II)$ always makes $\rho$ positive, and we can easily choose parameters to make it less than $2\kappa$.  In particular, this is automatically true for $\lambda = -1$, in which case the $(e)$ and $(i)$ vacua have identical gravitational physics.  Indeed, even for much more complicated models than those considered here, it follows from \eqref{eq:junction}, \eqref{eq:beta}, \eqref{eq:EOM}, and \eqref{eq:potential} that whenever $f_e=f_i$ we have $\alpha_e(r)= -\frac{\kappa r}{2}  = - \alpha_i(r)$ and the above conditions are satisfied identically.

Furthermore, \eqref{eq:ABC} gives
\begin{equation}
A = 1 - \rho^2/4, \ \ B = - \mu, \ \ C=0.
\end{equation}
and thus
\begin{equation}
\gamma : = - \mu_e\frac{A}{B} = A = 1 - \frac{\rho^2}4, \ \ \
\frac{\kappa \sqrt{C}}{\mu_e (1+
\lambda + \kappa^2)} =0, \ \ \ \frac{AC}{B^2} =0  \ \ \ {\rm and} \ \ \ \frac{B^2}{4C} = \infty.
\end{equation}
As a result, taking $\mu$ large at fixed $\rho$ easily satisfied the conditions of section \eqref{subsec:nose}.  Furthermore, using the above results \eqref{eq:Dg4Dt} simplifies to yield just
\begin{equation}
\label{eq:eratiowinside}
\frac{2t_{Ee} (r = \infty)}{\beta} =  f_D(\gamma)
\left( 1 + O(\mu^{-\frac{2}{D-1}})\right).
\end{equation}

Recall that $f_3(\gamma) = \frac{1}{2}$ and that for $D \ge 4$ the function $f_D(\gamma)$ decreases monotonically from $+\infty$ at $\gamma=0$ to $1/2$ at $\gamma=1$ as shown in figure \ref{fig:fD}.  As a result, a single domain wall of our form with $\mu_i=\mu_e$ always removes at least half of the Euclidean boundary on the $(e)$ side of the wall.  Since this is true for all $\lambda$, it must also hold on the $(i)$ side.  Indeed, since $\alpha_i = -\frac{\rho-2\kappa}{\rho}$, for the desired case $2\kappa > \rho$ one finds
\begin{equation}
\label{eq:iratiowinside}
\frac{2t_{Ei} (r = \infty)}{\beta} =  f_D(\tilde \gamma)\left( 1 + O(\mu^{-\frac{2}{D-1}})\right) \ \ \ {\rm with} \ \ \ \tilde \gamma = 1 + \rho^2/4\lambda = \frac{\lambda + 1 - \gamma}{\lambda},
\end{equation}
which apparently requires $-\lambda > \rho^2/4.$
As a result, adding two domain walls to each SAdS region must remove the entire AlAdS boundary.  So solutions of the form of figure \ref{fig:twochoices} (left) do not exist in the models studied thus far.

However, for $D=3$ this failure is marginal at all $\lambda, \kappa$ in case $(II)$, as regular Euclidean boundaries would have existed for any smaller value of \eqref{eq:eratiowinside}, \eqref{eq:iratiowinside}.  And for $D \ge 4$ it is marginal for $\gamma \approx \tilde \gamma \approx 1$, which in particular holds for $\lambda \approx -1$ with $\kappa$ small.  As a result, small alterations of the models considered thus far could potentially allow solutions of the desired form.

Indeed, since the relevant parameter involves the ratio of $t_{Ee}, t_{Ei}$ to the Euclidean period $\beta$, it is natural to study modifications that lower the temperature of the SAdS horizons.  The classic way to do so is by adding charge under an appropriate $U(1)$ gauge field.  To be concrete, let us focus on the case $D=4$, add a Maxwell field $F_{ab}$, and consider solutions with magnetic flux on the $S^2$ factor of the geometry\footnote{We choose $D=4$ both because of its familiarity and due to subtleties involving Maxwell fields for $D=3$.  For example, the charge contribution to the usual charged BTZ solutions grows logarithmically at large $r$.}.  In particular, we may take these to be the magnetically charged black holes of AdS$_4$  supergravity constructed in \cite{Toldo:2012ec,Klemm:2012yg} with the moduli tuned so that the dilaton and other scalars are independent of $r$.  Such solutions can be embedded in eleven-dimensional supergravity, so this provides a top-down model of the black holes backgrounds, if not of the domain walls.  For later use, we mention that doing so realizes the AdS$_4$ Maxwell potential as a Kaluza-Klein gauge field associated with reduction of the eleven-dimensional metric along a $U(1)$ fiber.  For simplicity, we take our domain walls to be uncharged under this Maxwell field and leave open for future investigation the question of whether they may also be embedded in a top-down model such as those studied in \cite{Maxfield:2014wea}.

As a result of adding magnetic flux, any $e$ region between a given pair of adjacent domain walls now becomes a piece of a $D=4$ magnetically charged AdS Reissner-Nordstr\"om (RNAdS) spacetime for which
\begin{equation}
\label{eq:femag}
f_e = r^2 +1 - \frac{\mu}{r} + \frac{Q^2}{r^2}.
\end{equation}
The basic formalism described in section \ref{TwoPiecesSpacetimes} will continue to apply, but with a modified effective potential $V_{\rm eff}$.  The key point, however, is that the new magnetic term in \eqref{eq:femag} falls off quickly at large $r$.   Indeed, if we introduce the horizon size $r_h$ of the corresponding uncharged black hole (defined by $0 = r^2_h + 1 - \frac{\mu}{r_h}$) and note that \eqref{eq:femag} will fail to vanish anywhere for $Q^2> \mu r_h$, one sees that the charge term in \eqref{eq:femag} can be neglected when $r \gg r_h$.  So the trajectory of domain walls with $r_{\min } \gg r_h$ is essentially unaltered by the addition of charge, and this is in particular true for the times $t_{Ee} (r = \infty), t_{Ei} (r = \infty)$ at which the walls reach the AlAdS boundary. In the limit of large $\mu$ and large $r_{\min }/r_h = \gamma^{-1/3}$, the change in the ratio \eqref{eq:eratiowinside} is thus dominated by the change in Euclidean period $\beta$.  And since $\beta$ can be made arbitrarily large by taking the black hole to be near extremality, we can easily lower the ratio $\frac{2t_{Ee} (r = \infty)}{\beta}$ below $\frac{1}{2}$, and similarly for $\frac{2t_{Ee} (r = \infty)}{\beta}$.  This is essentially the same mechanism employed to avoid self-intersections in \cite{Antonini:2019qkt}.

Now, the reader may be concerned that such solutions are finely tuned, in that large $r_{\min }/r_h$ requires $\gamma \rightarrow 0$ so that $f_4(\gamma)$ diverges. It may thus appear that our black hole must be very close to extremality in order for the desired solution to exist.  This may in some sense be true, but any fine tuning is only at the $O(1)$ level, and is not parametric in the Planck scale $\ell_p$.  Furthermore, as stated in \eqref{eq:fDexp} $f_4(\gamma)$ diverges at small $\gamma$ only as $\gamma^{-1/6}$, and thus as $\sqrt{r_{\min }/r_h}$.  So if we reduce the effects of the charge term to the 10\% level by taking $r_{\min }/r_h \sim 1/10$, we find $\gamma = 10^{-3}$ and $f_4(\gamma) \approx 2.11$.  The desired solutions then exist when $Q$ is large enough to decrease the black hole temperature by a bit more than of $4$ relative to its uncharged value. Since the temperature has a square root behavior near extremality, for a given $M$ this means that the charge should be within 5\% to 10\% of its extremal value.

The same will be true for the $i$ region if we make the corresponding choices for $\tilde \gamma$.  Taking both $\gamma$ and $\tilde \gamma$ to be small will require taking $\lambda$ to be near $-1$ and $\kappa \approx \rho \approx 2$, but again this fine-tuning is $O(1)$ and not parametric in $\ell_p$.

Periodic chains of the above form immediately define Euclidean wormholes.  We will discuss these further in section \ref{subsec:dominate}.  To turn a long chain  into a bag-of-gold requires us to end the chain after a finite number of units in each direction.   It is useful to first discuss the related two-boundary wormhole solutions in which the $t=0$ surface contain two disconnected pieces of the AlAdS boundary.  While these two pieces are connected through the Euclidean time direction along the boundary, in Lorentz signature the solution will become a wormhole with two disconnected boundaries.  To build such a solution from our chain, we need only omit both the left-most and right-most domain walls, so that the left-most and right-most RNAdS regions contain only a single domain wall each as in the left panel of figure \ref{fig:caseII-}.

A bag-of-gold solution would consist of roughly half of the long-wormhole solution just constructed.  Conservation of magnetic flux on the $S^2$ then requires either additional non-trivial topology or the addition of magnetically charged matter fields.    While the latter is natural, we opt for the former in order to avoid specifying further details.  In particular, we consider bags of gold that are ${\mathbb Z}_{2}$ quotients of the above long wormholes, where the (free) ${\mathbb Z}_{2}$ action simultaneously exchanges the right and left boundaries, acts as the anti-podal map on the $S^2$, and -- in order for the ${\mathbb Z}_{2}$ action to preserve the sign of the Maxwell field -- also acts as $\phi \rightarrow -\phi$ on the internal $U(1)$ Kaluza-Klein fiber mentioned above.  In other words, we take the central RNAdS-like region to in fact be a charged ${\mathbb RP}^2$ geon of the sort described in \cite{Louko:2004ej}.  In particular, it is worth mentioning that the full spacetime (including the internal dimensions) is non-orientable, but that is not a problem for a bulk theory like eleven-dimensional supergravity which describes at least the black hole sector of our model\footnote{Furthermore, if there is a second $U(1)$ Kaluza-Klein fiber, one may choose to invert it as well to give an orientable spacetime.}.

\subsection{Can Euclidean wormholes and large bags of gold dominate a path integral?}
\label{subsec:dominate}

Having argued that our models yield both Euclidean wormholes and Euclidean saddles that create arbitrarily large bags of gold, it is important to ask how their actions will compare with other possible saddles for the same path integral.  We now show that  -- within our model and with the assumed symmetries -- our bags of gold are in fact the {\it only} allowed saddles for properly chosen path integrals.  We therefore expect that they will dominate over other (less symmetric) saddles in our model.  We also comment briefly on issues going beyond our models in section \ref{sec:disc}, though a complete analysis is beyond the scope of this work.  After addressing bags of gold, we also discuss competing saddles for Euclidean wormholes.

In the above section we constructed bag-of-gold solutions as ${\mathbb Z}_2$ quotients of long two-boundary wormholes. In particular, the long wormholes have AlAdS boundary topology $S^1 \times S^{D-2}$ (where we now generalize from $D=4$ to arbitrary dimension $D$), and the ${\mathbb Z}_2$ acts on this boundary by $\theta \rightarrow -\theta$ on the $S^1$ and a simultaneous anti-podal map on the $S^{D-2}$.  The Euclidean bags-of-gold thus have boundary topology $S^1 \times S^{D-2}/{\mathbb Z}_2$.  Since any solution with boundary $S^1 \times S^{D-2}/{\mathbb Z}_2$ admits a ${\mathbb Z}_2$ cover with topology $S^1 \times S^{D-2}$, consideration of saddles that compete with our bags-of-gold is equivalent to considering saddles that compete with our long wormholes.  We find it simpler to focus on the latter.

We are thus interested in $S^1 \times S^{D-2}$ boundaries that are divided into alternating regions associated with the two distinct vacua in the bulk.  The vacua are naturally specified by boundary conditions as, in an Einstein-scalar model, they are associated with distinct asymptotic values for the scalar fields.  The transitions between two adjacent vacua along the AlAdS boundary will act as sources for bulk domain walls.  So if one fixes the metric and scalar sources on $S^1 \times S^{D-2}$, one can look for saddles that match the stated boundary conditions, and thus including a corresponding number of such walls.

\begin{figure}[t]
\centerline{\includegraphics[width=0.5\textwidth]{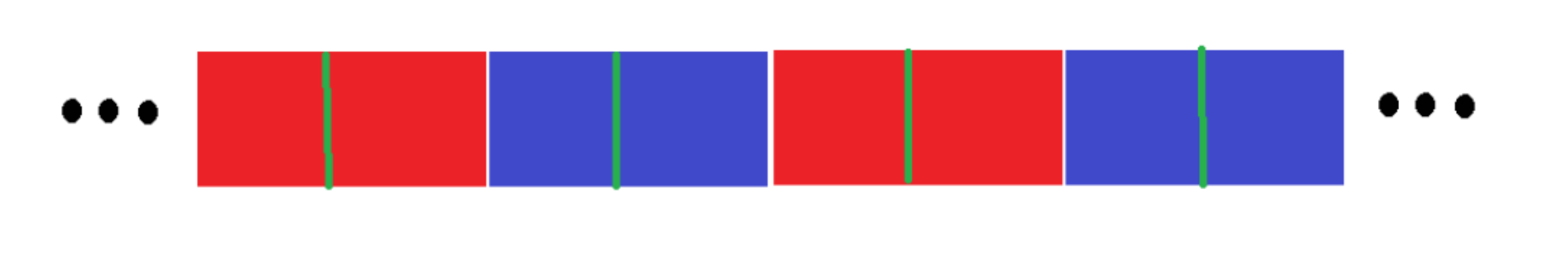}}
\caption{The AlAdS boundary features alternating regions of two distinct vacuua, shown here as red and blue. Dots on each side indicate that the pattern may continue.  Vertical cross-sections represent $S^{D-2}$-spheres for our solutions  The transitions between vacuua (white vertical lines) are sources of domain walls.  Our microcanonical path integrals include an energy constraint (vertical green lines) in each vacuum region.}
\label{fig:micro}
\end{figure}

However, we find it more convenient to use a form of the microcanonical path integral discussed in \cite{Marolf:2018ldl}.  This amounts to starting with a standard (canonical) path integral, perhaps with the standard metric on $S^1 \times S^{D-2}$  with some choice of sizes for the $S^1$ and $S^{D-2}$, and then adding a constraint that fixes the stress-energy flux (``the energy'')  $\int_{\Sigma_i} \sqrt{h} T_{ab}n^a \xi^b$ defined by the boundary stress tensor $T_{ab}$ through some set of surfaces $\Sigma_i$ with induced metric $h$ and normal $n^a$ in the direction defined by the Killing field $\xi^a$ along the $S^1$.  We will introduce one such energy constraint in every vacuum region as shown in figure \ref{fig:micro} above.

We will also fix the magnetic flux on the $S^{D-2}$.  In our model without charged matter, this requires only a single constraint.  Flux conservation then fixes the flux everywhere.  But in a more general model it would make sense to fix the flux separately in each vacuum region.

Within this model, maintaining spherical symmetry, fixing the energy, and also fixing the magnetic flux requires the bulk solution to be a piece of RNAdS.  The bulk must then consist of a collection of such pieces, each separated from adjacent ones by domain walls.  Furthermore, in stable theories any domain wall in a Euclidean solution must reach the AlAdS boundary.  And more explicitly, since $\lambda\approx -1$ the two vacuua are similar.  So all domain walls in our two-vacuum model have potentials $V_{\rm eff}$ with $A>0$, making it clear that all walls expand to reach the Euclidean AlAdS boundary.  In our model where domain-wall intersections are not allowed, we thus conclude that each RNAdS piece of the bulk must also reach the AlAdS boundary at a point where the boundary conditions transition between the two types of vacuum.

Furthermore, we assume that only one bulk domain wall reaches the AlAdS boundary at each such transition in the boundary conditions.  This is not something that can be determined directly from the thin wall model, but it is naturally guaranteed in appropriate more complete Einstein-scalar models.  There each transition should be described as a continuous change in the scalar boundary conditions from one vacuum to another over a finite piece of the AlAdS boundary. Thus each wall in fact has some finite thickness at the AlAdS boundary as determined by the boundary conditions.  Choosing the boundary conditions to vary monotonically from one vacuum to the next in a theory where domain walls are stable will then naturally yield a single (thickened) domain anchored to this part of the boundary.  In the same way, we see that each bulk RNAdS region must reach some finite piece of the AlAdS boundary, and must thus have parameters matching those of one of our boundary regions and thus fixed by our boundary conditions.

Now, recall that the trajectory of any domain wall between two such RNAdS regions is also fixed by the RNAdS parameters.  For most of the regions we will choose parameters so that $\frac{2t_E}{\beta}$ for such walls is $\frac{1}{2}-\epsilon$ for some small positive epsilon.  This constrains each RNAdS piece to have exactly one or two domain walls.  But we choose two regions that are diametrically opposite around the $S^1$, and which thus necessarily describe the same type of vacuum, in which we instead choose $\mu$ to be $\frac{2t_E}{\beta} = \frac{1}{2} + \epsilon,$ so that only one domain wall is allowed.  For small $\epsilon$, the local change in the bulk solution will be negligible; see appendix \ref{app:breakZ2} for more detailed comments on domain walls with $\mu_i>0$ with $\mu_i$ different from $\mu_e$.  For reasons that will become clear, we refer to these two special boundary regions as the endcaps.  To define a CFT state, we will choose to cut open the path integral along a pair of diametrically opposite $S^{D-2}$ spheres, with one $S^{D-2}$ in each endcap.

\begin{figure}[t]
\centerline{\includegraphics[width=0.5\textwidth]{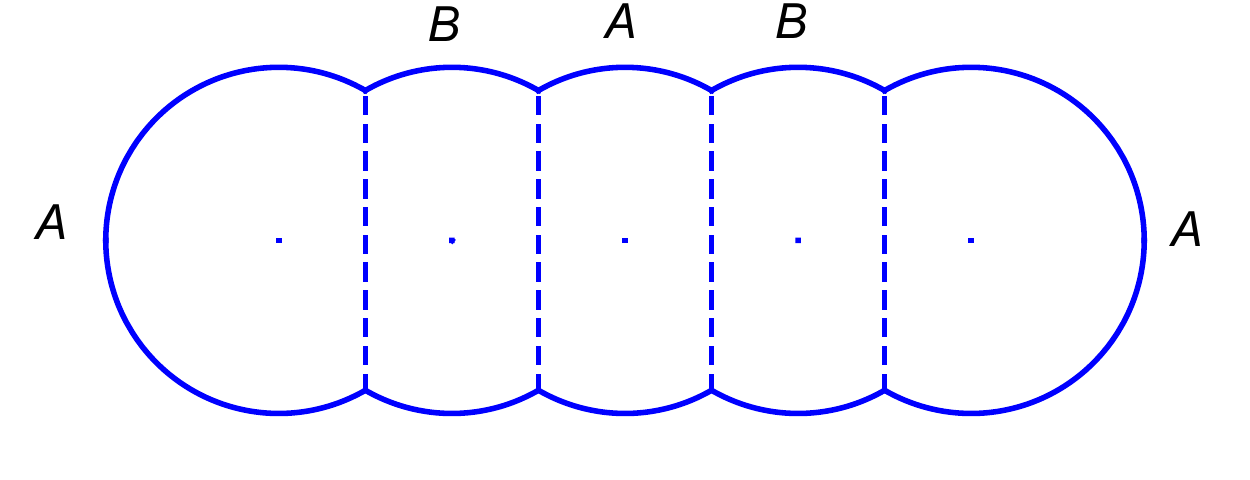}}
\caption{ Cartoon of bulk Euclidean spacetime for our long wormhole.  $A/B$ vacuum regions are marked on the boundary, and domain walls are shown as dashed lines.  The left-most and right-most $A$ regions are endcaps which admit only a single domain wall.  All other regions allow two.  This is the unique Euclidean spacetime compatible with such boundary conditions. }
\label{fig:LWEB}
\end{figure}

Let us call the two vacuua $A$ and $B$ and take the endcaps to be regions with boundary conditions appropriate to the $A$ vacuum.   Since only one domain wall can fit in this region, it must connect to both ends of this $A$-vacuum region as shown at e.g. the left end of figure \ref{fig:LWEB}.  In order to avoid both intersections and three-wall regions, the two adjacent type $B$ regions  (just to the right in \ref{fig:LWEB}) must then be connected by a single RNAdS piece with two walls.  Indeed, we are force to continue to pair up such boundary regions in this wall until we are left only with the final endcap, which is necessarily associated with another one-wall RNAdS piece; see again \ref{fig:LWEB}.  Thus we see that within our class of models (i.e., without charged matter or domain wall intersections) and with the chosen boundary conditions, our path integral admits only a single saddle.  It is natural to expect it to dominate even when less symmetric saddles are included.

In contrast, boundary conditions allowing two-boundary Euclidean wormholes described above, we always find disconnected saddles with which they can compete. For example, we find Euclidean wormholes whose boundaries are two disconnected copies of the $S^1 \times S^{D-2}$ discussed above, with each copy having its own set of alternative $A$ and $B$ regions, but now with all vacuum regions having parameters fixed so that $\frac{2t_E}{\beta} = \frac{1}{2}  \epsilon$; i.e., there are no special endcap regions.  Nevertheless, there is an allowed saddle given by two copies of the spacetime shown in figure \ref{fig:LWEB} in which two of the regions simply happen to have only a single domain wall.  Indeed, for $n$ vacuum regions there are $n/2$ such saddles, as any two diametrically opposed vacuum regions can be chosen to be special in this way.

Furthermore, while we have not computed the relevant actions, for $S^1 \times S^{D-2}$ boundaries one would expect the disconnected saddle to dominate.  One argument for this is based on the result of \cite{Marolf:2018ldl} that with an exact $S^1$ translational symmetry on the boundary, the microcanonical action is just $-1$ times the RT entropy. The disconnected solution will clearly require an RT surface with two disconnected components, while the RT surface in the connected saddle should have only one. And if the areas of these surfaces are fixed in a simple way by boundary parameters (as in standard black hole solutions), then the two component RT surface seems likely to have twice the area of the one-component surface. Generalizing this idea to boundaries that break translational symmetry on the $S^1$, and then further to arbitrary boundary manifolds with non-trivial fundamental group $\pi_1$, suggests that disconnected saddles may always dominate in this context.

This suggestion, however, motivates a closer look at Euclidean wormholes with spherical boundaries (so that $\pi _1 =  \varnothing $. Due to conservation of magnetic flux, this is not strictly possible in our models without introducing (magnetically) charged matter.  We thus leave it for future investigation.  But the idea that Euclidean wormholes with spherical boundaries might dominate certain path integrals is consistent with \cite{Yin:2007at} and \cite{Maxfield:2016mwh} (which showed that the they fail to dominate with non-spherical boundaries) and with \cite{Maldacena:2004rf} (which found low energy models where such wormholes appear to dominate but for which top-down constructions were not known).  Furthermore, we will give a rather generic construction of Euclidean wormholes below in section \ref{sec:disc}, suggesting that one can find fully fledged gauge/gravity dualities where such Euclidean wormholes do indeed dominate the low energy path integral.

\section{Discussion}
\label{sec:disc}

Our work above classified the possible spherically symmetric Euclidean solutions in which a thin domain wall of positive tension $\frac{D-2}{8\pi G_D} \kappa \ge 0$ separates an internal $(i)$  matter-free region with a regular origin ($r=0$) and cosmological constant $\lambda$ from an external $(e)$ matter-free region with negative cosmological constant unit AdS length scale.  We also discussed certain examples where there is no regular origin and the internal region instead contains a minimal surface on the $t=0$ slice.  We required our solutions to be smooth up to discontinuities in the extrinsic curvature at the domain wall.  Our goal was to better understand how asymptotically locally AdS (AlAdS) bulk spacetimes are described in a dual CFT by identifying the saddle points of bulk Euclidean path integrals for given AlAdS boundary conditions. In the bulk semi-classical limit, the dominant such saddle is then dual to the CFT state generated by the corresponding CFT path integral by an analogue of the arguments of \cite{Maldacena:2001kr}.    Particular goals included the study of AlAdS solutions with inflating bubbles, and also possible bag-of-gold solutions.  While we used a bottom-up approach, at least some top-down models of domain walls in AdS/CFT were found in \cite{Maxfield:2014wea}.

The first part of our study focused on models that allow inflating internal ($i$) bubbles (perhaps driven by domain wall inflation when $\lambda <0$) and a stable external ($e$) vacuum.  In this case, all Euclidean solutions satisfying the above assumptions are topologically $S^D$.  In particular, they have no AlAdS boundaries and so cannot be described as bulk saddle points of a dual CFT path integral.

The obstruction to finding good Euclidean solutions with AlAdS boundaries is essentially the same as that discussed by \cite{Farhi:1989yr} in the context of seeking instantons that mediate the nucleation of false vacuum bubbles by quantum tunneling from flat space.  Now, in that context, it has been suggested \cite{Farhi:1989yr,Fischler:1989se,Fischler:1990pk} that the process may nevertheless take place and that it is instead mediated by certain non-smooth saddles (see also \cite{Bachlechner:2016mtp,deAlwis:2019dkc}).  But by extending concerns already expressed in a related context by reference \cite{Fischler:1990pk}, we argued in section \ref{subsec:degenerate} that including such non-smooth saddles renders the tunneling rate ill-defined.  While there reamins much to understand about gravitational path integrals, and while further consideration of such issues would be useful, this observation suggests that only smooth saddles should be allowed.

Our analysis above has been limited to the thin wall approximation and to relatively simple classes of domain walls.  However, the basic obstruction to smooth saddles is simple to state more broadly:  Since Wick rotation changes the sign of $dt^2$, it turns any positive Lorentz signature acceleration driving inflation into a negative Euclidean signature acceleration driving collapse and trapping the domain wall in the interior so that it cannot be directly sourced at the Euclidean boundary.  And non-trivial source-free Euclidean solutions should not exist in theories with a stable vacuum.  This forbids Euclidean solutions with inflating bubbles in simple domain wall models.

As discussed at the end of section \ref{subsec:dS-SAdS}, it remains to investigate whether more complicated models and/or full Einstein-scalar field theories might allow inflating regions to be sourced in some indirect manner.  It would also be interesting to investigate whether it might shed light on the validity of degenerate saddles as in \cite{Fischler:1989se,Fischler:1990pk}.  However, extrapolating our results to such cases would suggest that AlAdS spacetimes with inflating bubbles may have no dual description in any dual CFT.  If so, it may be that such spacetimes are not in fact part of any complete theory of quantum gravity.  It is unclear to us whether this would then have further implications for discussions of exact or meta-stable de Sitter vacua (see e.g. \cite{Obied:2018sgi,Kachru:2003aw}), but it would be interesting to investigate further.

On the other hand, it should be noted that bubbles only inflate when they are larger than a certain critical size and that for $\lambda > -(\kappa-1)^2$ saddles also fail to create bubbles that are too small for this inflation to occur.  Furthermore, for small $\mu_e$, $\kappa $, $\lambda +1$ the stress tensor of the Lorentzian bubble solutions is small and the metric is perturbatively close to that of the $(e)$ AdS vacuum near $t=0$. In particular, viewing our thin wall solutions as approximations to those of a theory of gravity and a scalar field whose potential has multiple minima, one would expect to be able to construct the analogous Einstein-scalar saddles by adding boundary sources to the usual vacuum path integral through an analogue of the construction in \cite{Marolf:2017kvq}.  It would be interesting to understand if gravitational effects at non-zero Euclidean times in fact prevent the existence of such saddles, or whether some such saddles do exist in regimes where the thin wall approximation breaks down.

The second part of our study constructed Euclidean solutions that create bag-of-gold spacetimes.  We found a large class of models in which such solutions exist, at least at large mass parameter $\mu_e$.  In spacetimes with only a single domain wall, for $D > 3$ appendix \ref{subsec:bogssmall} shows the size of any bag of gold to bounded by a power of $\mu_e$. For $D=3$ one can in fact create arbitrarily large bags of gold at fixed $\mu_e$, but only by tuning the parameter $A$ to be small and taking $\kappa > 4/3$ (see again appendix \ref{subsec:bogssmall}).  As a result, the bag of gold is subject to an additional IR cutoff associated with the finite value of the internal ($i$) cosmological length scale $\ell_i = \ell_e /\sqrt{-\lambda} \approx
 \ell_e/(\kappa-1) < 3 \ell_e$ which again limits the entropy such bags may contain.

However, as described in section \ref{subsec:bogslarge}, expanding the class of models slightly to include magnetic charge allows the construction of arbitrarily large Euclidean bag-of-gold solutions that contain a correspondingly large number of domain walls.  Indeed, the solutions described in section \eqref{subsec:bogslarge} consist of a long chain of identical RNAdS regions, with each pair of adjacent regions separated by a domain wall.  We also argued that, at least within our class of models and preserving our symmetries, such Euclidean solutions give the unique saddles for our path integrals, which we took to be microcanonical in the sense of \cite{Marolf:2018ldl}.

As described in the introduction, such solutions create an interesting tension with the density of states in any dual CFT.  While the tension is very real, we would argue that there is no sharp contradiction.  Indeed, the semi-classical approximation to the bulk path integral is naturally considered to be an asymptotic expansion in small bulk Newton constant $G$.  A conservative perspective would thus be that it provides such an expansion for any given fixed {\it $G$-independent} bulk path integral.  In our context, this would mean that we should first fix AlAdS boundary conditions on the Euclidean metric and matter fields (and so fix the number $N$ of domain walls reaching the AlAdS boundary and the mass parameters $\mu$) and then take the limit $G \rightarrow 0$.  In particular, this perspective suggests that the semi-classical approximation need be valid only for $G\ll 1/N$.  Of course, it remains of great interest to understand which particular corrections to the semi-classical approximation become large for large $N$ and what effect they have on bulk physics.  We hope our solutions provide useful starting points for investigating such questions.

Even without filling our bags-of-gold with entropy, the results of \cite{Fu:2018kcp} suggest that our solutions may have interesting implications for the complexity equals action (CA) conjecture \cite{Brown:2015bva,Brown:2015lvg}.  Reference \cite{Fu:2018kcp} showed that a class of 2+1 bags-of-gold based on adding topology inside black holes had Wheeler-DeWitt patch actions that decrease with the size of the black hole interior.  With a finite UV cut-off, this result is in tension with the intrinsic positivity of complexity.  A resolution proposed there was that such spacetimes might lack CFT duals as they were not known to dominate any Euclidean path integrals.  But our spacetimes do appear to dominate such path integrals.  So corresponding negative contributions to Wheeler-DeWitt patch actions for these or other (perhaps $(2+1)$-dimensional) bag-of-gold solutions would pose a challenge to the CA conjecture.   This remains to be analyzed in detail but, if true, would revive the original tension noted in \cite{Fu:2018kcp}.

Using a construction closely related to our bags-of-gold, we also identified a new class of asymptotically AlAdS Euclidean wormhole solutions.  Indeed, in retrospect it seems likely that Euclidean wormholes exist in essentially any low energy theory of AdS gravity, and with essentially any boundary metric.  The point here is that time symmetric Euclidean wormholes Wick rotate to Lorentz signature closed cosmologies.  These are easy to construct, by simply choosing the desired topology and then adding enough radiation  (or gravitational waves) to satisfy the Hamiltonian constraint.  For example, if the metric on the time symmetric slice has everywhere non-negative Ricci scale (as for a metric sphere or a metric torus), the Hamiltonian constraint requires the sum of the energy densities from the cosmological constant and matter fields to be non-negative.  If we arrange for this energy to come from radiation (rather than from scalar field potentials), the Lorentz signature solution will clearly collapse.  Thus the Euclidean solution will expand and define a Euclidean wormhole. When the spacetime has non-contractible closed curves, we saw in section \ref{subsec:dominate} that -- at least in the microcanonical ensemble -- general arguments suggest that this wormhole will give only a subleading contribution to the path integral.   But with spherical topology, there may be potential for Euclidean wormholes to dominate.

This idea fits well with the results of \cite{Maldacena:2004rf,Yin:2007at,Maxfield:2016mwh}. In particular, \cite{Maldacena:2004rf} found low-energy effective theories of gravity in which two-boundary Euclidean wormholes with spherical boundaries appear to dominate the path integral.  It was unclear if top-down constructions could yield the particular  models studied, but since the above approach suggests that any top-down model will admit a broad class of Euclidean wormholes it is plausible that top-down models where Euclidean wormholes dominate can indeed be found.  This may have interesting implications for our understanding of AdS/CFT more generally, especially concerning any possible role of disorder \cite{Cotler:2016fpe}.  We hope to explore this further in the future.

There are, however, several issues that remain to be addressed for both classes of solutions.  One is that we have ignored the (magnetically) charged matter fields that one expects to be present in more realistic systems, and which in particular are required by the arguments of \cite{Harlow:2018jwu} and by appropriate versions of the weak gravity conjecture \cite{ArkaniHamed:2006dz}.  Such fields can cause Reissner-Nordstr\"om black holes to become unstable to growing scalar hair near extremality (see e.g. \cite{Hartnoll:2008kx}), and thus can modify the analysis.  However, since the detailed form of the charged black hole solution played little role in our analysis, and since the asymptotic form of all such solutions is identical so long as the conformal dimension of the charged scalars is not too small, we expect a similar analysis to hold even in models where such instabilities are present.  Indeed, we expect the existence of charged matter to make the construction of large bags of gold even easier, as one may then take the mass parameter (and thus $r_{\min }$) to increase by a constant ratio across each domain wall, and thus to increase exponentially as one moves further into the interior (see e.g. comments about $\mu_e\neq \mu_i$ in appendix \ref{app:breakZ2}).  It would then require only a logarithmic number of domain walls to create a bag-of-gold with entropy greater than the event horizon's $A/4G$.

Another such issue concerns the possibility of negative modes.  The fact that our saddles are the only ones for our models satisfying the stated symmetries and boundary conditions suggests that there will be no negative modes, but a detailed study remains to be performed.

Other possible concerns regarding our bags-of-gold include more quantum effects.  First, one might ask if fluctuations about our saddles might be large in the limit where are bags of gold become very large.  This is certainly true in some sense.  For example, as described in \cite{Marolf:2018ldl}, a microcanonical ensemble of small width $\Delta E$ is naturally associated with fluctuations in certain time correlations of size $\Delta t \sim 1/\Delta E$ that, if large, make the bulk far from any given classical metric.   Such fluctuations add in quadrature, and so the total fluctuation along a chain $n_{\rm chain}$ units long is proportional to $\sqrt{n_{\rm chain}}$.  Luckily, however, we need only a finite number of units each having horizon size $r_h$ to make a bag-of-gold large enough to allow bulk entropy greater than $S_{\rm BH} = r_h^{D-2}/4\ell_p^{D-2}$.  Since it is easy to fit an entropy $S_{\rm unit} = \frac{\ell^{D-2}}{r_h^{D-2}}$ into each unit with small back-reaction and small cost in energy, we require no more than $n_{\rm chain} \sim \frac{\ell^{D-2}}{\ell_p^{D-2}}$.  And taking the width of the microcanonical ensemble to be comparable to the width of the corresponding canonical ensemble, one finds in each unit that, relative to the corresponding Euclidean time period $\beta$,  the fluctuations satisfy $\frac{\Delta t}{\beta} \sim \frac{ \ell_p^{\frac{D-2}{2}}\ell}{r_h^{\frac{D-1}{2}}}$, so even after multiplying by $\sqrt{n_{\rm chain}}$ the time correlation fluctuations are still suppressed relative to $\beta$ by $\sqrt{\frac{\ell^{\frac{D-1}{2}}}{r_h^{\frac{D-1}{2}}}}$.

The remaining issue to explore is whether new complications arise when one considers not just the path integral to create a fixed bag-of-gold background, but to also actually fill the bag-of-gold with large entropy.  While we see no obstacles to doing so, a detailed analysis of this issue (perhaps following \cite{Marolf:2017kvq,Chen:2019ror,Haehl:2019fjz}) will be left for future work. The point here is that a single empty bag-of-gold is just a pure state, and does not by itself lead to tension with the Bekenstein-Hawking density of states\footnote{However,  if we can indeed create exponentially large bags of gold by taking the mass parameter to increase as we move inward by adding charged matter, such a tension can be created by taking an ensemble defined by saddles of our form with different mass parameters.}.  Furthermore, it was recently noted in \cite{Penington:2019npb,Almheiri:2019psf} that this distinction leads to important phase transitions for quantum extremal surfaces.  It follows that, at least in studying Renyi copies of our path integrals, saddles can exchange dominance depending on the amount of entropy in the bag-of-gold.  It would be extremely interesting to identify a similar phenomenon in saddles associated with the original state and to investigate their implications for the information problem.
Perhaps the saddles described here will provide fertile ground for future such investigations.

\section*{Acknowledgements}
We thank Ahmed Almheiri, Gary Horowitz, Veronika Hubeny, Per Kraus, Alexander Maloney, Henry Maxfield, Mukund Rangamani, Eva Silverstein, Stephen Shenker, and Douglas Stanford for related discussions over many years. We also thank Shanta de Alwis and Brian Swingle for more recent discussions. This work was supported in part by the U.S. National Science Foundation under grant PHY 1801805 and by the University of California.

\appendix

\section{No local minima of $V_\text{eff}$ for $A<0$}
\label{app:Vcd}

We now show that in cases where $V_\text{eff} \rightarrow -\infty$ at large $r$, $V_\text{eff}$ can be equal to zero at most twice. In such cases, the effective potential has the form
\begin{equation}
{V_{{\rm{eff}}}} = A{r^2} + 1 + \frac{B}{{{r^{D - 3}}}} - \frac{C}{{{r^{2D - 4}}}},
\end{equation}
where $A<0$, $C>0$, and $D\ge 3$. The sign of $B$ is indefinite. The first derivative and second derivative of the effective potential are
\begin{equation}
\label{eq:Vprime}
V_{{\rm{eff}}}' = 2Ar - \left( {D - 3} \right)\frac{B}{{{r^{D - 2}}}} + \left( {2D - 4} \right)\frac{C}{{{r^{2D - 3}}}},
\end{equation}
and
\begin{equation}
\label{eq:appendixA}
V_{{\rm{eff}}}'' = 2A + \left( {D - 3} \right)\left( {D - 2} \right)\frac{B}{{{r^{D - 1}}}} - \left( {2D - 4} \right)\left( {2D - 3} \right)\frac{C}{{{r^{2D - 2}}}}.
\end{equation}

Let $r_0$ be a point where ${\left. {V_{{\rm{eff}}}'} \right|_{r = {r_0}}} = 0$. At such $r_0$, \eqref{eq:Vprime} requires
\begin{equation}
\left( {D - 3} \right)\left( {D - 2} \right)\frac{B}{{r_0^{D - 1}}} = 2A\left( {D - 2} \right) + \left( {2D - 4} \right)\left( {D - 2} \right)\frac{C}{{r_0^{2D - 2}}}.
\end{equation}
Substituting this into \eqref{eq:appendixA} yields
\begin{equation}
\begin{aligned}
{\left. {V_{{\rm{eff}}}''} \right|_{r = {r_0}}} &= 2A - \left( {2D - 4} \right)\left( {2D - 3} \right)\frac{C}{{{r_0^{2D - 2}}}} + 2A\left( {D - 2} \right) + \left( {2D - 4} \right)\left( {D - 2} \right)\frac{C}{{r_0^{2D - 2}}}\\
&= 2A\left( {D - 1} \right) - \left( {2D - 4} \right)\left( {D - 1} \right)\frac{C}{{{r_0^{2D - 2}}}}\\
&< 0.
\end{aligned}
\end{equation}
As a result, $V_{\rm eff}$ has no local minima.  And since it is large and negative at both large and small $r$, it can have at most two zeros.

\section{Unique zero of $V_\text{eff}$ for $A> 0$}
\label{app:uniquezero}

We now show that in cases where $V_\text{eff} \rightarrow +\infty$ at large $r$, $V_\text{eff}$ can be equal to zero at most once. In such cases, the effective potential has the form
\begin{equation}
{V_{{\rm{eff}}}} = A{r^2} + 1 + \frac{B}{{{r^{D - 3}}}} - \frac{C}{{{r^{2D - 4}}}},
\end{equation}
where $A>0$, $C>0$, and $D\ge 3$. The sign of $B$ is indefinite. The first derivative of the effective potential is again given by \eqref{eq:Vprime}.

Let $r_0$ be a zero of $V_{\rm{eff}}$, so that
\begin{equation}
- \frac{B}{{r_0^{D - 2}}} = A{r_0} + \frac{1}{{{r_0}}} - \frac{C}{{r_0^{2D - 3}}}.
\end{equation}
Substituting this into \eqref{eq:Vprime} yields
\begin{equation}
\begin{aligned}
{\left. {V_{{\rm{eff}}}'} \right|_{r = {r_0}}} &= 2A{r_0} + \left( {2D - 4} \right)\frac{C}{{r_0^{2D - 3}}} + \left( {D - 3} \right)A{r_0} + \frac{{D - 3}}{{{r_0}}} - \left( {D - 3} \right)\frac{C}{{r_0^{2D - 3}}}\\
&= \left( {D - 1} \right)A{r_0} + \frac{{D - 3}}{{{r_0}}} + \left( {D - 1} \right)\frac{C}{{r_0^{2D - 3}}}\\
&> 0.
\end{aligned}
\end{equation}
Thus $V_\text{eff}$ is increasing whenever it crosses zero. In particular, once it becomes positive it cannot return to zero.   As a result, it can cross zero at most once.

\section{Constraints on the size of single-wall bags of gold}
\label{subsec:bogssmall}

As stated in the introduction, it is of interest to understand whether the bags of gold we create can become large enough for the bulk quantum fields inside to have entropy comparable to the Bekenstein-Hawking entropy $S_{\rm BH}$.  While our analysis is not exhaustive, we present some results below which suggest that this is not possible with a single domain wall and $\mu_i=0$.

First, for $D\ge 4$ we show in section \ref{subsec:D4bounds} that $r_{\min }$ is bounded above by either $(2\mu_e)^{\frac{1}{D-3}}$ or $(2\mu_e)^{\frac{1}{D-2}}$, whichever is greater.  While the bounds are not particularly strong, they show that the bags of gold do not become arbitrarily large at fixed $\mu_e$, and the for $D\ge 4$ there are bounds that are independent of $\lambda, \kappa$.  We then show in section \ref{subsec:D3bounds} that for $D=3$ taking $r_{\min }$ large at fixed $\mu_e$ requires tuning $\lambda $ to $-(\kappa -1)^2$ and taking $\kappa > 4/3$.  As a result, any large $r_{\min }$ limit requires fine-tuning and, furthermore, is subject to an additional IR cutoff associated with the finite value of the internal ($i$) cosmological length scale $\ell_i = \ell_e /\sqrt{-\lambda}= \ell_e/(\kappa-1) < 3 \ell_e$.  As a slight aside to our present goals, for completeness section \ref{subsec:large3DBOGs} then verifies that under these constraints $D=3$ saddles do in fact exist that create bags of gold with arbitrarily large $r_{\min}$.

The above bounds suggest that at large $\mu_e $ it will be difficult to generate bags of gold containing e.g. radiation with entropy exceeding the Bekenstein-Hawking entropy $S_{\rm BH}$ without tuning some property of the radiation.  For example, if the radiation is thermal, one may need to take the temperature to be parametrically large.  This raises the possibility of introducing uncontrolled Planck scale physics, and also raises the possibility that gravitational back-reaction from the radiation will become important.    While we have not carried out an exhaustive analysis of the possible scenarios, after some investigation we have certainly not located high-entropy regimes that are free of such issues.

Now, one might also seek bags of gold with entropy greater than $S_{\rm BH}$ by looking at small black holes.  There $S_{\rm BH}$ becomes small, so even a moderately-sized bag-of-gold would have higher entropy.  However, for fixed $\kappa $, $\lambda $ and $D\ge 4$, analyzing \eqref{eq:potential} at small $\mu_e$ gives $r_{\min } \propto \mu_e^{\frac{1}{D-2}}$.  And creating a bag-of-gold requires $\alpha_e(r_{\min }) < 0$, which from \eqref{eq:betas} implies $\frac{\mu_e}{r_{\min }^{D-1}} \frac{1}{1+\lambda + \kappa^2} < 1$.  So for $D\ge 4$ bags of gold do not arise at small $\mu_e$ without fine-tuning $\lambda, \kappa$.  The analysis for $D=3$ is similar, though there the zero-mass BTZ black hole has $\mu_e =1$ and a calculation shows that $\alpha_e(r_{\min }) > 0$ for all $\lambda, \kappa$, and that without fine-tuning $\lambda, \kappa$ the smallest bag-of-gold will have BTZ mass parameter $m_e = \mu_e-1$ of order $1$ so that $S_{\rm BH}$ will not be small.

\subsection{Bounds on $r_{\min }$ for $D\ge 4$}
\label{subsec:D4bounds}

As stated in the introduction, it is of interest to understand whether the bags of gold we create can become large inside a black hole of fixed surface area (here, fixed $\mu_e$). One notion of this size is set by $r_{\min }$, which from \eqref{eq:metric} is in fact the {\it maximum} radius of any $S^{D-2}$ of spherical symmetry inside the black hole.  We may thus equivalently ask if $r_{\min }$ can be large at fixed $\mu_e$. It turns out that it cannot.  In particular, while the large $\mu_e$ limit just studied yields $r_{\min } \rightarrow \infty$, it also takes $\mu_e$ large.  In fact, as we now show, for bags of gold it is possible to bound $r_{\min }$ from above whenever $\mu_e$ is fixed.

Let us begin by noting that $V_{\rm eff}$ becomes negative at any given value of $r$ when we take $C$ large at fixed values of $A,B$.  As a result, since $r_{\min }$ is the minimum at which $V_{\rm eff}$ becomes non-negative, in this limit $r_{\min }$ becomes large.  So at least one part of our task is to show that this limit cannot occur at fixed $\mu_e$.  Since $C = \frac{\mu_e^2}{4\kappa^2}$, we will need to show that $\kappa$ is bounded away from zero.

To do so, consider the value $r_{\alpha e}$ where $\alpha_e(r)=0$. Since bags of gold require $\alpha_e(r_{\min })<0$ and from \eqref{eq:betas} we see that $\alpha_e(r) >0$ for $r < r_{\alpha e}$, bags of gold must have $r_{\min} > r_{\alpha e}$, and thus
\begin{equation}
\label{eq:alphabound}
r_{\min }^{-(D-1)} < r_{\alpha e}^{-(D-1)} = \frac{1 + \lambda + \kappa^2}{\mu_e},
\end{equation}
where the last step used \eqref{eq:betas} to solve for $r_{\alpha e}^{-(D-1)}$.

On the other hand, \eqref{eq:sqrt} implies
\begin{equation}
\label{eq:sqrtg}
r_{\min }^{-(D-1)}  > \frac{B}{2C} + \sqrt{\frac{B^2}{4C^2} + \frac{A}{C}}.
\end{equation}
Combining \eqref{eq:ratios}, \eqref{eq:alphabound}, and \eqref{eq:sqrtg}, then requires $\lambda > - \kappa^2 $.  But bags of gold arise only in case $(II)$, for which $-(\kappa-1)^2 > \lambda$. These two inequalities are compatible only for $\kappa > 1/2$.  For bags of gold we have thus succeed in showing
\begin{equation}
\label{eq:Cbound}
C = \frac{\mu_e^2}{4 \kappa^2} < \mu_e^2.
\end{equation}

Now, another way to make $V_{\rm eff}$ very negative would be to take $B$ large and negative holding fixed $A$ and $C$.  But case $(II)$ also requires $\lambda \ge - (\kappa^2 +1)$, which implies
\begin{equation}
\label{eq:Bbound}
-B = \frac{\mu_e}{2\kappa^2}\left(\kappa^2 -1 - \lambda\right) < \mu_e.
\end{equation}

For $D>3$ a strict bound on $r_{\min }$ now follows quickly.  Defining $r_*$ by dropping the (positive) $Ar^2$ term from $V_{\rm eff}$ and setting the result to zero yields
\begin{equation}
\label{eq:rs}
1 = - \frac{B}{r_*^{D-3}} + \frac{C}{r_*^{2D-4}},
\end{equation}
and also that we must have $r_{\min } < r_*$.   But depending on which of the two terms on the right-hand-side of \eqref{eq:rs} are greater, we must also have either
\begin{equation}
\label{eq:Dg4}
r_*^{D-3} < -2B < 2 \mu_e {\text{ or }} r_*^{2D-4} < 2C < 2\mu_e^2.
\end{equation}
Note that the latter condition also implies $r_{\min } < (2\mu_e)^{\frac{1}{D-2}}$, which is the bound quoted in the introduction to this appendix.

\subsection{Bounds on $r_{\min}$ for $D=3$}
\label{subsec:D3bounds}

The case $D=3$ requires special treatment, as the first inequality in \eqref{eq:Dg4} then does not constrain $r_*$. But we will nevertheless derive a bound from the requirement that the domain wall trajectory has no self-intersections.  

We first observe that for $D=3$ the effective potential $V_{\rm eff}$ becomes $r^2$ times a quadratic in $r^{-2}$.  Thus we may write
\begin{equation}
\label{eq:Vfactor}
V_{\rm eff} = Ar^2 \left(1 - \frac{r_{+}^2}{r^2}\right)\left(1 + \frac{r_{-}^2}{r^2}\right),
\end{equation}
with
\begin{equation}
\label{eq:rmD=3}
r^{-2}_{+} = \frac{(1+B) + \sqrt{(1+B)^2 + 4AC}}{2C},\text{ }
r^{-2}_{-} = -\frac{(1+B) - \sqrt{(1+B)^2 + 4AC}}{2C},
\end{equation}
where $r_+^{-2}>0$, $r_-^{-2} > 0$, and $r_+ = r_{\min }$.

Now, from \eqref{eq:rmD=3} it is natural to expect that $A \rightarrow 0$ is a necessary condition for $r_{\min } = r_+ \rightarrow \infty$. Let us first carefully argue that this is indeed the case.  For $1+B \ge 0$ it is manifest, as \eqref{eq:rmD=3} and \eqref{eq:Cbound} then require
\begin{equation}
\label{eq:A1b}
r_{\min }^{-2} = r_+^{-2} \ge \frac{\sqrt{(1+B)^2 + 4AC}}{2C} \ge \sqrt{\frac{A}{C}} \ge \frac{\sqrt{A}}{\mu_e}.
\end{equation}
For $1 + B < 0$, we instead first note that \eqref{eq:Bbound} now implies $|1+B| = -1 - B < \mu_e -1$ and then use the fact that for any positive real numbers $a$ and $b$ one finds
\begin{equation}
\sqrt{b^2 + a^2} - b > \frac{a^2}{3b} {\text{ for }} b \ge a>0
\end{equation}
and
\begin{equation}
\sqrt{b^2 + a^2} - b > \frac{a}{3} {\text{ for }} a \ge b>0
\end{equation}
to conclude that $r_+^{-2} > \frac{1}{3}\frac{A}{|1+B|} > \frac{1}{3}\frac{A}{\mu_e-1} $ or $r_+^{-2} > \frac{1}{3}\sqrt{\frac{A}{C}} > \frac{\sqrt{A}}{3\mu_e}$.  In all cases $r_{\min } \rightarrow \infty$  with $\mu_e$ bounded requires $A \rightarrow 0$, and at sufficiently large $r_{\min }$  (with $\mu_e$ fixed) one in fact finds
\begin{equation}
\label{eq:Abound}
A < \frac{3\mu_e}{r_{\min }^2} = \frac{3(r_h^2 +1)}{r_{\min}^2}.
\end{equation}
For future use we also note that \eqref{eq:rmD=3} yields
\begin{equation}
\frac{\mu_e^2}{r_{\min}^2} > \mu_e^2 \frac{1+B}{2C} = 4\kappa^2 (1+B),
\end{equation}
so, since $\kappa > \frac{1}{2}$, taking $r_{\min}\rightarrow \infty $ requires taking $1+B$ to zero or a negative value.

With the above observations in hand, we will use \eqref{eq:tEEOM} to study possible self-intersections.  Note that since $A$ is small at large $r_{\min }$, we may write
\begin{equation}
1 + \kappa^2 + \lambda = 2 \kappa (1 + O(A)),
\end{equation}
where the overall sign on the right is fixed by the requirement that we remain in case $(II)$. Since $\mu_e$ is fixed and $\kappa > \frac{1}{2}$ as described above, for all $r > r_{\min }$ we find
\begin{equation}
\alpha_e(r) = -r \left(1 + O(A)\right) \left(1 + O(r_{\min }^{-2})\right) = -r \left(1 + O(r_{\min }^{-2})\right) .
\end{equation}
Similarly, we have
\begin{equation}
f_e(r) = r^2  \left(1 + O(r_{\min }^{-2})\right).
\end{equation}
As a result, defining $\tilde r =r/ r_{\min } = r/r_+$ and choosing the sign appropriate to moving outward from $r_{\min }$ to $r=+\infty$ with increasing $t_{Ee}$ we see that \eqref{eq:tEEOM} yields
\begin{equation}
\label{eq:D3tEe}
\begin{aligned}
\frac{dt_{Ee}}{d \tilde r} & = \frac{1}{A^{1/2} r_+ \tilde r \sqrt{\tilde r^2-1}}  \frac{1}{\sqrt{1 + \frac{r_-^2}{r_+^2 \tilde r^2}}} \left(1 + O(A) + O(r_{\min }^{-2}) \right) \\ &>  \frac{1}{A^{1/2} r_+ \tilde r \sqrt{\tilde r^2-1}}  \frac{1}{\sqrt{1 + \frac{r_-^2}{r_+^2}}} \left(1 + O(r_{\min }^{-2}) \right),
\end{aligned}
\end{equation}
where in the last step we used the fact that ${1 + \frac{r_-^2}{r_+^2 \tilde r^2}}$ is a decreasing function of $\tilde r$ together with $\tilde r \ge 1$. Since self-intersections arise when $t_{Ee}(r = \infty) - t_{Ee}(r_{\min })$ exceeds half the period $\beta$ of Euclidean time, recalling that $\frac{d}{d \tilde r} \left(\text{arcsec} \left(\tilde r\right)\right) = \frac{1}{\tilde r \sqrt{\tilde r^2-1}}$, setting $\epsilon = A (1 + \frac{r_-^2}{r_+^2})$,  integrating \eqref{eq:D3tEe} with $t_{Ee} =0$ at $r_{\min }$, and forbidding self-intersections yields
\begin{equation}
\frac{\beta}{2} \ge t_{Ee}(r= \infty) \ge \frac{\pi}{2r_+\sqrt{\epsilon}}(1 + O(r_{\min }^{-2})),
\end{equation}
or
\begin{equation}
\label{eq:Aepsilon}
\frac{\pi^2}{\beta^2} (1 + O(r_{\min }^{-2})) < r_+^2 \epsilon = A\left(r_+^2 + r_-^2\right) = \sqrt{(1+B)^2 + 4AC},
\end{equation}
where in the last step follows from \eqref{eq:rmD=3}.

Since the BTZ period satisfies $\frac{\pi^2}{\beta^2} =\frac{\mu_e-1}{4}$, recalling that any limit where $r_{\min } \rightarrow \infty$ must have $A \rightarrow 0$ and  (since $A \rightarrow 0$ and $\kappa > \frac{1}{2}$ prevent $B$ from diverging) there must be a limit point $B_0$ of $B$ with $1+B_0 \le 0$, we may use \eqref{eq:Aepsilon} to write
\begin{equation}
\label{eq:Bepsilon}
\frac{\mu_e-1}{4}  \rightarrow -(1+B_0) = \left(1 - \frac{1}{\kappa}\right)\mu_e - 1,
\end{equation}
and thus
\begin{equation}
\label{eq:kappabound}
\kappa \ge \frac{4}{3}\frac{\mu_e}{\mu_e-1} > \frac{4}{3}.
\end{equation}
The condition $A \rightarrow 0$ and remaining in case $(II)$ then requires
\begin{equation}
\label{eq:lambdabound}
\lambda \rightarrow - (\kappa -1)^2 < -\frac{1}{9}.
\end{equation}
In particular, $\lambda$ is bounded away from zero and the internal ($i$) cosmological length scale must remain finite.

\subsection{With fine tuning, $D=3$ bags of gold can have large $r_{\min}$}
\label{subsec:large3DBOGs}

\begin{figure}[t]
\centerline{\includegraphics[width=0.45\textwidth]{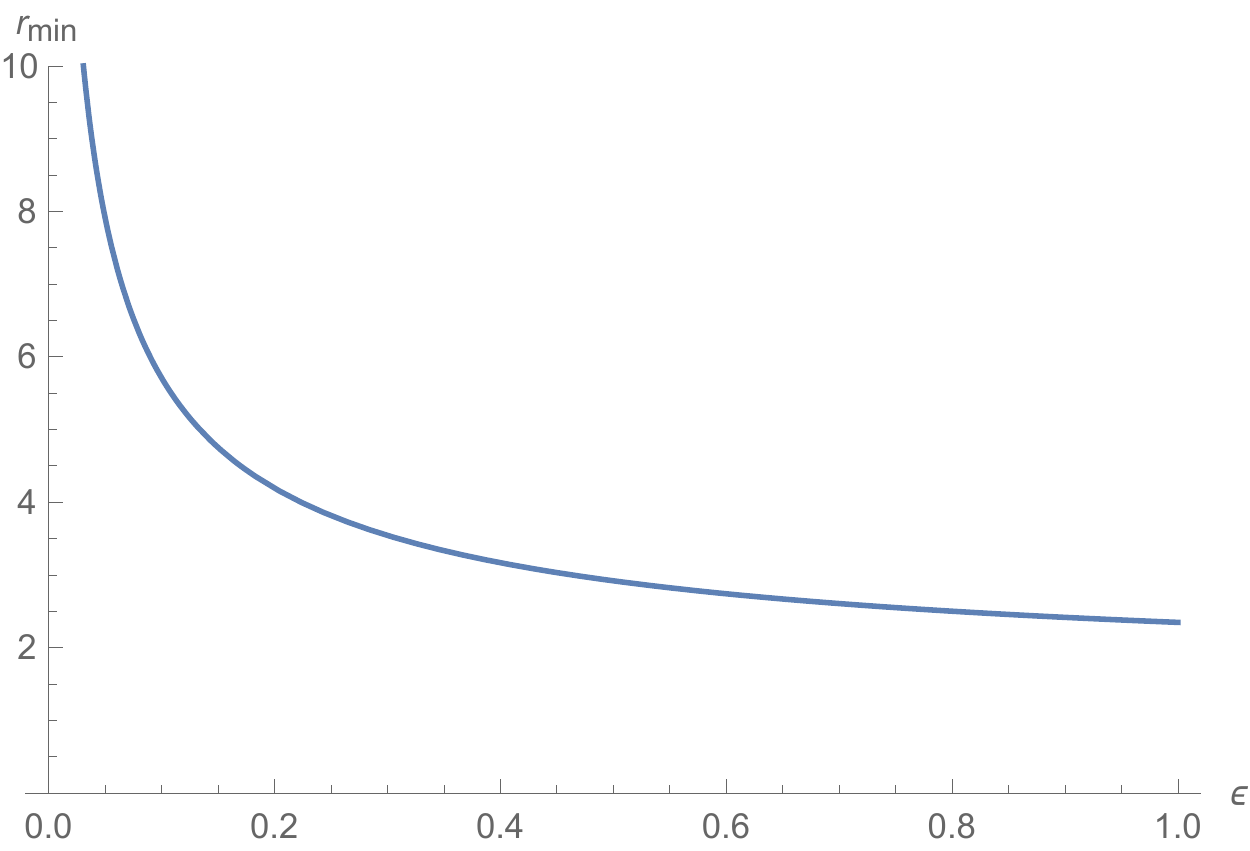} \hspace{.5cm} \includegraphics[width=0.45\textwidth]{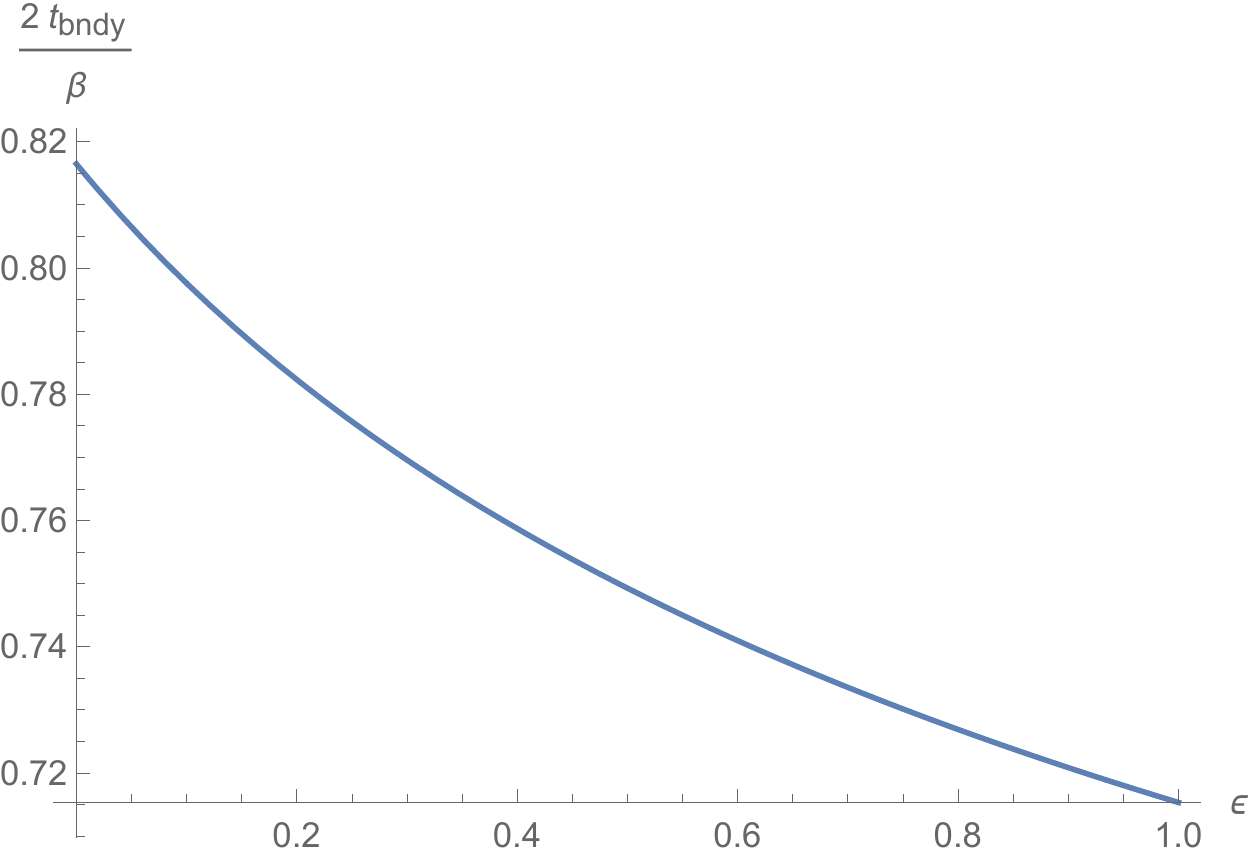}}
\caption{$D=3$ Numerical results for $r_{\min}$ and $\frac{2t_{Ee}(r=\infty)}{\beta_e}$ with $\lambda = -1 - \varepsilon$, $\varepsilon>0$, $\kappa =2$, and $\mu_e = 5$.  We see that $r_{\min }$ diverges as $\epsilon \rightarrow 0$.  Since $\frac{2t_{Ee}(r=\infty)}{\beta_e} <1$, the solutions are free of self-intersections.    These choices of parameters lie in case $(II)$, and we also find $\alpha_e(r_{\min}) <0$.  So as $\epsilon \rightarrow 0$ we find saddles that create bags of gold with arbitrarily large $r_{\min}$. }
\label{fig:D3LargeBOGs}
\end{figure}

The analysis above also allows us to readily show that in $D=3$ one can find bags of gold with arbitrarily large $r_{\min}$ at fixed $\mu_e$ by taking $A \rightarrow 0$ and $B \rightarrow B_0$ with $1+B_0 < 0$. To do so, note that \eqref{eq:rmD=3} shows that $r_{\min} = r_+$ becomes large in this limit with
\begin{equation}
\label{eq:rpmlim}
r_+^2 = - \frac{1+B_0}{A} \left((1 + O\left(\frac{A}{(1+B_0)^2} \right) \right).
\end{equation}
The first line of \eqref{eq:D3tEe} then yields
\begin{equation}
\label{eq:D3tEeworks}
\frac{dt_{Ee}}{d \tilde r}  < \frac{1}{A^{1/2} r_+ \tilde r \sqrt{\tilde r^2-1}}  \left(1 + O(A) + O(r_{\min }^{-2}) \right).
\end{equation}
We may then integrate this result as above to find
\begin{equation}
t_{Ee}(r= \infty) < \frac{\pi}{2r_{\min }\sqrt{A}}(1 + O(r_{\min }^{-2})) \rightarrow \frac{\pi}{\sqrt{-2(1+B_0)}}.
\end{equation}
So choosing the right-hand side to be less than $\frac{\beta}{2}$ makes our solution free of self-intersections.  In other words, we find saddles that create $D=3$ bags of gold with $r_{\min } \rightarrow \infty$ in any fixed-$\mu_e$ limit where $\lambda$ approaches $-(\kappa-1)^2$ from below and  the first inequality in \eqref{eq:kappabound} holds.  As a check, some numerical results for such large bags of gold are shown in figure \ref{fig:D3LargeBOGs}.

\section{Results for $\mu_i \neq \mu_e$, $\mu_i >0$}
\label{app:breakZ2}

We record here a few results concerning the case $\mu_i \neq \mu_e$, $\mu_i >0$.  These serve mostly to demonstrate the continuity at $\mu_i=\mu_e$, and to set the stage for future more detailed future investigations.

Recall that for SAdS regions with $\mu_e=\mu_i$ we find $\frac{2t_{Ee}(r=\infty)}{\beta} = f_D(\gamma)$ and $\frac{2t_{Ei}(r=\infty)}{\beta}= f_D(\tilde \gamma)$, with $f_3(\gamma)=\frac{1}{2}$ and $f_D(\gamma)$ decreasing monotonically from infinity at $\gamma=0$ to $\frac{1}{2}$ at $\gamma=1$.  This prohibits adding two such domain walls to a given SAdS region as in figure \ref{fig:twochoices} (left),  but the failure is marginal in $D=3$ and near $\gamma=1$ for all $D$.

In the main text we deal with this failure by adding magnetic charge.  However, it is also interesting to explore cases with $\mu_i \neq \mu_e$.  We do so briefly below.

For $\mu_i$ near $\mu_e$ and the qualitative form of the solutions for each domain wall will be similar to that discussed in section \ref{subsec:bogslarge}.  In particular, at $r=r_{\min }$ we will have negative $\alpha_e$ and positive $\alpha_i$, so the region between two adjacent domain walls will contain an SAdS horizon as desired.

\begin{figure}[t]
\centerline{\includegraphics[width=0.5\textwidth]{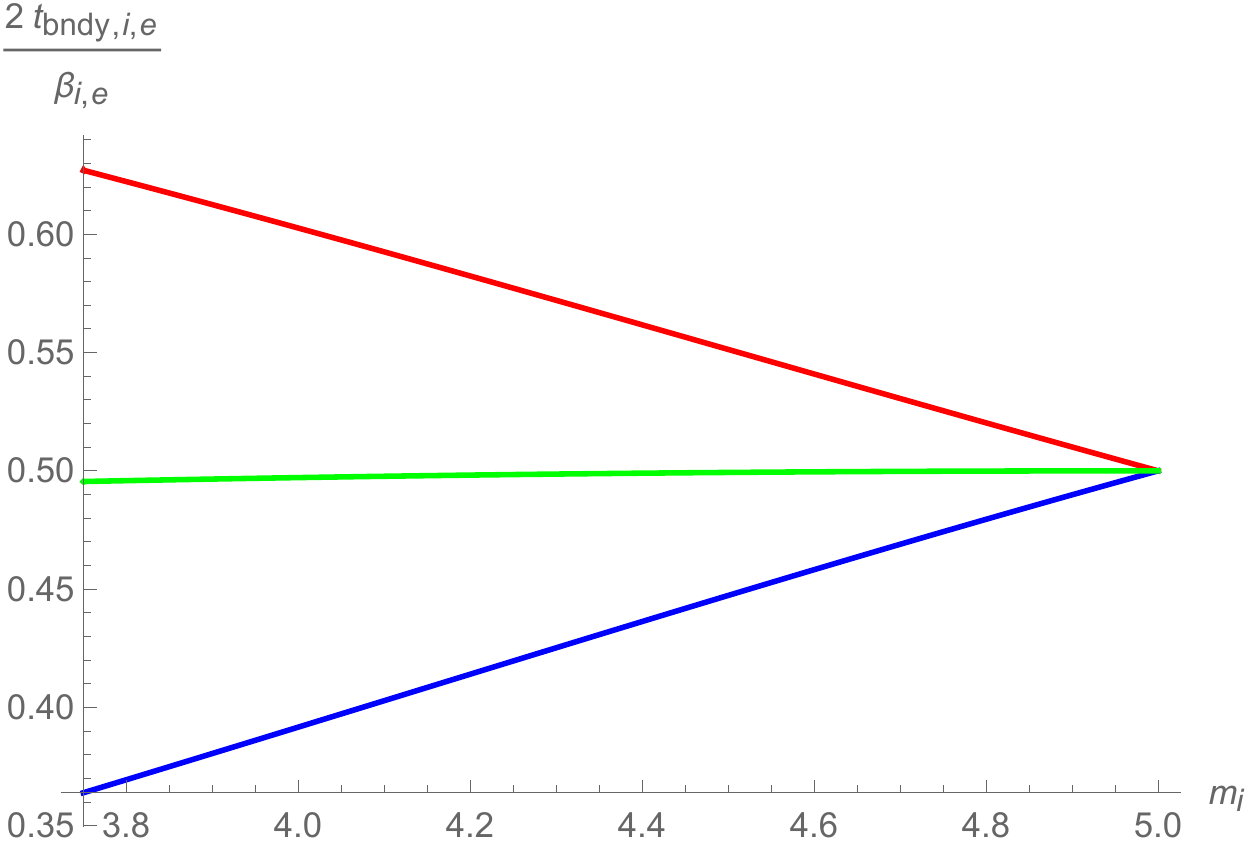}}
\caption{$D=3$ Numerical results for $\frac{2t_{Ee}(r=\infty)}{\beta_e}$ (red), $\frac{2t_{Ee}(r=\infty)}{\beta_e}$ (blue), and the average of the two (green) with $\lambda=-1$ and $\kappa=1/2$ as a function of $m_i=\mu_i-1$ for $m_e = \mu_e-1 = 5$.  In fact, the ratio depends only on $\frac{m_e}{m_i} = \frac{\mu_e-1}{\mu_i-1}$ for $\frac{m_e}{m_i} <1$.   Results for other values of $\kappa$ are similar. For $\lambda =-1$ the system is symmetric under exchanging the labels $e$ and $i$, so for $\frac{m_e}{m_i} >1$ one can instead find $\frac{2t_{Ee}(r=\infty)}{\beta_e}$ from the blue curve and $\frac{2t_{Ei}(r=\infty)}{\beta_i}$ from the red curve.  All ratios are $\frac{1}{2}$ and $\frac{m_e}{m_i}=1$, but the average (green) is smaller at other ratios. }
\label{fig:D3asym}
\end{figure}

The critical issue is then only the behavior of $\frac{2t_{Ee}(r=\infty)}{\beta_e}$, $\frac{2t_{Ei}(r=\infty)}{\beta_i}$ for each wall.  Let us begin with $D=3$.  Since both of these quantities are $\frac{1}{2}$ at $\mu_i = \mu_e$, for general $\mu_i,\mu_e$ it is natural to expect that one will be larger and the other smaller, corresponding to a non-zero first order term in the Taylor expansion about $\mu_i = \mu_e$.   For $D=3$ it is useful to introduce the parameters $m_e = \mu_e - 1$ and $m_i = \mu_i -1$ to match the standard conventions for BTZ black holes where empty AdS$_3$ has mass $m=-1$.  Having done so, for any $\lambda$ one may scale parameters much as in section \ref{subsec:nose} to show that $\frac{2t_{Ee}(r=\infty)}{\beta_e}$ depends only on the ratio $\tilde m = \frac{m_i}{m_e}$.  While we have not been able to obtain a closed form expression for $\frac{2t_{Ee}(r=\infty)}{\beta_e}$ as a function of $\tilde m$, for any $\lambda$ one may generalize the calculation \eqref{eq:3Dlimt} to arbitrary $\tilde m$ and expand in powers of $\tilde m-1$ to find integrals that can again be performed analytically\footnote{Most conveniently with help from Mathematica.}.  The results for $D=3$ may then be written in the form
\begin{equation}
\label{eq:3Dratio2}
\frac{2t_{Ee}(r=\infty)}{\beta_e} = \frac{1}{2} + \frac{1-\tilde m}{4\kappa } + \frac{3(1+\lambda +\kappa ^2)(1-\tilde m)^2}{32\kappa ^3} + O((1-\tilde m)^3).
\end{equation}
As expected, the linear term is non-zero and is positive for $\tilde m <1$; i.e., the ratio exceeds $\frac{1}{2}$ on the side of the domain wall with larger $\mu$.  Sample numerical results are also shown in figure \ref{fig:D3asym}.

Without invoking magnetic charge or otherwise altering our simple model, to find a region like that shown at left in figure \ref{fig:twochoices} having both two domain walls and a non-trivial AlAdS boundary the mass parameter $\mu$ for $D=3$  must thus behave monotonically as one passes from each SAdS region to the next.  The details of how $\mu$ evolves as we move inward are constrained only by the need only ensure that the decrease in $\frac{2t_{E}(r=\infty)}{\beta}$ below $\frac{1}{2}$ on the larger mass side more than compensates for the increase in $\frac{2t_{E}(r=\infty)}{\beta}$ above $\frac{1}{2}$ on the smaller mass side.
Note that using $1-\tilde m = (\tilde m^{-1} -1) - (\tilde m^{-1}-1)^2 + O(\tilde m^{-1}-1)^3$, one finds from \eqref{eq:3Dratio2} that the average of
\eqref{eq:3Dratio2} at $\frac{m_i}{m_e} = \tilde m$ and at $\frac{m_i}{m_e} = \tilde m^{-1}$ is
\begin{equation}
\label{eq:3Dratioav}
\begin{aligned}
& \frac{1}{2}\left(\frac{2t_{Ee}(r=\infty)}{\beta_e}|_{\frac{m_i}{m_e} = \tilde m} + \frac{2t_{Ee}(r=\infty)}{\beta_e}|_{\frac{m_i}{m_e} = \tilde m^{-1}} \right) \\
= & \frac{1}{2} - \frac{(5k^2 -3(1+\lambda))(1-\tilde m)^2}{32k^3} + O((1-\tilde m)^3).
\end{aligned}
\end{equation}
This average is thus less than $\frac{1}{2}$ for $\kappa^2 > 3(1+\lambda)$ and $\tilde m$ close to $1$; see also figure \ref{fig:D3asym} for a numerical check.  As a result, with parameters chosen as above it suffices to take all walls to have the same ratio $\frac{\mu_i}{\mu_e}$, with each SAdS region then serving as the interior $(i)$ for one wall and then as the exterior $(e)$ for the next.

Numerical computations for $D>4$ and $\lambda$ near $-1$ give similar results, though cases with $\mu_i \approx \mu_e$ can have values near $\frac{1}{2}$ only for small $\kappa$.  In general the results depend separately on $\mu_i$ and $\mu_e$, but in the limit of large $\mu_e$ they again depend only on the ratio $\frac{\mu_i}{\mu_e}$.  This can again be seen by repeating the sort of analysis described in section \ref{subsec:nose}, but is also simply a manifestation of the well-known facts that large $\mu$ SAdS black holes are effectively planar black holes, and that two planar black holes for differing mass parameters $\mu$ in fact have the same bulk geometry and differ only by the normalization of the time translation; i.e., in the limit $\mu \rightarrow \infty$ the geometries lose any notion of intrinsic scale.  In practice, as usual, this amounts to being about to drop the $1$ in $f_e$, $f_i$, and $V_{\rm eff}$.  In all dimensions, numerical results show that for two walls with the same mass ratio $\frac{\mu_i}{\mu_e} \approx 1$ at large $\mu$ the average of $\frac{2t_{Ee}(r=\infty)}{\beta_e}$ , $\frac{2t_{Ei}(r=\infty)}{\beta_i}$ is again less than the large $\mu$ value at $\frac{\mu_i}{\mu_e} =1$.  So, at least at small enough $\kappa$,  at large $\mu$ it again suffices to use the same ratio $\frac{\mu_i}{\mu_e}$ across each domain wall\footnote{Interestingly, there appears to be a sharp transition at $\kappa=1$ in all dimensions $D>3$.  For smaller values of $\kappa$, the average just described becomes numerically small -- and certainly smaller than $1/2$ -- when $\frac{\mu_i}{\mu_e} \rightarrow 0$ (or equivalently $\frac{\mu_i}{\mu_e} \rightarrow \infty$).  But this is not the case for $2>\kappa >1$.}.

As noted above, the mass parameter $\mu$ must behave monotonically as we go inward.  Since regularity at the center of the bag-of-gold requires $\mu=0$, one should thus expect that $\mu$ must decrease monotonically.   In fact, the case $\mu=0$ requires special treatment as (for $D>4$) it clearly is not large and since (for $D=3$) it corresponds to $m<0$. But numerically we do indeed find  $\frac{2t_{Ee}(r=\infty)}{\beta_e} > \frac{1}{2}$ in regions where our numerics is stable\footnote{In any case, in a chain with fixed ratio $\frac{\mu_i}{\mu_e} > 1$ the parameter $\mu$ will increase exponentially and quickly become very large.  As discussed in section \ref{subsec:nose}, one can then show analytically for $D=3$, $4$, and $5$ that this ratio is not less than $\frac{1}{2}$.}. In particular, we may take $\mu$ to decrease by an appropriate fixed factor $\frac{\mu_i}{\mu_e} <1$ as we move inward past each successive domain wall.

\bibliographystyle{jhep}
	\cleardoublepage

\renewcommand*{\bibname}{References}

\bibliography{biblio}

\providecommand{\href}[2]{#2}\begingroup\raggedright\begin{thebibliography}{10}

\bibitem{Jafferis:2015del}
D.~L. Jafferis, A.~Lewkowycz, J.~Maldacena and S.~J. Suh, \emph{{Relative
  entropy equals bulk relative entropy}},
  \href{http://dx.doi.org/10.1007/JHEP06(2016)004}{\emph{JHEP} {\bfseries 06}
  (2016) 004}, [\href{https://arxiv.org/abs/1512.06431}{{\ttfamily
  1512.06431}}].

\bibitem{Dong:2016eik}
X.~Dong, D.~Harlow and A.~C. Wall, \emph{{Reconstruction of Bulk Operators
  within the Entanglement Wedge in Gauge-Gravity Duality}},
  \href{http://dx.doi.org/10.1103/PhysRevLett.117.021601}{\emph{Phys. Rev.
  Lett.} {\bfseries 117} (2016) 021601},
  [\href{https://arxiv.org/abs/1601.05416}{{\ttfamily 1601.05416}}].

\bibitem{Faulkner:2017vdd}
T.~Faulkner and A.~Lewkowycz, \emph{{Bulk locality from modular flow}},
  \href{http://dx.doi.org/10.1007/JHEP07(2017)151}{\emph{JHEP} {\bfseries 07}
  (2017) 151}, [\href{https://arxiv.org/abs/1704.05464}{{\ttfamily
  1704.05464}}].

\bibitem{Wheeler}
J.~Wheeler, \emph{{Geometrodynamics and the Issue of the Final State}},  in
  \emph{Relativity, Groups, and Topology, 1963 Les Houches Lectures} (B.~S.
  DeWitt and C.~M. DeWitt, eds.).
\newblock Gordon and Breach, New York, 1964.

\bibitem{Marolf:2008tx}
D.~Marolf, \emph{{Black Holes, AdS, and CFTs}},
  \href{http://dx.doi.org/10.1007/s10714-008-0749-7}{\emph{Gen.Rel.Grav.}
  {\bfseries 41} (2009) 903--917},
  [\href{https://arxiv.org/abs/0810.4886}{{\ttfamily 0810.4886}}].

\bibitem{Hsu:2009kv}
S.~D.~H. Hsu and D.~Reeb, \emph{{Monsters, black holes and the statistical
  mechanics of gravity}},
  \href{http://dx.doi.org/10.1142/S0217732309031624}{\emph{Mod. Phys. Lett.}
  {\bfseries A24} (2009) 1875--1887},
  [\href{https://arxiv.org/abs/0908.1265}{{\ttfamily 0908.1265}}].

\bibitem{Almheiri:2018xdw}
A.~Almheiri, \emph{{Holographic Quantum Error Correction and the Projected
  Black Hole Interior}},  \href{https://arxiv.org/abs/1810.02055}{{\ttfamily
  1810.02055}}.

\bibitem{Freivogel:2005qh}
B.~Freivogel, V.~E. Hubeny, A.~Maloney, R.~C. Myers, M.~Rangamani and
  S.~Shenker, \emph{{Inflation in AdS/CFT}},
  \href{http://dx.doi.org/10.1088/1126-6708/2006/03/007}{\emph{JHEP} {\bfseries
  03} (2006) 007}, [\href{https://arxiv.org/abs/hep-th/0510046}{{\ttfamily
  hep-th/0510046}}].

\bibitem{Harlow:2014yka}
D.~Harlow, \emph{{Jerusalem Lectures on Black Holes and Quantum Information}},
  \href{http://dx.doi.org/10.1103/RevModPhys.88.015002}{\emph{Rev. Mod. Phys.}
  {\bfseries 88} (2016) 015002},
  [\href{https://arxiv.org/abs/1409.1231}{{\ttfamily 1409.1231}}].

\bibitem{Marolf:2017jkr}
D.~Marolf, \emph{{The Black Hole information problem: past, present, and
  future}}, \href{http://dx.doi.org/10.1088/1361-6633/aa77cc}{\emph{Rept. Prog.
  Phys.} {\bfseries 80} (2017) 092001},
  [\href{https://arxiv.org/abs/1703.02143}{{\ttfamily 1703.02143}}].

\bibitem{Penington:2019npb}
G.~Penington, \emph{{Entanglement Wedge Reconstruction and the Information
  Paradox}},  \href{https://arxiv.org/abs/1905.08255}{{\ttfamily 1905.08255}}.

\bibitem{Almheiri:2019psf}
A.~Almheiri, N.~Engelhardt, D.~Marolf and H.~Maxfield, \emph{{The entropy of
  bulk quantum fields and the entanglement wedge of an evaporating black
  hole}},  \href{https://arxiv.org/abs/1905.08762}{{\ttfamily 1905.08762}}.

\bibitem{Wald:1984}
R.~M. Wald, \emph{General Relativity}.
\newblock The University of Chicago Press, 1984.

\bibitem{Shenker:2013yza}
S.~H. Shenker and D.~Stanford, \emph{{Multiple Shocks}},
  \href{http://dx.doi.org/10.1007/JHEP12(2014)046}{\emph{JHEP} {\bfseries 12}
  (2014) 046}, [\href{https://arxiv.org/abs/1312.3296}{{\ttfamily 1312.3296}}].

\bibitem{Maldacena:2001kr}
J.~M. Maldacena, \emph{{Eternal black holes in anti-de Sitter}}, {\emph{JHEP}
  {\bfseries 0304} (2003) 021},
  [\href{https://arxiv.org/abs/hep-th/0106112}{{\ttfamily hep-th/0106112}}].

\bibitem{Krasnov:2000zq}
K.~Krasnov, \emph{{Holography and Riemann surfaces}},
  \href{http://dx.doi.org/10.4310/ATMP.2000.v4.n4.a5}{\emph{Adv. Theor. Math.
  Phys.} {\bfseries 4} (2000) 929--979},
  [\href{https://arxiv.org/abs/hep-th/0005106}{{\ttfamily hep-th/0005106}}].

\bibitem{Krasnov:2003ye}
K.~Krasnov, \emph{{Black hole thermodynamics and Riemann surfaces}},
  \href{http://dx.doi.org/10.1088/0264-9381/20/11/319}{\emph{Class. Quant.
  Grav.} {\bfseries 20} (2003) 2235--2250},
  [\href{https://arxiv.org/abs/gr-qc/0302073}{{\ttfamily gr-qc/0302073}}].

\bibitem{Balasubramanian:2014hda}
V.~Balasubramanian, P.~Hayden, A.~Maloney, D.~Marolf and S.~F. Ross,
  \emph{{Multiboundary Wormholes and Holographic Entanglement}},
  \href{http://dx.doi.org/10.1088/0264-9381/31/18/185015}{\emph{Class. Quant.
  Grav.} {\bfseries 31} (2014) 185015},
  [\href{https://arxiv.org/abs/1406.2663}{{\ttfamily 1406.2663}}].

\bibitem{Maxfield:2014kra}
H.~Maxfield, \emph{{Entanglement entropy in three dimensional gravity}},
  \href{http://dx.doi.org/10.1007/JHEP04(2015)031}{\emph{JHEP} {\bfseries 04}
  (2015) 031}, [\href{https://arxiv.org/abs/1412.0687}{{\ttfamily 1412.0687}}].

\bibitem{Marolf:2015vma}
D.~Marolf, H.~Maxfield, A.~Peach and S.~F. Ross, \emph{{Hot multiboundary
  wormholes from bipartite entanglement}},
  \href{http://dx.doi.org/10.1088/0264-9381/32/21/215006}{\emph{Class. Quant.
  Grav.} {\bfseries 32} (2015) 215006},
  [\href{https://arxiv.org/abs/1506.04128}{{\ttfamily 1506.04128}}].

\bibitem{Maxfield:2016mwh}
H.~Maxfield, S.~Ross and B.~Way, \emph{{Holographic partition functions and
  phases for higher genus Riemann surfaces}},
  \href{http://dx.doi.org/10.1088/0264-9381/33/12/125018}{\emph{Class. Quant.
  Grav.} {\bfseries 33} (2016) 125018},
  [\href{https://arxiv.org/abs/1601.00980}{{\ttfamily 1601.00980}}].

\bibitem{Marolf:2017shp}
D.~Marolf, M.~Rota and J.~Wien, \emph{{Handlebody phases and the polyhedrality
  of the holographic entropy cone}},
  \href{http://dx.doi.org/10.1007/JHEP10(2017)069}{\emph{JHEP} {\bfseries 10}
  (2017) 069}, [\href{https://arxiv.org/abs/1705.10736}{{\ttfamily
  1705.10736}}].

\bibitem{Kourkoulou:2017zaj}
I.~Kourkoulou and J.~Maldacena, \emph{{Pure states in the SYK model and
  nearly-$AdS_2$ gravity}},  \href{https://arxiv.org/abs/1707.02325}{{\ttfamily
  1707.02325}}.

\bibitem{Marolf:2017vsk}
D.~Marolf and J.~Wien, \emph{{The Torus Operator in Holography}},
  \href{http://dx.doi.org/10.1007/JHEP01(2018)105}{\emph{JHEP} {\bfseries 01}
  (2018) 105}, [\href{https://arxiv.org/abs/1708.03048}{{\ttfamily
  1708.03048}}].

\bibitem{Alberghi:1999kd}
G.~L. Alberghi, D.~A. Lowe and M.~Trodden, \emph{{Charged false vacuum bubbles
  and the AdS / CFT correspondence}},
  \href{http://dx.doi.org/10.1088/1126-6708/1999/07/020}{\emph{JHEP} {\bfseries
  07} (1999) 020}, [\href{https://arxiv.org/abs/hep-th/9906047}{{\ttfamily
  hep-th/9906047}}].

\bibitem{Banks:2000fe}
T.~Banks, \emph{{Cosmological breaking of supersymmetry?}},
  \href{http://dx.doi.org/10.1142/S0217751X01003998}{\emph{Int. J. Mod. Phys.}
  {\bfseries A16} (2001) 910--921},
  [\href{https://arxiv.org/abs/hep-th/0007146}{{\ttfamily hep-th/0007146}}].

\bibitem{Bousso:2000nf}
R.~Bousso, \emph{{Positive vacuum energy and the N bound}},
  \href{http://dx.doi.org/10.1088/1126-6708/2000/11/038}{\emph{JHEP} {\bfseries
  11} (2000) 038}, [\href{https://arxiv.org/abs/hep-th/0010252}{{\ttfamily
  hep-th/0010252}}].

\bibitem{Banks:2001yp}
T.~Banks and W.~Fischler, \emph{{M theory observables for cosmological
  space-times}},  \href{https://arxiv.org/abs/hep-th/0102077}{{\ttfamily
  hep-th/0102077}}.

\bibitem{Witten:2001kn}
E.~Witten, \emph{{Quantum gravity in de Sitter space}},  in \emph{{Strings
  2001: International Conference Mumbai, India, January 5-10, 2001}}, 2001.
\newblock \href{https://arxiv.org/abs/hep-th/0106109}{{\ttfamily
  hep-th/0106109}}.

\bibitem{Fischler:2001yj}
W.~Fischler, A.~Kashani-Poor, R.~McNees and S.~Paban, \emph{{The Acceleration
  of the universe, a challenge for string theory}},
  \href{http://dx.doi.org/10.1088/1126-6708/2001/07/003}{\emph{JHEP} {\bfseries
  07} (2001) 003}, [\href{https://arxiv.org/abs/hep-th/0104181}{{\ttfamily
  hep-th/0104181}}].

\bibitem{Hellerman:2001yi}
S.~Hellerman, N.~Kaloper and L.~Susskind, \emph{{String theory and
  quintessence}},
  \href{http://dx.doi.org/10.1088/1126-6708/2001/06/003}{\emph{JHEP} {\bfseries
  06} (2001) 003}, [\href{https://arxiv.org/abs/hep-th/0104180}{{\ttfamily
  hep-th/0104180}}].

\bibitem{Strominger:2001pn}
A.~Strominger, \emph{{The dS / CFT correspondence}},
  \href{http://dx.doi.org/10.1088/1126-6708/2001/10/034}{\emph{JHEP} {\bfseries
  10} (2001) 034}, [\href{https://arxiv.org/abs/hep-th/0106113}{{\ttfamily
  hep-th/0106113}}].

\bibitem{Strominger:2001gp}
A.~Strominger, \emph{{Inflation and the dS / CFT correspondence}},
  \href{http://dx.doi.org/10.1088/1126-6708/2001/11/049}{\emph{JHEP} {\bfseries
  11} (2001) 049}, [\href{https://arxiv.org/abs/hep-th/0110087}{{\ttfamily
  hep-th/0110087}}].

\bibitem{Alishahiha:2004md}
M.~Alishahiha, A.~Karch, E.~Silverstein and D.~Tong, \emph{{The dS/dS
  correspondence}}, \href{http://dx.doi.org/10.1063/1.1848341}{\emph{AIP Conf.
  Proc.} {\bfseries 743} (2004) 393--409},
  [\href{https://arxiv.org/abs/hep-th/0407125}{{\ttfamily hep-th/0407125}}].

\bibitem{Dong:2010pm}
X.~Dong, B.~Horn, E.~Silverstein and G.~Torroba, \emph{{Micromanaging de Sitter
  holography}},
  \href{http://dx.doi.org/10.1088/0264-9381/27/24/245020}{\emph{Class. Quant.
  Grav.} {\bfseries 27} (2010) 245020},
  [\href{https://arxiv.org/abs/1005.5403}{{\ttfamily 1005.5403}}].

\bibitem{Freivogel:2006xu}
B.~Freivogel, Y.~Sekino, L.~Susskind and C.-P. Yeh, \emph{{A Holographic
  framework for eternal inflation}},
  \href{http://dx.doi.org/10.1103/PhysRevD.74.086003}{\emph{Phys. Rev.}
  {\bfseries D74} (2006) 086003},
  [\href{https://arxiv.org/abs/hep-th/0606204}{{\ttfamily hep-th/0606204}}].

\bibitem{Cooper:2018cmb}
S.~Cooper, M.~Rozali, B.~Swingle, M.~Van~Raamsdonk, C.~Waddell and D.~Wakeham,
  \emph{{Black Hole Microstate Cosmology}},
  \href{http://dx.doi.org/10.1007/JHEP07(2019)065}{\emph{JHEP} {\bfseries 07}
  (2019) 065}, [\href{https://arxiv.org/abs/1810.10601}{{\ttfamily
  1810.10601}}].

\bibitem{Antonini:2019qkt}
S.~Antonini and B.~Swingle, \emph{{Cosmology at the end of the world}},
  \href{https://arxiv.org/abs/1907.06667}{{\ttfamily 1907.06667}}.

\bibitem{deAlwis:2019dkc}
S.~P. De~Alwis, F.~Muia, V.~Pasquarella and F.~Quevedo, \emph{{Quantum
  Transitions Between Minkowski and de Sitter Spacetimes}},
  \href{https://arxiv.org/abs/1909.01975}{{\ttfamily 1909.01975}}.

\bibitem{Jackiw:1984je}
R.~Jackiw, \emph{{Lower Dimensional Gravity}},
  \href{http://dx.doi.org/10.1016/0550-3213(85)90448-1}{\emph{Nucl. Phys.}
  {\bfseries B252} (1985) 343--356}.

\bibitem{Teitelboim:1983ux}
C.~Teitelboim, \emph{{Gravitation and Hamiltonian Structure in Two Space-Time
  Dimensions}},
  \href{http://dx.doi.org/10.1016/0370-2693(83)90012-6}{\emph{Phys. Lett.}
  {\bfseries 126B} (1983) 41--45}.

\bibitem{Saad:2018bqo}
P.~Saad, S.~H. Shenker and D.~Stanford, \emph{{A semiclassical ramp in SYK and
  in gravity}},  \href{https://arxiv.org/abs/1806.06840}{{\ttfamily
  1806.06840}}.

\bibitem{Saad:2019lba}
P.~Saad, S.~H. Shenker and D.~Stanford, \emph{{JT gravity as a matrix
  integral}},  \href{https://arxiv.org/abs/1903.11115}{{\ttfamily 1903.11115}}.

\bibitem{Maldacena:2004rf}
J.~M. Maldacena and L.~Maoz, \emph{{Wormholes in AdS}},
  \href{http://dx.doi.org/10.1088/1126-6708/2004/02/053}{\emph{JHEP} {\bfseries
  02} (2004) 053}, [\href{https://arxiv.org/abs/hep-th/0401024}{{\ttfamily
  hep-th/0401024}}].

\bibitem{Betzios:2019rds}
P.~Betzios, E.~Kiritsis and O.~Papadoulaki, \emph{{Euclidean Wormholes and
  Holography}}, \href{http://dx.doi.org/10.1007/JHEP06(2019)042}{\emph{JHEP}
  {\bfseries 06} (2019) 042},
  [\href{https://arxiv.org/abs/1903.05658}{{\ttfamily 1903.05658}}].

\bibitem{Coleman:1988cy}
S.~R. Coleman, \emph{{Black Holes as Red Herrings: Topological Fluctuations and
  the Loss of Quantum Coherence}},
  \href{http://dx.doi.org/10.1016/0550-3213(88)90110-1}{\emph{Nucl. Phys.}
  {\bfseries B307} (1988) 867--882}.

\bibitem{Giddings:1988cx}
S.~B. Giddings and A.~Strominger, \emph{{Loss of Incoherence and Determination
  of Coupling Constants in Quantum Gravity}},
  \href{http://dx.doi.org/10.1016/0550-3213(88)90109-5}{\emph{Nucl. Phys.}
  {\bfseries B307} (1988) 854--866}.

\bibitem{Giddings:1988wv}
S.~B. Giddings and A.~Strominger, \emph{{Baby Universes, Third Quantization and
  the Cosmological Constant}},
  \href{http://dx.doi.org/10.1016/0550-3213(89)90353-2}{\emph{Nucl. Phys.}
  {\bfseries B321} (1989) 481--508}.

\bibitem{Farhi:1989yr}
E.~Farhi, A.~H. Guth and J.~Guven, \emph{{Is It Possible to Create a Universe
  in the Laboratory by Quantum Tunneling?}},
  \href{http://dx.doi.org/10.1016/0550-3213(90)90357-J}{\emph{Nucl. Phys.}
  {\bfseries B339} (1990) 417--490}.

\bibitem{Fischler:1989se}
W.~Fischler, D.~Morgan and J.~Polchinski, \emph{{Quantum Nucleation of False
  Vacuum Bubbles}},
  \href{http://dx.doi.org/10.1103/PhysRevD.41.2638}{\emph{Phys. Rev.}
  {\bfseries D41} (1990) 2638}.

\bibitem{Fischler:1990pk}
W.~Fischler, D.~Morgan and J.~Polchinski, \emph{{Quantization of False Vacuum
  Bubbles: A Hamiltonian Treatment of Gravitational Tunneling}},
  \href{http://dx.doi.org/10.1103/PhysRevD.42.4042}{\emph{Phys. Rev.}
  {\bfseries D42} (1990) 4042--4055}.

\bibitem{Bachlechner:2016mtp}
T.~C. Bachlechner, \emph{{Inflation Expels Runaways}},
  \href{http://dx.doi.org/10.1007/JHEP12(2016)155}{\emph{JHEP} {\bfseries 12}
  (2016) 155}, [\href{https://arxiv.org/abs/1608.07576}{{\ttfamily
  1608.07576}}].

\bibitem{Marolf:2018ldl}
D.~Marolf, \emph{{Microcanonical Path Integrals and the Holography of small
  Black Hole Interiors}},
  \href{http://dx.doi.org/10.1007/JHEP09(2018)114}{\emph{JHEP} {\bfseries 09}
  (2018) 114}, [\href{https://arxiv.org/abs/1808.00394}{{\ttfamily
  1808.00394}}].

\bibitem{Coleman:1980aw}
S.~R. Coleman and F.~De~Luccia, \emph{{Gravitational Effects on and of Vacuum
  Decay}}, \href{http://dx.doi.org/10.1103/PhysRevD.21.3305}{\emph{Phys. Rev.}
  {\bfseries D21} (1980) 3305}.

\bibitem{Israel:1966rt}
W.~Israel, \emph{{Singular hypersurfaces and thin shells in general
  relativity}}, \href{http://dx.doi.org/10.1007/BF02710419,
  10.1007/BF02712210}{\emph{Nuovo Cim.} {\bfseries B44S10} (1966) 1}.

\bibitem{Blau:1986cw}
S.~K. Blau, E.~I. Guendelman and A.~H. Guth, \emph{{The Dynamics of False
  Vacuum Bubbles}},
  \href{http://dx.doi.org/10.1103/PhysRevD.35.1747}{\emph{Phys. Rev.}
  {\bfseries D35} (1987) 1747}.

\bibitem{Horowitz:1991fr}
G.~T. Horowitz, \emph{{Topology change in general relativity}},  in
  \emph{{Recent developments in theoretical and experimental general
  relativity, gravitation and relativistic field theories. Proceedings, 6th
  Marcel Grossmann Meeting, Kyoto, Japan, June 23-29, 1991. Pts. A, B}},
  pp.~1167--1181, 1991.
\newblock \href{https://arxiv.org/abs/hep-th/9109030}{{\ttfamily
  hep-th/9109030}}.

\bibitem{Marolf:1993ij}
D.~M. Marolf, \emph{{Pull back invariant matter couplings}},
  \href{http://dx.doi.org/10.1088/0264-9381/11/1/023}{\emph{Class. Quant.
  Grav.} {\bfseries 11} (1994) 239--251},
  [\href{https://arxiv.org/abs/gr-qc/9306007}{{\ttfamily gr-qc/9306007}}].

\bibitem{Toldo:2012ec}
C.~Toldo and S.~Vandoren, \emph{{Static nonextremal AdS4 black hole
  solutions}}, \href{http://dx.doi.org/10.1007/JHEP09(2012)048}{\emph{JHEP}
  {\bfseries 09} (2012) 048},
  [\href{https://arxiv.org/abs/1207.3014}{{\ttfamily 1207.3014}}].

\bibitem{Klemm:2012yg}
D.~Klemm and O.~Vaughan, \emph{{Nonextremal black holes in gauged supergravity
  and the real formulation of special geometry}},
  \href{http://dx.doi.org/10.1007/JHEP01(2013)053}{\emph{JHEP} {\bfseries 01}
  (2013) 053}, [\href{https://arxiv.org/abs/1207.2679}{{\ttfamily 1207.2679}}].

\bibitem{Maxfield:2014wea}
T.~Maxfield and S.~Sethi, \emph{{Domain Walls, Triples and Acceleration}},
  \href{http://dx.doi.org/10.1007/JHEP08(2014)066}{\emph{JHEP} {\bfseries 08}
  (2014) 066}, [\href{https://arxiv.org/abs/1404.2564}{{\ttfamily 1404.2564}}].

\bibitem{Louko:2004ej}
J.~Louko, R.~B. Mann and D.~Marolf, \emph{{Geons with spin and charge}},
  \href{http://dx.doi.org/10.1088/0264-9381/22/7/016}{\emph{Class. Quant.
  Grav.} {\bfseries 22} (2005) 1451--1468},
  [\href{https://arxiv.org/abs/gr-qc/0412012}{{\ttfamily gr-qc/0412012}}].

\bibitem{Yin:2007at}
X.~Yin, \emph{{On Non-handlebody Instantons in 3D Gravity}},
  \href{http://dx.doi.org/10.1088/1126-6708/2008/09/120}{\emph{JHEP} {\bfseries
  09} (2008) 120}, [\href{https://arxiv.org/abs/0711.2803}{{\ttfamily
  0711.2803}}].

\bibitem{Obied:2018sgi}
G.~Obied, H.~Ooguri, L.~Spodyneiko and C.~Vafa, \emph{{De Sitter Space and the
  Swampland}},  \href{https://arxiv.org/abs/1806.08362}{{\ttfamily
  1806.08362}}.

\bibitem{Kachru:2003aw}
S.~Kachru, R.~Kallosh, A.~D. Linde and S.~P. Trivedi, \emph{{De Sitter vacua in
  string theory}},
  \href{http://dx.doi.org/10.1103/PhysRevD.68.046005}{\emph{Phys.Rev.}
  {\bfseries D68} (2003) 046005},
  [\href{https://arxiv.org/abs/hep-th/0301240}{{\ttfamily hep-th/0301240}}].

\bibitem{Marolf:2017kvq}
D.~Marolf, O.~Parrikar, C.~Rabideau, A.~Izadi~Rad and M.~Van~Raamsdonk,
  \emph{{From Euclidean Sources to Lorentzian Spacetimes in Holographic
  Conformal Field Theories}},
  \href{http://dx.doi.org/10.1007/JHEP06(2018)077}{\emph{JHEP} {\bfseries 06}
  (2018) 077}, [\href{https://arxiv.org/abs/1709.10101}{{\ttfamily
  1709.10101}}].

\bibitem{Fu:2018kcp}
Z.~Fu, A.~Maloney, D.~Marolf, H.~Maxfield and Z.~Wang, \emph{{Holographic
  complexity is nonlocal}},
  \href{http://dx.doi.org/10.1007/JHEP02(2018)072}{\emph{JHEP} {\bfseries 02}
  (2018) 072}, [\href{https://arxiv.org/abs/1801.01137}{{\ttfamily
  1801.01137}}].

\bibitem{Brown:2015bva}
A.~R. Brown, D.~A. Roberts, L.~Susskind, B.~Swingle and Y.~Zhao,
  \emph{{Holographic Complexity Equals Bulk Action?}},
  \href{http://dx.doi.org/10.1103/PhysRevLett.116.191301}{\emph{Phys. Rev.
  Lett.} {\bfseries 116} (2016) 191301},
  [\href{https://arxiv.org/abs/1509.07876}{{\ttfamily 1509.07876}}].

\bibitem{Brown:2015lvg}
A.~R. Brown, D.~A. Roberts, L.~Susskind, B.~Swingle and Y.~Zhao,
  \emph{{Complexity, action, and black holes}},
  \href{http://dx.doi.org/10.1103/PhysRevD.93.086006}{\emph{Phys. Rev.}
  {\bfseries D93} (2016) 086006},
  [\href{https://arxiv.org/abs/1512.04993}{{\ttfamily 1512.04993}}].

\bibitem{Cotler:2016fpe}
J.~S. Cotler, G.~Gur-Ari, M.~Hanada, J.~Polchinski, P.~Saad, S.~H. Shenker
  et~al., \emph{{Black Holes and Random Matrices}},
  \href{http://dx.doi.org/10.1007/JHEP09(2018)002,
  10.1007/JHEP05(2017)118}{\emph{JHEP} {\bfseries 05} (2017) 118},
  [\href{https://arxiv.org/abs/1611.04650}{{\ttfamily 1611.04650}}].

\bibitem{Harlow:2018jwu}
D.~Harlow and H.~Ooguri, \emph{{Constraints on Symmetries from Holography}},
  \href{http://dx.doi.org/10.1103/PhysRevLett.122.191601}{\emph{Phys. Rev.
  Lett.} {\bfseries 122} (2019) 191601},
  [\href{https://arxiv.org/abs/1810.05337}{{\ttfamily 1810.05337}}].

\bibitem{ArkaniHamed:2006dz}
N.~Arkani-Hamed, L.~Motl, A.~Nicolis and C.~Vafa, \emph{{The String landscape,
  black holes and gravity as the weakest force}},
  \href{http://dx.doi.org/10.1088/1126-6708/2007/06/060}{\emph{JHEP} {\bfseries
  06} (2007) 060}, [\href{https://arxiv.org/abs/hep-th/0601001}{{\ttfamily
  hep-th/0601001}}].

\bibitem{Hartnoll:2008kx}
S.~A. Hartnoll, C.~P. Herzog and G.~T. Horowitz, \emph{{Holographic
  Superconductors}},
  \href{http://dx.doi.org/10.1088/1126-6708/2008/12/015}{\emph{JHEP} {\bfseries
  0812} (2008) 015}, [\href{https://arxiv.org/abs/0810.1563}{{\ttfamily
  0810.1563}}].

\bibitem{Chen:2019ror}
H.~Z. Chen and M.~Van~Raamsdonk, \emph{{Holographic CFT states for localized
  perturbations to AdS black holes}},
  \href{http://dx.doi.org/10.1007/JHEP08(2019)062}{\emph{JHEP} {\bfseries 08}
  (2019) 062}, [\href{https://arxiv.org/abs/1903.00972}{{\ttfamily
  1903.00972}}].

\bibitem{Haehl:2019fjz}
F.~M. Haehl, E.~Mintun, J.~Pollack, A.~J. Speranza and M.~Van~Raamsdonk,
  \emph{{Nonlocal multi-trace sources and bulk entanglement in holographic
  conformal field theories}},
  \href{http://dx.doi.org/10.1007/JHEP06(2019)005}{\emph{JHEP} {\bfseries 06}
  (2019) 005}, [\href{https://arxiv.org/abs/1904.01584}{{\ttfamily
  1904.01584}}].

\end{thebibliography}\endgroup

\end{document}